\pdfoutput=1
\documentclass[prd,onecolumn,notitlepage,eqsecnum,nofootinbib,floatfix,superscriptaddress]{revtex4-1}
\usepackage{amssymb}
\usepackage{amsmath}
\usepackage{amsfonts} 
\usepackage{bm}
\usepackage{color} 
\usepackage{graphicx}
\usepackage[bookmarks=false]{hyperref} 
\hypersetup{pdfstartview=FitH,pdfhighlight=/O,colorlinks=false}
 
\DeclareFontFamily{OT1}{rsfs}{} 
\DeclareFontShape{OT1}{rsfs}{m}{n}{<-7> rsfs5 
    <7-10> rsfs7 <10-> rsfs10}{}   
\DeclareMathAlphabet{\scr}{OT1}{rsfs}{m}{n} 
\newcommand{\dslash}{\delta^{\!\!\!\!-}\!} 
\newcommand{\V}{{\scr V}} 
\newcommand{\B}{{\cal B}} 
 
\newcommand{\C}{{\cal C}} 
\renewcommand{\S}{{\cal S}} 
\newcommand{\T}{{\cal T}} 
\newcommand{\N}{{\cal N}} 

\newcommand{\be}{\begin{equation}}
\newcommand{\ee}{\end{equation}}
\newcommand{\bea}{\begin{eqnarray}}
\newcommand{\eea}{\end{eqnarray}}
\newcommand{\beq}{\begin{equation}}
\newcommand{\eeq}{\end{equation}}
\newcommand{\beqa}{\begin{eqnarray}}
\newcommand{\eeqa}{\end{eqnarray}}
\newcommand{\beqar}{\begin{eqnarray*}}
\newcommand{\eeqar}{\end{eqnarray*}}

\newcommand{\labell}[1]{\label{#1}} 

\newcommand{\eg}{{\it e.g.,}\ } 
\newcommand{\ie}{{\it i.e.,}\ }
\newcommand{\reef}[1]{(\ref{#1})}

\newcommand{\mt}[1]{\textrm{\tiny #1}}

\allowdisplaybreaks
\begin{document}
\title{Gravitational action with null boundaries}
\author{Luis Lehner}
\author{Robert C.\ Myers}
\affiliation{Perimeter Institute for Theoretical Physics, Waterloo,
  Ontario N2L 2Y5, Canada}
\author{Eric Poisson}
\affiliation{Department of Physics, University of Guelph, Guelph,
  Ontario N1G 2W1, Canada} 
\author{Rafael D.\ Sorkin}
\affiliation{Perimeter Institute for Theoretical Physics, Waterloo,
  Ontario N2L 2Y5, Canada} 

\begin{abstract}
We present a complete discussion of the boundary term in the action
functional of general relativity when the boundary includes null
segments in addition to the more usual timelike and spacelike
segments. We confirm that ambiguities appear in the contribution from
a null segment, because it depends on an arbitrary choice of
parametrization for the generators. We also show that similar
ambiguities appear in the contribution from a codimension-two surface
at which a null segment is joined to another (spacelike, timelike, or
null) segment. The parametrization ambiguity can be tamed by insisting 
that the null generators be affinely parametrized; this forces each
null contribution to the boundary action to vanish, but leaves intact
the fredom to rescale the affine parameter by a constant factor on
each generator.  Once a choice of parametrization is made, the
ambiguity in the joint contributions can be eliminated by formulating
well-motivated rules that ensure the additivity of the gravitational
action. Enforcing these rules, we calculate the time rate of change of
the action when it is evaluated for a so-called ``Wheeler-deWitt
patch'' of a black hole in asymptotically-anti de Sitter space. We
recover a number of results cited in the literature, obtained with a
less complete analysis.    
\end{abstract}

\maketitle
\tableofcontents

\section{Introduction and summary 
\labell{sec:intro} }

The action functional for the gravitational field in general
relativity, the famous Hilbert-Einstein action, is given simply (in
the absence of a cosmological constant) by the spacetime integral of
the Ricci scalar. But it has long been recognized that a well-defined  
variational principle for a finite domain of spacetime must also
involve a contribution from the domain's
boundary~\cite{PhysRevLett.28.1082, PhysRevD.15.2752}. In the typical
context in which the boundary consists of timelike and spacelike
hypersurfaces, the boundary action is given by the surface integral of 
the trace of the extrinsic curvature. When the intersection between
two segments of the boundary is not smooth, the
extrinsic curvature is singular and the boundary action acquires
additional contributions from the
intersection~\cite{JR,Hayward:1993my}. While all this is well-known,
the case in which the boundary includes segments of null hypersurfaces has 
received very little attention in the literature. Indeed, to our
knowledge the contribution of a null boundary to the gravitational
action has only been examined recently in   
Refs.~\cite{Neiman:2012fx} and \cite{Parattu:2015gga}; the 
second reference, in particular, offers a detailed account of the
variational principle of general relativity in the presence of null
boundaries. But these works do not consider the contribution to the 
gravitational action coming from a nonsmooth intersection between a
null segment of the boundary with another (spacelike, timelike, or 
null) segment. It appeared important to us to fill this gap, and to
provide a complete account of the boundary term in the action
functional of general relativity when the boundary includes null
segments, in addition to the more usual timelike and spacelike
segments. 

The desire for completeness was not the sole motivation for
undertaking this work. We were also motivated by a desire to better  
understand the calculations supporting the recent ``complexity equals
action'' conjecture of Brown et al \cite{Brown:2015bva, Brown:2015lvg}, which
was made in the context of the AdS/CFT correspondence
\cite{revue}. This proposal emerged from previous studies attempting
to understand the growth of the Einstein-Rosen bridge for AdS black
holes in terms of circuit complexity in the dual boundary CFT
\cite{Susskind:2014rva,Stanford:2014jda,Susskind:2014jwa,Susskind:2016tae}. As
we will describe, the calculations on the gravity side which support
this conjecture rely in an essential way on evaluating the
gravitational action for regions with null boundaries. On the CFT
side, the conjecture considers the complexity $C$ of the quantum state
$|\psi(t)\rangle$ on a particular time slice of the boundary conformal
field theory. Loosely, we may think of $C$ as the minimum number of
quantum gates required to produce $|\psi\rangle$ 
from a particular reference state --- see \cite{Brown:2015lvg} for further details. 
The conjecture then relates $C$ to the gravitational action
$I$ evaluated for a corresponding region in the dual (asymptotically)
anti-de Sitter spacetime, known as a ``Wheeler-deWitt (WdW)
patch.'' The WdW patch is the region enclosed by past and future light
sheets sent into the bulk spacetime from the time slice on the
boundary, where $C$ is to be evaluated. An example is illustrated in 
Fig.~\ref{fig:WdW0}, and the conjecture states that 
$C = I/(\pi \hbar)$.

A particularly interesting case in which to examine this conjecture is that of
an eternal black hole in anti de Sitter space
\cite{Brown:2015bva, Brown:2015lvg}. In this case, the quantum state
$|\psi(t_{\rm L}, t_{\rm R})\rangle$ depends on two times $t_{\rm L}$
and $t_{\rm R}$, \ie the time on each of the asymptotic boundaries on
either side of the Einstein-Rosen bridge (the left and right
boundaries in a conformal diagram).\footnote{The two boundaries of an
  eternal AdS black hole correspond to the original CFT and its
  thermofield double, and the bulk geometry is then dual to a
  purification of a thermal density matrix involving these two CFTs
  \cite{eternal}.} The corresponding WdW patch is displayed in 
Fig.~\ref{fig:WdW0}. In part, the ``complexity equals
action'' conjecture was motivated by the expectation that
the complexity in this situation should increase linearly in time (for
a very long initial period), and the observation that this property is
shared by the action of the Wheeler-deWitt patch. In particular, it
was found that at late times \cite{Brown:2015bva, Brown:2015lvg} 
\begin{equation} 
\frac{dI}{dt} = 2\, M 
\labell{dAdt} 
\end{equation} 
for a Schwarzschild-AdS black hole, where $M$ is the total
mass-energy of the spacetime, and $t$ stands for one of the boundary
times (\ie $t_{\rm L}$ or $t_{\rm R}$), with the other time being held
fixed. Similar results for other spacetimes, all indicating that $I$ 
increases linearly with $t$ at late times, were reported by Brown et
al in support of the conjecture. 

\begin{figure} 
\includegraphics[width=.5\linewidth]{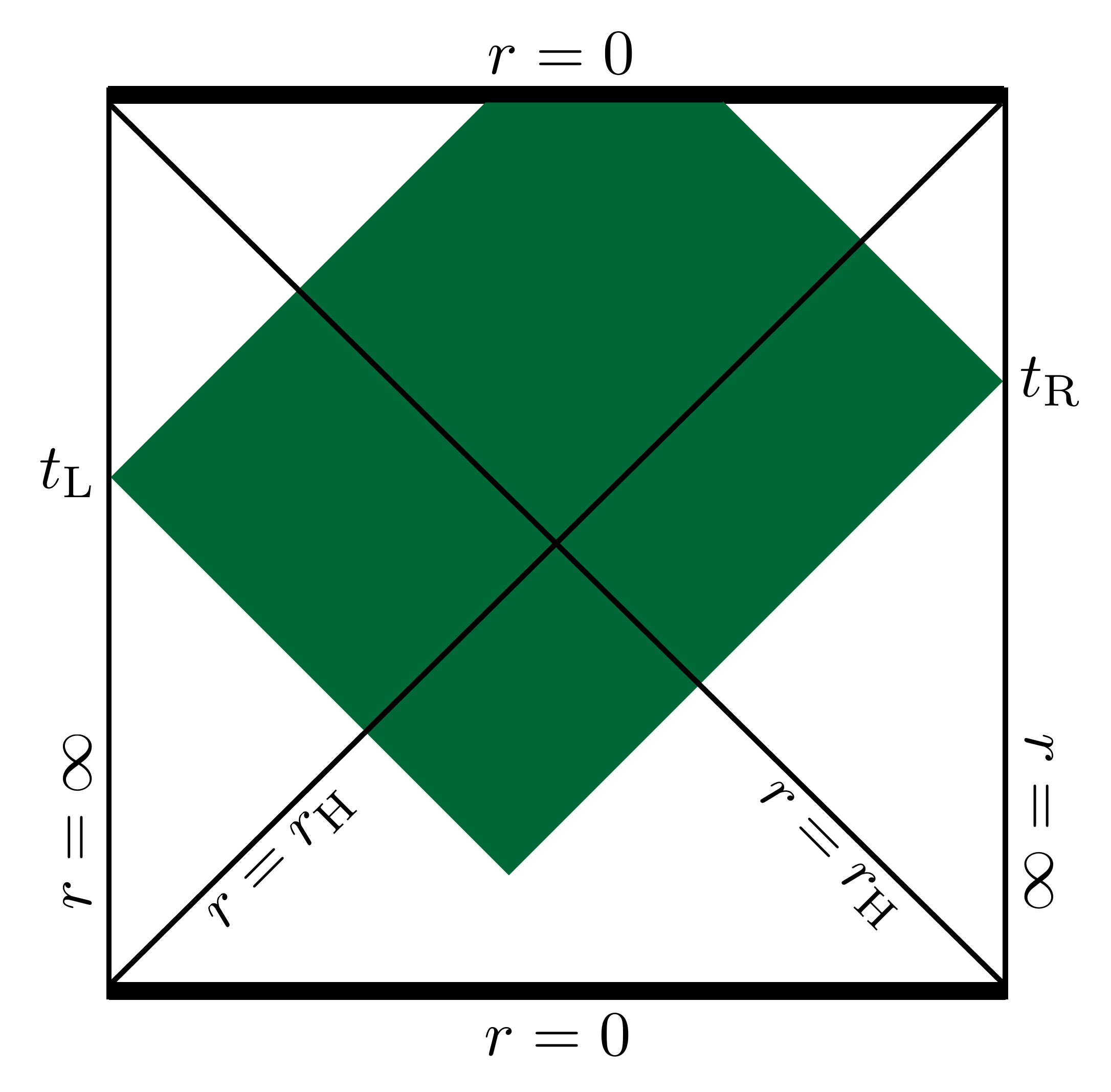}
\caption{Wheeler-deWitt patch of an eternal Schwarzschild-anti de
  Sitter black hole. The patch is defined by a future light cone
  originating inside the white-hole horizon and reaching the left
  boundary at time $t_{\rm L}$ and the right boundary at time 
  $t_{\rm R}$. This light cone is joined to a past light cone
  converging to the future singularity.}    
\labell{fig:WdW0} 
\end{figure} 

One of our main purposes in this paper is to critically examine how
Eq.~(\ref{dAdt}) was obtained: We question the methods by which
$dI/dt$ was calculated in \cite{Brown:2015bva, Brown:2015lvg}, we
identify what we take to be a more rigorous approach, and we
recalculate $dI/dt$ according to these methods. We perform the
calculations for both a Schwarzschild and 
Reissner-Norstr\"om black hole in anti-de Sitter space, and we find that
our results for $dI/dt$ precisely agree with those reported by Brown et al
\cite{Brown:2015bva, Brown:2015lvg}. This agreement, in spite of the very different
methods used in the calculation, may seem at first sight a surprising outcome. The
mechanism behind the agreement will be discussed in detail in
Sec.~\ref{sec:comp}. 

There are two main reasons to suspect the methods adopted by
Brown et al, and hence to be skeptical of their results for
$dI/dt$. First, the Wheeler-deWitt patch has a boundary that includes
segments of null hypersurfaces, and the familiar boundary term in the
gravitational action (the Gibbons-Hawking-York $K$ term 
\cite{PhysRevLett.28.1082,PhysRevD.15.2752}) is ill-defined for such
segments; it applies only to spacelike or timelike segments of the
boundary. One might attempt to evade this problem by evaluating the
boundary contribution for a null segment by approaching the
hypersurface through a sequence of timelike or spacelike
surfaces. However, as we will show in Appendix \ref{sec:limit}, in
general this limiting procedure is ambiguous and does not yield a
unique answer. The second key problem is that the 
boundary of the Wheeler-deWitt patch also includes codimension-two
surfaces --- joints --- at which different boundary surfaces
intersect; for example, in Fig.~\ref{fig:WdW0}, the spacelike portion of the
boundary at (or rather near) $r=0$ is joined to null segments
extending towards the two asymptotic AdS regions. Because the boundary
is not smooth at such joints, we should expect them to make separate
contributions to the gravitational action, in spite of the higher
codimensionality of these surfaces. Indeed, that nonsmooth portions of
the boundary contribute to the action was demonstrated by Hayward
\cite{Hayward:1993my} in the case of joints between timelike and spacelike
surfaces.\footnote{Similar joint contributions were found for the
  Regge calculus action with Euclidean signature in \cite{JR}.}
One might attempt to define the joint contributions for an
intersection involving a null segment by applying a limiting procedure
to the Hayward terms, but as we show in Appendix \ref{sec:limit}, this
yields a divergent result. 

The first issue, of correctly assigning a boundary contribution to the
gravitational action when the boundary includes a null segment, was
recently examined by Neiman~\cite{Neiman:2012fx} and given a
much more thorough analysis by Parattu et al~\cite{Parattu:2015gga}. 
The correct boundary term is identified by a careful consideration of
the variational principle for general relativity, which keeps track of
all terms that are pushed to the boundary when an integration by parts
is carried out. In the case of a timelike or spacelike segment, this
exercise reveals the Gibbons-Hawking-York $K$ term, \ie the trace of the
extrinsic curvature integrated over the boundary segment \cite{PhysRevLett.28.1082,PhysRevD.15.2752}. In the case
of a null segment, Parattu et al show that the boundary term is given
by an integral of the form\footnote{These authors also include a term involving
the expansion $\Theta$ of the null generators. As we discuss below, this term
is not required because it depends only on the surface's intrinsic geometry.} $\int \kappa\, dS d\lambda$, in which
$\lambda$ is the parameter running on the null generators of the
hypersurface, $dS$ is an area element on the cross-sections 
$\lambda = \mbox{constant}$, and $\kappa(\lambda)$ measures the
failure of $\lambda$ to be an affine parameter\footnote{This can 
also be interpreted as (a component of) the extrinsic curvature of the null 
segment~\cite{PhysRevD.62.104025, Gomez:2000kn}.}: if the vector field 
$k^\alpha$ is tangent to the null generators, then 
$k^\beta \nabla_\beta k^\alpha = \kappa\, k^\alpha$. This expression for  
the boundary term reveals a striking fact: its value depends on the 
parametrization of the null generators, and it can be altered at will
by a change of parametrization. For example, the boundary term
vanishes when $\lambda$ is chosen to be an affine parameter. This
observation implies that in general, the gravitational action is 
ambiguous when it is evaluated for a region of spacetime that is
bounded in part by a segment of null hypersurface.

The second issue, the proper accounting of contributions from joints, 
was examined by Hayward \cite{Hayward:1993my} in the context of
timelike and spacelike surfaces, but his treatment does not apply to
null surfaces. We consider such situations in this paper, and evaluate
the contribution of null joints to the gravitational action. We recall
that Hayward's conclusion was that when (say) two spacelike segments 
are joined together, the boundary term in the action acquires a
contribution of the form $\int \eta\, dS$, where $\eta$ is the
rapidity parameter relating the two unit normals by a Lorentz
transformation, and $dS$ is a surface element on the joint. On the
other hand, when a spacelike, timelike, or null segment is joined to a
null segment of the boundary, we find below that the contribution to
the boundary action is of the similar form $\int a\, dS$, where $a$ is 
a quantity tied to the description of the null hypersurface.\footnote{As this paper was nearing completion, we learned that similar junction terms were proposed in \cite{Booth:2001gx}.} More 
precisely, if the null segment is described by the equation 
$\Phi = 0$, with $\Phi$ a scalar function in the spacetime, and if its null
normal is given by $k_\alpha = -\mu \partial_\alpha \Phi$ in the
adopted parametrization, with $\mu$ another scalar, then 
$a = \ln\mu$. This contribution to the action is also ambiguous,
because $a$ can be changed at will by a redefinition of the function
$\Phi$.

These observations imply that the computation of the gravitational
action for a Wheeler-deWitt patch is plagued with ambiguities: The 
contribution to the action from each null segment of the boundary
depends arbitrarily on the choice of parametrization for the
generators, and the contribution from each joint between a null 
segment and another (spacelike, timelike, or null) segment is also
arbitrary. These ambiguities may seem to be problematic for the
``complexity equals action" conjecture, but we note that the
complexity is also expected to be ambiguous --- we return to
this point in Sec.~\ref{discuss}. In any event, at a pragmatic
level, the ambiguities must be tamed before $I$ can be computed and 
featured in a critical examination of the conjecture. Let us add that the
ambiguities apply only to the gravitational action evaluated for a
given region of a given spacetime --- the on-shell action; they are
evaded when the action is varied in an implementation of the
variational principle for general relativity.

To see in more concrete terms what the gravitational action looks  
like when it is evaluated for a region $\V$ of spacetime whose
boundary $\partial \V$ is broken up into a number of segments, we
consider (in some fixed spacetime) the region illustrated in
Fig.~\ref{fig:V+dV}. The boundary includes four spacelike segments, 
four null segments, and the joints between them. For this region,
the gravitational action takes the form of 
\begin{equation} 
S := 16\pi G_\mt{N}\, I = \int_{\V} (R-2\Lambda)\, dV + S_{\partial \V}, 
\labell{equation} 
\end{equation} 
where $G_\mt{N}$ is Newton's gravitational constant, $R$ the Ricci
scalar, $\Lambda$ the cosmological constant, $dV$ an invariant volume
element in $\V$, and where the boundary term is given explicitly given by 
\begin{align} 
S_{\partial \V} &= 2 \int_{\S_1} K\, d\Sigma 
+ 2 \int_{\S_2} K\, d\Sigma  
- 2 \int_{\S_3} K\, d\Sigma 
- 2 \int_{\S_4} K\, d\Sigma 
\nonumber \\ & \quad 
+ 2 \int_{\N_1} \kappa\, dS d\lambda 
+ 2 \int_{\bar{\N}_2} \kappa\, dS d\lambda 
- 2 \int_{\bar{\N}_3} \kappa\, dS d\lambda 
- 2 \int_{\N_4} \kappa\, dS d\lambda 
\nonumber \\ & \quad 
+ 2 \oint_{\B_{11}} a\, dS 
- 2 \oint_{\B_{12}} \eta\, dS 
+ 2 \oint_{\B_{22}} a\, dS 
- 2 \oint_{\B_{24}} a\, dS  
\nonumber \\ & \quad 
+ 2 \oint_{\B_{44}} a\, dS 
+ 2 \oint_{\B_{34}} \eta\, dS 
+ 2 \oint_{\B_{33}} a\, dS 
- 2 \oint_{\B_{13}} a\, dS, 
\end{align} 
in terms of quantities introduced previously. The sign in front of
each integral will be explained in the technical sections of the
paper. We recall that the contribution from each null segment is
ill-defined because it depends on the choice of $\lambda$ (which
implies a choice of $\kappa$), and that except for $\B_{12}$ and
$\B_{34}$, the contribution from each joint is also ill-defined
because of the freedom to redefine $a$.  
  
\begin{figure} 
\includegraphics[width=.7\linewidth]{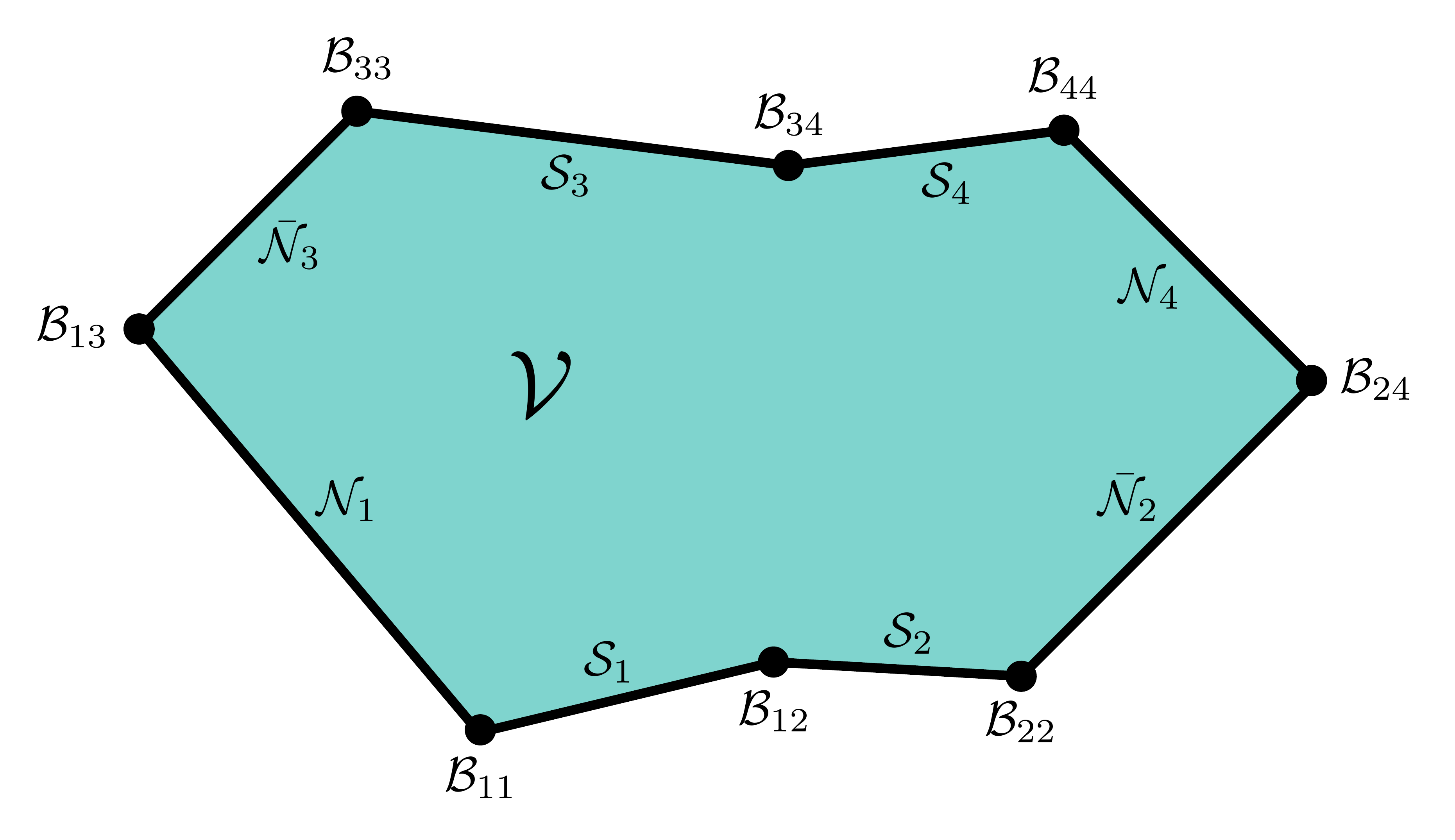}
\caption{A region $\V$ of spacetime with its broken boundary 
  $\partial \V$. The boundary consists of four spacelike segments
  $\S_1$, $\S_2$, $\S_3$, $\S_4$, and four null segments $\N_1$,
  $\bar{\N}_2$, $\bar{\N}_3$, $\N_4$. These are joined at
  codimension-two surfaces denoted $\B_{jk}$.     
\labell{fig:V+dV} }
\end{figure} 

The gravitational action only becomes well-defined when rules are
introduced to specify $\kappa(\lambda)$ on each null segment, and $a$
on each null joint. An attractive choice of parametrization suggests
itself: By ensuring that the generators of each null segment are
affinely parametrized, we can set $\kappa = 0$ and eliminate all such 
contributions to $S_{\partial \V}$. A well-motivated rule to assign
$a$ on each null joint emerges when the influence of these joints on
the variational principle is duly considered, and when the on-shell
gravitational action is required to be properly
additive, in the sense that $S(\V)=S(\V_1)+S(\V_2)$ when a spacetime
region $\V$ is subdivided into two subregions, $\V_1$ and $\V_2$. 
\footnote{Additivity beyond the on-shell action, for example
  in the context of a path integral for quantum gravity, cannot be
  guaranteed. This is because the metric may not be sufficiently
  smooth across a null boundary to ensure that $\lambda$ is an affine
  parameter on both sides of the boundary. A further obstruction to
  addivity would arise from an extra {\it imaginary} contribution to
  the action, which has been argued to exist
  in~\cite{Sorkin:1974pya,Sorkin:1975ah,1997CQGra..14..179L} -- see also
  \cite{Farhi:1989yr,Neiman:2012fx,Neiman:2013ap,Neiman:2013lxa,Neiman:2013taa}. 
  However, this contribution is typically neglected and we do so here as
  well.} 
Our rule goes as follows: When a null segment is joined to a spacelike
or timelike surface, we set $a = \ln|n \cdot k|$, where $n^\alpha$ is
the unit normal to the spacelike or timelike surface, $k^\alpha$ is
the normal to the null hypersurface, and  $n \cdot k :=
g_{\alpha\beta} n^\alpha k^\beta$ is 
their inner product; when a null segment is joined to another null
segment, we set instead  $a = \ln|\frac{1}{2} k \cdot \bar{k}|$, where
$k^\alpha$ and $\bar{k}^\alpha$ are the null normals. Along with an
appropriate specification of signs, this rule ensures that the
on-shell gravitational action is properly additive.\footnote{It should
  be noted that there are exceptions to this statement: As Brill and
  Hayward have demonstrated \cite{1994PhRvD..50.4914B}, the action may
  not be additive when a joint $\B$ possesses a timelike 
direction; we exclude such cases from our considerations.}   

The assignment $\kappa = 0$ and the joint rules for $a$ eliminate the
ambiguities from the gravitational action, except for a remaining
freedom to rescale $\lambda$ by a constant factor. We can remove this
final ambiguity for the WdW patch by imposing a normalization
condition on the null normals at the asymptotic AdS boundary. With
these rules in place, we can finally turn to the task of evaluating
$S$ for a Wheeler-deWitt patch of a Schwarzschild-AdS black hole, and
calculate its rate of change $dS/dt$. As stated previously and
described in detail below, we arrive at precisely the same result
first obtained by Brown et al \cite{Brown:2015lvg}: 
$dS/dt = 32\pi G_\mt{N}\,M$, or $dI/dt = 2M$. 

In the remainder of the paper we offer a detailed account of the
results summarized above. We begin in Sec.~\ref{contributionstoaction}
with a thorough description of the variational principle in general
relativity, when $\partial\V$ consists of spacelike, timelike, and
null hypersurface segments\footnote{A treatment by complementary methods 
will appear in \cite{prepraf}.}. After some preliminary remarks in
Sec.~\ref{background} we review the well-understood case of  
timelike and spacelike segments in Sec.~\ref{sec:nonnull}, before
moving on to the null case in Sec.~\ref{sec:null}. In
Sec.~\ref{sec:closed-nonnull} we form a closed boundary by joining
spacelike and timelike segments, and obtain Hayward's expression for
the $\eta$-terms in the gravitational action; his rules are summarized in
Sec.~\ref{sec:TSjoints}. In Sec.~\ref{sec:closed-null} we form a
closed boundary by joining spacelike and null segments, and derive the
appropriate joint contributions to the action, our own $a$-terms; the
rules for null joints are summarized in Sec.~\ref{sec:Njoints}. The
additivity rules for $a$ are formulated in Sec.~\ref{sec:additivity},
which concludes the section. While much of the material contained in
Sec.~\ref{contributionstoaction} is known from the literature, we
consider that a self-contained and complete account enhances the
clarity of the presentation. We also take the opportunity to fill in
some of the technical details left implicit in Hayward's 
work~\cite{Hayward:1993my}, and to provide minor improvements on the  
developments of Parattu et al~\cite{Parattu:2015gga}.  

In Sec.~\ref{sec:dSdt} we revisit the calculation of $dS/dt$ for a
Wheeler-deWitt patch of a spacetime describing an eternal black hole 
in asymptotically anti-de Sitter space. We begin with an uncharged
black hole in Secs.~\ref{sec:SadS}, \ref{sec:WdW}, and
\ref{sec:cal-deltaS}, compare our calculations to those of Brown et al
\cite{Brown:2015lvg} in Sec.~\ref{sec:comp}, and turn to the charged
case in Sec.~\ref{sec:charged}. In Appendix~\ref{sec:limit} we 
provide an analysis of the ambiguous nature of the gravitational
action when the boundary includes a null segment. We show that in
general, evaluating the boundary action for a null segment through the
limit of a sequence of (say) timelike surfaces produces an ill-defined
result. A notable exception to this statement arises when the limit is
a stationary null surface, \eg a Killing horizon; in this case the
limit is unique. The theme is pursued further in
Appendix~\ref{nocando}, in which we show that the 
parametrization ambiguity of the gravitational action can be
eliminated by adding a suitable counterterm; the consequences of this
observation will be explored in a forthcoming publication
\cite{prep}. Finally, we conclude in Appendix~\ref{action-manual} with
an {\em Action User's Manual} that provides a concise guide on how
each relevant contribution to the action, and in particular, the sign
of each contribution, are evaluated.  

\section{Volume, surface, and joint contributions to the gravitational
  action \labell{contributionstoaction} }

For the developments of this section, we focus our attention on a 
spacetime of $d=4$ dimensions; the generalization to higher-dimensional
spacetimes is immediate. To keep track of the sign of different
contributions to the action, we adopt the convention according to which
all timelike and null normal vectors are future-directed, and all spacelike
normals are outward-directed with respect to the region of interest. 
Further, we recall that the quantity $S$ considered
below is related to the usual gravitational action $I$ by $S = 16\pi G_\mt{N} \,I$, as defined in Eq.~\reef{equation}.  

\subsection{Background and previous results}
\labell{background}

It is well-known that the Hilbert-Einstein action defined on a
four-dimensional domain $\V$, 
\begin{equation} 
 S_{\V} := \int_{\V} (R-2\Lambda)\, \sqrt{-g}\,d^4 x, 
\end{equation} 
must be supplemented by a boundary term $S_{\partial \V}$ in order to
give rise to a variational principle in which only the metric
variation $\delta g^{\mu\nu}$ (and not its derivatives) is
required to vanish on the boundary $\partial \V$. The reason is that
variation of $S_{\V}$ produces the expression 
\begin{equation} 
\delta S_{\V} = \int_{\V} \bigl( G_{\mu\nu} 
+ \Lambda g_{\mu\nu} \bigr) \delta g^{\mu\nu}\, \sqrt{-g} d^4 x  
+ \oint_{\partial \V} \dslash v^\mu\, d\Sigma_\mu, 
\end{equation} 
with an additional boundary term that must be properly disposed of. We
have introduced 
\begin{equation} 
\dslash v^\mu := g^{\alpha\beta} \delta \Gamma^\mu_{\alpha\beta} 
- g^{\alpha\mu} \delta \Gamma^\beta_{\alpha\beta}\,, 
\labell{dv-def} 
\end{equation} 
and $d\Sigma_\mu$ is an outward-directed surface element on 
$\partial \V$. The slash in the middle of the $\delta$ symbol in
$\dslash v^\mu$ reminds us that this infinitesimal quantity is {\it not} the
variation of another quantity $v^\mu$.  

The manipulations carried out in the following subsections reveal that
in all the cases considered, 
\begin{equation} 
\oint_{\partial \V} \dslash v^\mu\, d\Sigma_\mu   
= -\delta S_{\partial \V}, 
\labell{boundaryterm-def} 
\end{equation} 
where $S_{\partial \V}$ is a suitable boundary term, whose variation
reproduces the expression of the left-hand side when the induced
metric on $\partial \V$ is held fixed. With this result established, the
gravitational action is properly identified with  
$S_{\V} + S_{\partial \V}$, and its variation yields 
\begin{equation} 
\delta \bigl( S_{\V} + S_{\partial \V} \bigr) 
= \int_{\V} \bigl( G_{\mu\nu} + \Lambda g_{\mu\nu} \bigr) 
\delta g^{\mu\nu}\, \sqrt{-g} d^4 x. 
\end{equation} 

The boundary $\partial \V$ is usually constructed from spacelike and
timelike hypersurface segments, and in this case the manipulations
that lead to the identification of $S_{\partial \V}$ are
well-known. The most complete version of this computation was
presented by Hayward~\cite{Hayward:1993my}, who paid careful attention
to situations in which $\partial \V$ is not a smooth
hypersurface. Specifically, Hayward examined cases in which a
(timelike or spacelike) segment of $\partial \V$ is joined to another
(timelike or spacelike) segment at a two-dimensional surface, in such
a way that the normal vector field is discontinuous at the joint. He
showed that in general, such joints contribute to the boundary
action. We reproduce Hayward's computations below, and provide details
that were left out of his paper. 

The boundary $\partial \V$ can also include segments of null
hypersurfaces. This case was not given much attention in the
literature, with the notable recent exceptions of
Neiman~\cite{Neiman:2012fx} and 
Parattu et al.~\cite{Parattu:2015gga}.
We revisit these constructions here, providing a more complete
treatment. Unlike Neiman, who took the null
generators of the hypersurface segments to be affinely parametrized,
we allow the generators to be arbitrarily parametrized. This
generalization reveals the important fact that the boundary action
evaluated on a null segment depends on the choice of parameter and is
therefore ill-defined in general. And unlike Parattu et al, who
did allow for an arbitrary parametrization but did not consider the
joints with other surfaces, we pay close attention to the joint that
arises when a null segment is joined to a spacelike, timelike, or null 
segment. Our manipulations pertaining to a given null segment also
offer a minor improvement on the treatment provided 
by Parattu et al: while their derivation requires the normal
vector to $\partial \V$ to be given an extension off the hypersurface
so as to define its derivatives in all directions, our derivation
involves only tangential derivatives and therefore does not require
such an extension. 

\subsection{Spacelike/timelike segment
\labell{sec:nonnull} }

We begin with the well-studied task of evaluating $\int \dslash
v^\mu\, d\Sigma_\mu$ on a spacelike or timelike segment of 
$\partial \V$. We denote this segment by $\Sigma$, and imagine that it
is bounded by the two-surfaces $\B_1$ and $\B_2$. When $\Sigma$ is
spacelike, $\B_1$ represents an inner boundary, and $\B_2$ an outer
boundary. When $\Sigma$ is timelike, $\B_1$ represents a past
boundary, and $\B_2$ a future boundary. 

\subsubsection{Preliminaries} 

We first import some helpful results from Secs.~3.1, 3.2, and 3.4 
of Ref.~\cite{2004rtmb.book.....P}. (We will make frequent use of
results obtained in this book, and we shall refer to it as the 
{\it Toolkit}.) The hypersurface $\Sigma$ is described by the 
relation $\Phi(x^\alpha) = 0$ for some scalar field $\Phi$. When
$\Sigma$ is spacelike, $\Phi$ is taken to increase toward the future
across the hypersurface; when it is timelike, $\Phi$ increases
outward. The hypersurface can also be described by parametric
equations $x^\alpha = x^\alpha(y^a)$, in which $y^a$ are
intrinsic coordinates on the hypersurface. The unit normal is  
\begin{equation} 
n_\alpha = \epsilon\, \mu\, \partial_\alpha \Phi, 
\end{equation} 
where $\epsilon := n_\alpha n^\alpha = \pm 1$ and $\mu :=
|g^{\alpha\beta} \partial_\alpha \Phi \partial_\beta
\Phi|^{-1/2}$. The equation implies that when $\Sigma$ is spacelike,
$n^\alpha$ is a future-directed vector; when $\Sigma$ is timelike,
$n^\alpha$ points to the outside. The vectors 
\begin{equation} 
e^\alpha_a := \frac{\partial x^\alpha}{\partial y^a} 
\end{equation} 
are tangent to $\Sigma$ and orthogonal to $n_\alpha$. We define
$e^a_\alpha := h^{ab} g_{\alpha\beta} e^\beta_b$. The induced 
metric on the hypersurface is 
\begin{equation} 
h_{ab} := g_{\alpha\beta} e^\alpha_a e^\beta_b, 
\end{equation} 
and we denote its determinant by $h$ and its inverse by $h^{ab}$. The
completeness relation for the inverse metric is given by 
\begin{equation} 
g^{\alpha\beta} = \epsilon\,n^\alpha n^\beta + h^{\alpha\beta}, \qquad 
h^{\alpha\beta} := h^{ab} e^\alpha_a e^\beta_b. 
\end{equation} 
The directed surface element on $\Sigma$ is 
\begin{equation} 
d\Sigma_\alpha = \epsilon\, n_\alpha\, d\Sigma, \qquad 
d\Sigma := |h|^{1/2}\, d^3y,  
\end{equation} 
with the convention that 
$d\Sigma_\alpha \propto \partial_\alpha \Phi$, with a positive factor
of proportionality. The extrinsic curvature of the hypersurface is
defined by   
\begin{equation} 
K_{ab} := e^\alpha_a e^\beta_b \nabla_\alpha n_\beta, 
\end{equation} 
and we recall the identity 
\begin{equation} 
e^\alpha_a \nabla_\alpha e^\beta_b 
= \Gamma^c_{ab} e^\beta_c - \epsilon\, K_{ab} n^\beta 
\end{equation} 
for the derivatives of the tangent vectors; the Christoffel symbols
$\Gamma^c_{ab}$ are those constructed from $h_{ab}$. 

\subsubsection{Variation of geometric quantities} 

We next perform a variation $\delta g^{\alpha\beta}$ of the metric,
and see how various geometric quantities defined on $\Sigma$ respond
to the variation. In this exercise it is understood that the
description of the hypersurface is unchanged during the variation, so
that the equations $\Phi = 0$ and $x^\alpha = x^\alpha(y^a)$ keep
their original form. This implies that the tangent vectors
$e^\alpha_a$ are unaffected by the variation. A variation of the
metric, however, induces a variation of $n_\alpha$, which is given by  
\begin{equation} 
\delta n_\alpha = \frac{\delta\mu}{\mu}\, n_\alpha, 
\qquad 
\frac{\delta \mu}{\mu} = -\frac{1}{2} \epsilon\,
n_\alpha n_\beta \delta g^{\alpha\beta}. 
\labell{deltan} 
\end{equation} 
There is also a change in $e^a_\alpha$: the relation 
$\delta^a_{\ b} = e^a_\alpha e^\alpha_b$ implies that 
$0 = e^\alpha_b \delta e^a_\alpha$, so that   
\begin{equation} 
\delta e^a_\alpha = \dslash A^a n_\alpha 
\end{equation} 
for some infinitesimal quantity $\dslash A^a$. The relation 
$0 = e^a_\alpha n^\alpha$ implies that $0 = e^a_\alpha \delta n^\alpha  
+ \epsilon \dslash A^a$, and $\epsilon = n_\alpha n^\alpha$ implies
that $0 = n_\alpha \delta n^\alpha + \epsilon \,\delta \mu/\mu$. We
have obtained 
\begin{equation} 
\delta n^\alpha = -\frac{\delta \mu}{\mu} n^\alpha 
- \epsilon \dslash A^a e^\alpha_a
\end{equation} 
for the variation of $n^\alpha$. 

The quantity $\dslash A^a$ can be expressed in a number of ways. We 
have 
\begin{equation} 
\dslash A^a = \epsilon\, n^\alpha \delta e^a_\alpha 
= - \epsilon\, e^a_\alpha \delta n^\alpha, 
\end{equation} 
and combining the second form with the identity $\delta n^\alpha =
n_\beta \delta g^{\alpha\beta} + g^{\alpha\beta} \delta n_\beta$ gives
\begin{equation} 
\dslash A^a = -\epsilon\, e^a_\alpha n_\beta \delta g^{\alpha\beta}. \
\labell{deltaA} 
\end{equation} 
This shows that $\dslash A^a$ is associated with the variation of the 
normal-tangent components of the inverse metric. Further, these results reveal
that $\dslash A^a$ is {\it not} the variation of a quantity $A^a$.  

The completeness relation for the inverse metric implies that 
\begin{equation} 
\delta g^{\alpha\beta} 
= -2\epsilon\, \frac{\delta \mu}{\mu}\, n^\alpha n^\beta 
- \dslash A^a \bigl( e^\alpha_a n^\beta + n^\alpha e^\beta_a \bigr) 
+ \delta h^{ab}\, e^\alpha_a e^\beta_b. 
\end{equation} 
This expression confirms that $\delta \ln\mu$ represents the
variation of the normal-normal component of the inverse metric,
$\dslash A^a$ the variation of the normal-tangent components, and
shows that the variation of the purely tangential components is
captured by $\delta h^{ab}$. 

We next work out two expressions for $\delta K$, the variation of the
trace of the extrinsic curvature. For the first, we begin with $K =
h^{ab} K_{ab}$ and write $\delta K = K_{ab} \delta h^{ab} 
+ h^{ab} \delta K_{ab}$. Recalling the definition of the extrinsic
curvature, we have that 
\begin{align} 
\delta K_{ab} &= e^\alpha_a e^\beta_b 
\bigl( \nabla_\alpha \delta n_\beta  
- n_\mu \delta \Gamma^\mu_{\alpha\beta} \bigr) 
\nonumber \\ 
&= e^\alpha_a e^\beta_b \biggl[ 
  \nabla_\alpha\bigl( \delta \ln\mu \bigr)\, n_\beta 
+(\delta \ln\mu) \nabla_\alpha n_\beta 
- n_\mu \delta \Gamma^\mu_{\alpha\beta} \biggr] 
\nonumber \\ 
&= \frac{\delta\mu}{\mu} K_{ab} 
- e^\alpha_a e^\beta_b n_\mu 
\delta \Gamma^\mu_{\alpha\beta}, 
\end{align} 
so that 
\begin{equation} 
\delta K = K_{ab} \delta h^{ab} 
+ \frac{\delta\mu}{\mu} K 
- h^{\alpha\beta} n_\mu 
\delta \Gamma^\mu_{\alpha\beta}. 
\labell{deltaK1} 
\end{equation} 

For the second expression for $\delta K$, we begin with 
$\delta K = h^\alpha_{\ \beta} \nabla_\alpha n^\beta$ which implies
\begin{equation} 
\delta K = \bigl( \delta h^\alpha_{\ \beta} \bigr) 
  \nabla_\alpha n^\beta 
+ h^\alpha_{\ \beta} \nabla_\alpha \delta n^\beta 
+ n^\alpha h^\beta{}_{\mu} \delta \Gamma^\mu_{\alpha\beta}.
\end{equation} 
To evaluate the first term, we write $h^\alpha_{\ \beta} = e^\alpha_a
e^a_\beta$, take the variation to get $\delta h^\alpha_{\ \beta} =
e^\alpha_a n_\beta \dslash A^a$, and combine this with $\nabla_\alpha
n^\beta$ to get zero, because $e^\alpha_a n_\beta \nabla_\alpha n^\beta
= \frac{1}{2} e^\alpha_a \nabla_\alpha(n_\beta n^\beta) = 0$. The
second term requires more work. We have 
\begin{align} 
h^\alpha_{\ \beta} \nabla_\alpha \delta n^\beta &= 
e^\alpha_a e^a_\beta \nabla_\alpha \delta n^\beta 
\nonumber \\ 
&= e^\alpha_a \nabla_\alpha \bigl( e^a_\beta \delta n^\beta) 
- e^\alpha_a \bigl( \nabla_\alpha e^a_\beta \bigr)\delta n^\beta 
\nonumber \\ 
&= -\epsilon\, \partial_a \dslash A^a
+ \epsilon\, \dslash A^b e^\alpha_a e^\beta_b \nabla_\alpha e^a_\beta
+ \frac{\delta \mu}{\mu} e^\alpha_a n^\beta \nabla_\alpha e^a_\beta. 
\end{align} 
The second term involves 
\begin{equation} 
e^\alpha_a e^\beta_b \nabla_\alpha e^a_\beta = 
e^\alpha_a \nabla_\alpha \bigl( e^\beta_b e^a_\beta \bigr) 
- e^\alpha_a e^a_\beta \nabla_\alpha e^\beta_b 
= e^\alpha_a \nabla_\alpha \bigl( \delta^a_{\ b} \bigr) 
- e^a_\beta \Gamma^c_{ab} e^\beta_c 
= -\Gamma^c_{cb}, 
\end{equation} 
and the third term involves 
\begin{equation} 
e^\alpha_a n^\beta \nabla_\alpha e^a_\beta = 
e^\alpha_a \nabla_\alpha \bigl( n^\beta e^a_\alpha \bigr) 
- e^\alpha_a e^a_\beta \nabla_\alpha n^\beta = -K. 
\end{equation} 
Collecting results, we have obtained 
\begin{equation} 
h^\alpha_{\ \beta} \nabla_\alpha \delta n^\beta
= -\epsilon\, D_a \dslash A^a - \frac{\delta \mu}{\mu} K,
\end{equation} 
where $D_a$ is the covariant-derivative operator compatible with the
induced metric $h_{ab}$. Our second expression for $\delta K$ is
therefore 
\begin{equation} 
\delta K = -\epsilon\, D_a \dslash A^a - \frac{\delta \mu}{\mu} K 
+ n^\alpha h^\beta{}_{\mu}\, \delta \Gamma^\mu_{\alpha\beta}.
\labell{deltaK2} 
\end{equation} 

\subsubsection{Boundary term} 

We may now evaluate 
\begin{equation} 
\int_\Sigma \dslash v^\mu\, d\Sigma_\mu 
= \int_\Sigma \epsilon\, \dslash v^\mu n_\mu\, d\Sigma 
\end{equation} 
when $\Sigma$ is a spacelike or timelike hypersurface. We have that 
\begin{equation} 
\dslash v^\mu n_\mu = \bigl( g^{\alpha\beta} n_\mu 
- n^\alpha \delta^\beta_{\ \mu} \bigr) \delta\Gamma^\mu_{\alpha\beta} 
= \bigl( h^{\alpha\beta} n_\mu - n^\alpha h^\beta_{\ \mu} \bigr) 
\delta\Gamma^\mu_{\alpha\beta}, 
\end{equation} 
where the completeness relation was used to go from the first
expression to the second. Invoking next Eqs.~(\ref{deltaK1}) and
(\ref{deltaK2}), we obtain 
\begin{equation} 
\dslash v^\mu n_\mu = -2\delta K - \epsilon\, D_a \dslash  A^a 
+ K_{ab} \delta h^{ab},  
\end{equation} 
so that 
\begin{equation} 
\int_\Sigma \dslash v^\mu\, d\Sigma_\mu 
= \int_\Sigma \epsilon\, \bigl( -2 \delta K 
+ K_{ab} \delta h^{ab} \bigr)\, d\Sigma 
- \oint_{\B_2} \dslash A^a\, dS_a
+ \oint_{\B_1} \dslash A^a\, dS_a,  
\labell{deltaS-nonnull} 
\end{equation} 
where $dS_a$ is a surface element on ${\cal B}_1$ and ${\cal B}_2$,
the boundaries of $\Sigma$.  

We now require the variation $\delta h^{ab}$ to vanish on
$\Sigma$,\footnote{This is, of course, the usual boundary condition
  for Einstein's general relativity, \ie the intrinsic geometry is held fixed on
  the (timelike and spacelike) boundary surfaces.} and see that  
the former expression becomes  
\begin{align} 
\int_\Sigma \dslash v^\mu\, d\Sigma_\mu 
&= -2 \epsilon \int_\Sigma \delta K\, d\Sigma 
- \oint_{\B_2} \dslash A^a\, dS_a
+ \oint_{\B_1} \dslash A^a\, dS_a  
\nonumber \\ 
&= \delta \biggl( -2 \epsilon \int_\Sigma K\, d\Sigma \biggr)   
- \oint_{\B_2} \dslash A^a\, dS_a
+ \oint_{\B_1} \dslash A^a\, dS_a.   
\labell{deltaS-nonnullfinal} 
\end{align} 
If $\dslash A^a$ were the variation of a quantity $A^a$, we could take 
the variation sign outside the $\B_1$ and $\B_2$ integrals
and identify a boundary term $S_\Sigma$ for the spacelike or timelike
segment. But $\dslash A^a$ is not the variation of anything by itself, and these
manipulations will not go through until we join segments together to
form a closed hypersurface $\partial \V$ --- see section
\ref{sec:closed-nonnull} below. 

\subsection{Null segment 
\labell{sec:null} }

We next turn to the task of evaluating $\int \dslash
v^\mu\, d\Sigma_\mu$ on a null segment of $\partial \V$. We again
denote this segment by $\Sigma$, and take it to be bounded in the past
by a two-surface $\B_1$ and in the future by a two-surface
$\B_2$. 

\subsubsection{Preliminaries} 

To handle the case of a null hypersurface we follow the methods
reviewed in Sec.~3.1 of the 
{\it Toolkit}~\cite{2004rtmb.book.....P}. We describe the 
hypersurface by the relation $\Phi(x^\alpha) = 0$ for some scalar
$\Phi$, with the convention that $\Phi$ increases toward the future. 
The hypersurface can also be described by the parametric equations 
$x^\alpha = x^\alpha(\lambda,\theta^A)$, where $\theta^A$ is 
constant on each null generator spanning the hypersurface, while
$\lambda$ is a parameter on each generator. The null normal to the
hypersurface is 
\begin{equation} 
k_\alpha = -\mu\, \partial_\alpha \Phi, 
\labell{k-def} 
\end{equation} 
where $\mu$ is a (positive definite) scalar function on $\Sigma$; the minus sign ensures that
$k^\alpha$ is a future-directed vector. The definition of the
intrinsic coordinates $(\lambda,\theta^A)$ implies that the vectors  
\begin{equation} 
k^\alpha = \frac{\partial x^\alpha}{\partial \lambda}, \qquad 
e^\alpha_A = \frac{\partial x^\alpha}{\partial \theta^A} 
\labell{ke-def} 
\end{equation} 
are tangent to the hypersurface\footnote{We emphasize that for a null
  boundary segment $\Sigma$, the normal $k_\alpha$ is orthogonal to
  the surface but $k^\alpha$ is tangent to it. Further, our
  convention is that $k^\alpha$ is future directed and hence
  $k_\alpha$ is past directed.} and  $k^\alpha = g^{\alpha\beta}
k_\beta$ 
is orthogonal to the spacelike vectors $e^\alpha_A$. The null vector
satisfies the geodesic equation
\begin{equation} 
k^\beta \nabla_\beta k^\alpha = \kappa\, k^\alpha, 
\labell{kappa-def1} 
\end{equation} 
with $\kappa(\lambda,\theta^A)$ measuring the failure of $\lambda$ to
be an affine parameter on the null generators. The vector basis is
completed with a second null vector $N^\alpha$, which is transverse to
the hypersurface, orthogonal to $e^\alpha_A$, and which we choose to normalize by 
$k_\alpha N^\alpha = -1$. This allows us to write 
\begin{equation} 
\kappa = -N_\alpha\, k^\beta \nabla_\beta k^\alpha.  
\labell{kappa-def2} 
\end{equation} 

We let 
\begin{equation} 
\gamma_{AB} := g_{\alpha\beta}\, e^\alpha_A e^\beta_B 
\end{equation} 
be an induced metric on $\Sigma$, noting that a displacement on the
hypersurface comes with the line element 
$ds^2 = \gamma_{AB}\, d\theta^A d\theta^B$. In this description, which  
exploits the congruence of null generators to construct a system of
adapted intrinsic coordinates $(\lambda,\theta^A)$, the induced metric
is not merely degenerate but explicitly two-dimensional. We let
$\gamma^{AB}$ denote the matrix inverse to $\gamma_{AB}$, and 
$\gamma := \mbox{det}[\gamma_{AB}]$. We also introduce 
$e^A_\alpha := \gamma^{AB} g_{\alpha\beta} e^\beta_B$.  

The completeness relation for the inverse metric is given by  
\begin{equation} 
g^{\alpha\beta} = -k^\alpha N^\beta - N^\alpha k^\beta 
+ \gamma^{\alpha\beta}, \qquad 
\gamma^{\alpha\beta} := \gamma^{AB} e^\alpha_A e^\beta_B. 
\labell{completeness}
\end{equation} 
According to Eq.~(3.20) of the {\it Toolkit}, the directed surface
element on $\Sigma$ is   
\begin{equation} 
d\Sigma_\alpha = -k_\alpha \sqrt{\gamma}\, d^2\theta d\lambda, 
\end{equation} 
with the convention that $d\Sigma_\alpha \propto \partial_\alpha
\Phi$, with a positive factor of proportionality. 

The two-tensor 
\begin{equation} 
B_{AB} := e^\alpha_A e^\beta_B\, \nabla_\alpha k_\beta
\labell{B-def} 
\end{equation} 
governs the behavior of the congruence of null generators. It is
typically decomposed as 
\begin{equation} 
B_{AB} = \frac{1}{2} \Theta\, \gamma_{AB} + \sigma_{AB}\,, 
\end{equation} 
with $\Theta := \gamma^{AB} B_{AB}$ measuring the rate of expansion of
the congruence, and $\sigma_{AB}$ its rate of shear. We have that 
\begin{equation} 
\Theta = \frac{1}{\sqrt{\gamma}}\, \frac{\partial
  \sqrt{\gamma}}{\partial \lambda}, 
\labell{Theta-def} 
\end{equation} 
which indicates that $\Theta$ is the relative rate of change of
$\sqrt{\gamma}\, d^2\theta$, the cross-sectional area of a bundle of
null generators. 

We conclude these preliminary remarks with a discussion of the
arbitrariness involved in the description of null
hypersurfaces. First, the parametrization of the null generators is
arbitrary, and this implies that $\lambda$ can be redefined
independently on each generator, 
$\lambda \to \bar{\lambda}(\lambda,\theta^A)$. The tangent vector
$k^\alpha$, therefore, is defined up to a multiplicative factor that
can vary arbitrarily over the hypersurface. Second, the function
$\Phi$ is itself arbitrary, and could be replaced by a different
function that also vanishes on the hypersurface, 
$\Phi \to \bar{\Phi}({\Phi})$. For a given vector field $k_\alpha$
corresponding to a given choice of parametrization, the freedom to
change $\Phi$ corresponds to the freedom to change $\mu$ by 
an arbitrary multiplicative factor in Eq.~(\ref{k-def}). Note that these two
sources of arbitrariness are independent from one another: For a fixed 
choice of $\Phi$, a reparametrization changes $k^\alpha$ by an
arbitrary multiplicative factor, which is then inherited by $\mu$
through Eq.~(\ref{k-def}); for a fixed choice of parametrization and
$k^\alpha$, a change of $\Phi$ corresponds to a change of $\mu$ by an
independent multiplicative factor.    
 
\subsubsection{Variation of geometric quantities} 

We next perform a variation $\delta g_{\alpha\beta}$ of the metric,  
and see how various geometric quantities defined on $\Sigma$ respond 
to the variation. It is again understood that the description of the
hypersurface is unchanged during the variation, so that the equations 
$\Phi = 0$ and $x^\alpha = x^\alpha(\lambda,\theta^A)$ keep their
original form. This implies that the tangent vectors $k^\alpha$ and
$e^\alpha_A$ are unaffected by the variation. We also assume that the
hypersurface stays null during the variation, and therefore impose  
\begin{equation} 
\delta \bigl( g_{\alpha\beta} k^\alpha k^\beta \bigr) 
= k^\alpha k^\beta \delta g_{\alpha\beta} = 0 
\end{equation} 
in our manipulations. We further assume
that the vectors $k^\alpha$ and $e^\alpha_A$ stay orthogonal during
the variation, so that 
\begin{equation} 
\delta \bigl( g_{\alpha\beta} e^\alpha_A k^\beta \bigr) 
= e^\alpha_A k^\beta \delta g_{\alpha\beta} = 0.  
\end{equation} 
At a later stage we shall impose the additional restriction that the 
variation of the induced two-metric $\gamma_{AB} := g_{\alpha\beta}
e^\alpha_A e^\beta_B$ should vanish, completing to six the count of
fixed metric components on the hypersurface (the same count as for a
timelike or spacelike boundary surface).  

The statements that $\delta k^\alpha = 0$ and $\delta e^\alpha_A = 0$,
together with the relations $k^\alpha k_\alpha = 0$ and 
$e^\alpha_A k_\alpha = 0$, imply that $k^\alpha \delta k_\alpha = 0$ 
and $e^\alpha_A \delta k_\alpha = 0$, which means that 
\begin{equation} 
\delta k_\alpha = \delta a\, k_\alpha 
\labell{deltak} 
\end{equation} 
for some $\delta a$. There is actually a quantity $a$ whose variation
is $\delta a$. To see this, recall the relation $k_\alpha = -\mu
\nabla_\alpha \Phi$, which implies that $\delta k_\alpha = (\delta
\ln\mu) k_\alpha$, so that 
\begin{equation} 
a = \ln\mu. 
\labell{a-vs-mu} 
\end{equation} 
This quantity will play a very important role below. A similar
calculation reveals that   
\begin{equation} 
\delta e^A_\alpha = \dslash a^A\, k_\alpha
\end{equation} 
for some infinitesimal quantity $\dslash a^A$. 

Variation of $\nabla_\alpha k^\mu$ yields 
\begin{equation} 
\delta \bigl( \nabla_\alpha k^\mu \bigr) 
= k^\beta \delta \Gamma^\mu_{\alpha\beta}, 
\labell{delta-gradk1} 
\end{equation}  
and a short calculation also reveals that 
\begin{equation} 
\delta \bigl( \nabla_\alpha k_\beta \bigr) 
= \bigl( \nabla_\alpha \delta a \bigr) k_\beta
+ \delta a\, \nabla_\alpha k_\beta 
- k_\mu \delta \Gamma^\mu_{\alpha\beta}. 
\labell{delta-gradk2} 
\end{equation} 
For each one of these identities it is understood that $k^\mu$ or 
$k_\alpha$ is differentiated in the directions tangent to the
hypersurface. 

Equations (\ref{kappa-def1}) and (\ref{delta-gradk1}) immediately
imply that 
\begin{equation} 
\delta \kappa = -k^\alpha k^\beta N_\mu 
\delta \Gamma^\mu_{\alpha\beta}. 
\labell{delta-kappa1} 
\end{equation} 
Alternatively, we can write Eq.~(\ref{kappa-def1}) in the form
$k^\alpha \nabla_\alpha k_\beta = \kappa k_\beta$, and construct the
variation using Eq.~(\ref{delta-gradk2}). This yields 
\begin{equation} 
\delta \kappa = k^\alpha \nabla_\alpha \delta a 
+ k^\alpha N^\beta k_\mu \delta \Gamma^\mu_{\alpha\beta}. 
\labell{delta-kappa2} 
\end{equation} 
 
We next examine the variation of $\Theta$, the rate of expansion of
the congruence of null generators. In the first version of this
calculation we write $\Theta = \gamma^{AB} B_{AB}$ and express the
variation as $\delta \Theta = B_{AB} \delta \gamma^{AB} + \gamma^{AB}
\delta B_{AB}$. Eqs.~(\ref{B-def}) and (\ref{delta-gradk2}) imply
that 
\begin{equation} 
\delta B_{AB} = \delta a\, B_{AB} - e^\alpha_A e^\beta_B k_\mu 
\delta \Gamma^\mu_{\alpha\beta},
\end{equation} 
and taking the trace returns 
\begin{equation} 
\delta \Theta = B_{AB} \delta \gamma^{AB} + \Theta\, \delta a
- \gamma^{\alpha\beta} k_\mu \delta \Gamma^\mu_{\alpha\beta}. 
\labell{delta-Theta1} 
\end{equation}  
In the second version of the calculation we write instead 
\begin{equation} 
\Theta = \gamma^\alpha_{\ \mu} \nabla_\alpha k^\mu 
= e^\alpha_A e^A_\mu \nabla_\alpha k^\mu 
\end{equation} 
and take the variation using Eq.~(\ref{delta-gradk1}). We have 
\begin{align} 
\delta \Theta &= e^\alpha_A \bigl( \delta e^A_\alpha \bigr) 
\nabla_\alpha k^\mu + e^\alpha_A e^A_\mu \delta 
\bigl( \nabla_\alpha k^\mu \bigr) 
\nonumber \\ 
&= e^\alpha_A \bigl( \dslash a^A \bigr) k_\mu \nabla_\alpha k^\mu 
+ e^\alpha_A e^A_\mu k^\beta \delta \Gamma^\mu_{\alpha\beta}. 
\end{align} 
The first term vanishes, and we end up with 
\begin{equation} 
\delta \Theta = k^\alpha \gamma^\beta_{\ \mu} 
\delta \Gamma^\mu_{\alpha\beta}. 
\labell{delta-Theta2} 
\end{equation} 

\subsubsection{Boundary term
\labell{subsec:null-boundary-term} } 

We may now evaluate $\int_\Sigma \dslash v^\mu\, d\Sigma_\mu$ 
when $\Sigma$ is a null hypersurface. We recall
Eq.~(\ref{dv-def}) and write  
\begin{equation} 
\dslash v^\mu k_\mu = \bigl( g^{\alpha\beta} k_\mu 
- k^\alpha g^\beta_{\ \mu} \bigr) 
\delta \Gamma^\mu_{\alpha\beta}, 
\end{equation} 
in which we insert the completeness relation
(\ref{completeness}). After some simple algebra we arrive at  
\begin{equation} 
\dslash v^\mu k_\mu = \bigl( k^\alpha k^\beta N_\mu 
- k^\alpha N^\beta k_\mu 
+ \gamma^{\alpha\beta} k_\mu 
- k^\alpha \gamma^\beta_{\ \mu} \bigr) 
\delta \Gamma^\mu_{\alpha\beta}. 
\end{equation} 
We next use Eqs.~(\ref{delta-kappa1}), (\ref{delta-kappa2}),
(\ref{delta-Theta1}), and (\ref{delta-Theta2}) to replace each term
involving $\delta \Gamma^\mu_{\alpha\beta}$ by variations of
quantities defined on the hypersurface. We obtain 
\begin{equation} 
\dslash v^\mu k_\mu = k^\alpha \nabla_\alpha \delta a 
+ \Theta\, \delta a - 2 \delta(\kappa + \Theta) 
+ B_{AB} \delta \gamma^{AB}. 
\end{equation} 
With Eq.~(\ref{Theta-def}) this equation becomes 
\begin{equation} 
\dslash v^\mu k_\mu = 
\frac{1}{\sqrt{\gamma}} \frac{\partial}{\partial \lambda} 
\bigl( \sqrt{\gamma}\, \delta a \bigr) 
- 2 \delta(\kappa + \Theta) 
+ B_{AB} \delta \gamma^{AB}, 
\end{equation} 
and this shall be our final expression for $\dslash v^\mu k_\mu$.  

We have found that the hypersurface integral is given by  
\begin{equation} 
\int_\Sigma \dslash v^\mu\, d\Sigma_\mu 
= \int_\Sigma \Bigl[ 2\delta(\kappa+\Theta) 
- B_{AB} \delta \gamma^{AB} \Bigr] \sqrt{\gamma}\, 
d^2\theta d\lambda
- \int_\Sigma \frac{\partial}{\partial \lambda} \bigl( \delta a\, 
\sqrt{\gamma}\, d^2\theta \bigr)\,d\lambda\,.  
\end{equation} 
Incorporating our assumption that $\Sigma$ is bounded in the future by
a two-surface $\B_2$ and in the past by a two-surface $\B_1$, this is 
\begin{equation} 
\int_\Sigma \dslash v^\mu\, d\Sigma_\mu 
=  \int_\Sigma \Bigl[ 2\delta(\kappa+\Theta) 
- B_{AB} \delta \gamma^{AB} \Bigr] \sqrt{\gamma} 
d^2\theta d\lambda
- \oint_{\B_2} \delta a\, \sqrt{\gamma}\, d^2\theta 
+ \oint_{\B_1} \delta a\, \sqrt{\gamma}\, d^2\theta\,.   
\labell{deltaS-full} 
\end{equation} 
This can be expressed in a different form by manipulating the 
$\delta \Theta$ term. Because $\Theta = \partial_\lambda
\ln\sqrt{\gamma}$ we have that 
\begin{align} 
\int_\Sigma \delta \Theta\, \sqrt{\gamma} d^2\theta d\lambda 
&= \int_\Sigma \partial_\lambda (\delta \ln\sqrt{\gamma})
\sqrt{\gamma}\, d^2\theta d\lambda 
\nonumber \\ 
&= -\int_\Sigma (\partial_\lambda \sqrt{\gamma})\, 
\delta \ln\sqrt{\gamma}\, d^2\theta d\lambda 
+ \oint_{\B_2} \sqrt{\gamma}\, \delta \ln\sqrt{\gamma}\, d^2\theta 
- \oint_{\B_1} \sqrt{\gamma}\, \delta \ln\sqrt{\gamma}\, d^2\theta 
\nonumber \\ 
&= -\int_\Sigma \Theta \delta \sqrt{\gamma}\, d^2\theta d\lambda 
+ \oint_{\B_2} \delta \sqrt{\gamma}\, d^2\theta
- \oint_{\B_1} \delta \sqrt{\gamma}\, d^2\theta. 
\end{align} 
In the second and third terms the variation sign can be taken out of
the integral, and in the first term we can write $\delta \sqrt{\gamma} 
= -\frac{1}{2} \sqrt{\gamma} \gamma_{AB} \delta \gamma^{AB}$. This
yields 
\begin{equation} 
\int_\Sigma \delta \Theta\, \sqrt{\gamma} d^2\theta d\lambda =
\frac{1}{2} \int_\Sigma \Theta\, \gamma_{AB} \delta \gamma^{AB}\, 
\sqrt{\gamma}\, d^2\theta d\lambda
+ \delta {\cal A}_2 - \delta {\cal A}_1\,,  
\end{equation} 
where ${\cal A}_j := \oint_{\cal B_j} \sqrt{\gamma} d^2\theta$ is the
area of the two-surface ${\cal B}_j$. Substitution within
Eq.~(\ref{deltaS-full}) gives 
\begin{equation} 
\int_\Sigma \dslash v^\mu\, d\Sigma_\mu = 
\int_\Sigma \Bigl[ 2\delta \kappa
- \bigl( B_{AB} - \Theta \gamma_{AB} \bigr) \delta \gamma^{AB} \Bigr]
\sqrt{\gamma}  d^2\theta d\lambda
+ \delta {\cal A}_2 - \oint_{\B_2} \delta a\, \sqrt{\gamma} d^2\theta 
- \delta {\cal A}_1 + \oint_{\B_1} \delta a\, \sqrt{\gamma} d^2\theta.    
\labell{deltaS-alt} 
\end{equation} 

If we now assume that $\delta \gamma^{AB} = 0$ on $\Sigma$ (as part of
the variational conditions on the null surface --- see the discussion
above), the result simplifies to 
\begin{align} 
\int_\Sigma \dslash v^\mu\, d\Sigma_\mu&=  
2 \int_\Sigma \delta\kappa\, \sqrt{\gamma} d^2\theta d\lambda
- \oint_{\B_2} \delta a\, \sqrt{\gamma} d^2\theta 
+ \oint_{\B_1} \delta a\, \sqrt{\gamma} d^2\theta  
\nonumber \\ 
&= \delta \biggl( 2 \int_\Sigma \kappa\, \sqrt{\gamma} 
d^2\theta d\lambda
- \oint_{\B_2} a\, \sqrt{\gamma} d^2\theta 
+ \oint_{\B_1} a\, \sqrt{\gamma} d^2\theta \biggr). 
\labell{deltaS-nullfinal} 
\end{align} 
This computation reveals the existence of a boundary term 
\begin{equation} 
S_\Sigma = -2 \int_\Sigma \kappa\, \sqrt{\gamma}  
d^2\theta d\lambda
+ \oint_{\B_2} a\, \sqrt{\gamma} d^2\theta 
- \oint_{\B_1} a\, \sqrt{\gamma} d^2\theta
\labell{bond44}
\end{equation} 
for a segment $\Sigma$ of a null hypersurface. We note that by virtue
of Eq.~(\ref{Theta-def}), the condition $\delta \gamma^{AB} = 0$
automatically implies that $\delta \Theta = 0$, and this term 
was therefore eliminated in Eq.~(\ref{deltaS-nullfinal}). We note that
our boundary term \reef{bond44} differs from the one given in
\cite{Parattu:2015gga} by a term proportional to $\Theta$, since the
$\delta \Theta$ term was retained there in their final expression for
$\int_\Sigma \dslash v^\mu\, d\Sigma_\mu$.  

Our expression for $S_\Sigma$ pertains to an isolated segment of null
hypersurface. This segment, however, is only part of a closed
boundary $\partial\V$ of a finite domain $\V$ of spacetime. In particular, $\Sigma$
will be joined to other (spacelike, timelike, or null) segments
comprising $\partial\V$ at $\B_1$ and $\B_2$, and hence we should
expect additional contributions at these joints coming from the
neighbouring segments. We will show below in
Sec.~\ref{sec:closed-null} that with the addition of these
contributions, $S_\Sigma$ becomes   
\begin{equation} 
S_\Sigma(\mbox{joined}) = 
-2 \int_\Sigma \kappa\, \sqrt{\gamma}  d^2\theta d\lambda
+ 2 \oint_{\B_2} a\, \sqrt{\gamma} d^2\theta 
- 2 \oint_{\B_1} a\, \sqrt{\gamma} d^2\theta\,. 
\labell{S-joined} 
\end{equation} 
That is, the joint terms at $\B_1$ and $\B_2$ acquire a factor of
two. Furthermore, the joining of segments forces $a$ to take the
specific form $a = \ln|n \cdot k| + a_0$ when $\Sigma$ is joined to a
spacelike or timelike surface with unit normal $n^\alpha$, or of the
form $a = \ln(-k \cdot \bar{k}) + a_0$ when it is joined to another
null surface with normal $\bar{k}^\alpha$. Here, a dot indicates an
inner product between vectors, for example $n \cdot k 
:= g_{\alpha\beta} n^\alpha k^\beta$, and $a_0$ is an arbitrary
quantity that satisfies $\delta a_0 = 0$.   

It is striking that the value of $S_\Sigma(\mbox{joined})$ is
ill-defined, first because it depends on the choice of parametrization
for the null generators, and second because it depends on the choice
of $a_0$. However, the variation of the boundary term is
well-defined, because the parametrization and $a_0$ are fixed while
taking the variation. For a stationary null hypersurface there exists
a preferred parametrization $\lambda^*$ defined such that 
$\kappa$, $\gamma_{AB}$, and $a$ are all independent
of $\lambda^*$. An example of this is a Killing horizon, for which
$k^\alpha$ can be identified with the Killing vector $\xi^\alpha$
evaluated on the horizon, $\Phi$ is chosen to be equal to $\xi_\alpha
\xi^\alpha$, and then $a = -\ln(2\kappa)$. In this case
$S_\Sigma(\mbox{joined})$ reduces to    
\begin{equation} 
S^*_\Sigma(\mbox{joined}) 
= -2 \kappa^* {\cal A} (\lambda^*_2 - \lambda^*_1),
\labell{S-stationary} 
\end{equation} 
where ${\cal A} := \int \sqrt{\gamma} d^2\theta$ is the
cross-sectional area of the hypersurface.  Note that the two joint terms in Eq.~\reef{S-joined} have canceled here because the cross-sections are invariant under the Killing flow, \ie ${\cal A}_1={\cal A}_2={\cal A}$ and $a_1=a_2$.

\subsubsection{Reparametrizations
\labell{subsub:reparam} }

It is instructive to work out what happens to
$S_\Sigma(\mbox{joined})$ when the parametrization of each generator
is changed from $\lambda$ to 
$\bar{\lambda} = \bar{\lambda}(\lambda,\theta^A)$. The effect of this 
transformation on the various geometrical quantities was deduced in 
Ref.~\cite{vega-poisson-massey:11}. Defining 
$e^{-\beta} := \partial\bar{\lambda}/\partial \lambda$, we have that 
\begin{equation} 
\bar{k}^\alpha = e^\beta k^\alpha, \qquad 
\bar{\gamma}_{AB} = \gamma_{AB}, \qquad 
\bar{B}_{AB} = e^\beta B_{AB}, \qquad 
\bar{\kappa} = e^\beta (\kappa + \partial_\lambda \beta). 
\end{equation} 
The first relation implies that $\bar{a} = a + \beta$ (assuming that
$\Phi$ is not changed during the reparametrization), and the third 
gives $\bar{\Theta} = e^\beta \Theta$. Inserting this within 
$S_\Sigma(\mbox{joined})$ yields
\begin{align} 
\bar{S}_\Sigma(\mbox{joined}) &= S_\Sigma(\mbox{joined}) 
- 2 \int_\Sigma \partial_\lambda \beta\, \sqrt{\gamma} d^2\theta
d\lambda 
+ 2 \oint_{\B_2} \beta\, \sqrt{\gamma} d^2\theta 
- 2 \oint_{\B_1} \beta\, \sqrt{\gamma} d^2\theta 
\nonumber\\ 
&= S_\Sigma(\mbox{joined}) 
+ 2\int_\Sigma \Theta \beta\, \sqrt{\gamma} d^2\theta d\lambda, 
\labell{barS} 
\end{align} 
with the second expression following from the first after an
integration by parts. As expected, in general the value of
$S_\Sigma(\mbox{joined})$ for a given spacetime is not invariant under
a reparametrization of the null generators. An exception arises in the
case of a stationary hypersurface, for which $\Theta = 0$. In this
case the boundary term is invariant under a reparametrization, and 
it will therefore return the same value as in Eq.~\reef{S-stationary}
irrespective of the parameterization of the null generators (so long
as the choice of $\Phi$ is fixed).  

\subsubsection{Redefinition of $\Phi$} 

We have pointed out that $S_\Sigma(\mbox{joined})$ is ill-defined
because it depends on the choice of parameter $\lambda$, and also
because it depends on the choice of function $\Phi$ that describes the
hypersurface. To conclude this discussion, we describe the change to
$S_\Sigma(\mbox{joined})$ that results when we perform the redefinition   
\begin{equation} 
\Phi \to \bar{\Phi}(\Phi)\,, \labell{redux}
\end{equation} 
assuming that $\bar{\Phi} = 0$ when $\Phi = 0$. We keep the
parametrization fixed during this operation, so that $k_\alpha$ is
unchanged as a vector field on the hypersurface. It is easy to see
that under the redefinition \reef{redux}, $k_\alpha$ is re-expressed as  
\begin{equation} 
k_\alpha = -\bar{\mu}\, \partial_\alpha \bar{\Phi}, \qquad 
\bar{\mu} := \mu \frac{d\Phi}{d\bar{\Phi}}. 
\end{equation} 
This implies that $a := \ln\mu$ is changed to 
\begin{equation} 
\bar{a} = a + \ln \frac{d\Phi}{d\bar{\Phi}}\,;  
\end{equation} 
this change is actually in the $a_0$ piece of $a$, since the remaining
piece --- given by $\ln|n \cdot k|$ or $\ln(-k\cdot \bar{k})$ --- is
fixed for a given parametrization. The boundary action becomes 
\begin{equation} 
\bar{S}_\Sigma(\mbox{joined}) = S_\Sigma(\mbox{joined}) 
+ 2\oint_{\B_2} \ln \frac{d\Phi}{d\bar{\Phi}}\, \sqrt{\gamma} d^2\theta 
- 2\oint_{\B_2} \ln \frac{d\Phi}{d\bar{\Phi}}\, \sqrt{\gamma} d^2\theta. 
\end{equation} 
This shows that the value of $S_\Sigma(\mbox{joined})$ for a given 
spacetime is not invariant under a redefinition of $\Phi$.  

\subsection{Closed hypersurface: Timelike and spacelike segments
\labell{sec:closed-nonnull} }

\begin{figure} 
\includegraphics[width=.5\linewidth]{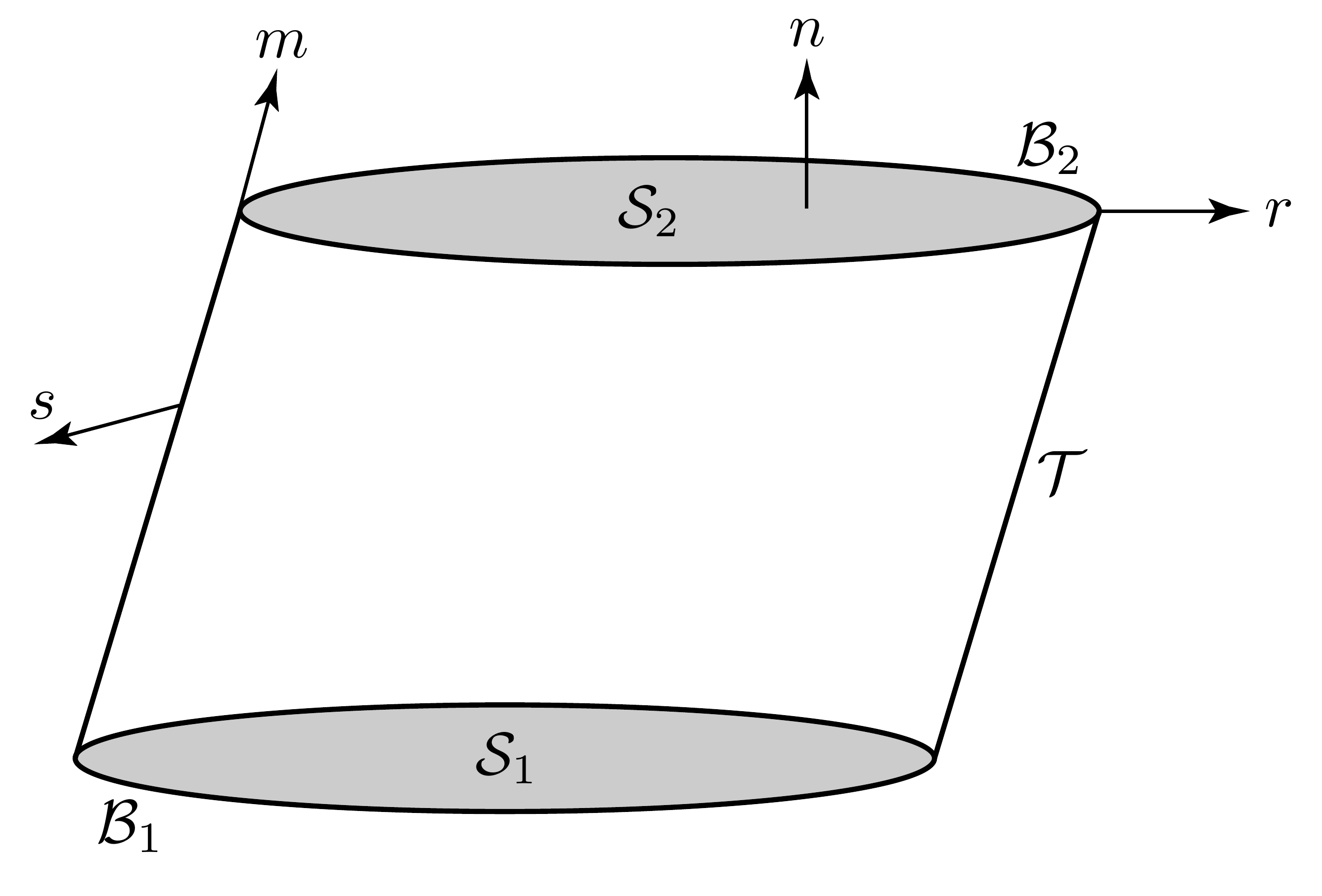}
\caption{Domain $\V$ bounded by a closed hypersurface 
 $\partial \V$ consisting of a timelike segment $\T$ and two
 spacelike segments $\S_1$ and $\S_2$. The intersection between $\T$ 
 and $\S_j$ is the closed two-surface $\B_j$.  \labell{fig1} }
\end{figure} 
 
In sections \ref{sec:nonnull} and \ref{sec:null}, we derived boundary terms for the gravitational action \reef{equation}. However, as we noted there, our analysis only examined isolated (spacelike, timelike, or null) boundary segments, which are implicitly part of a closed boundary $\partial\V$ of a finite domain $\V$ of spacetime. As a result, we were unable to give a complete description of the boundary contributions arising at the joints between neighbouring segments. We repair this deficiency in the next sections by focusing on the intersection of various boundary segments.
 
We begin in this section by forming a closed hypersurface $\partial \V$ with a timelike segment $\T$ joined to two spacelike segments $\S_1$ (in
the past) and $\S_2$ (in the future), as illustrated in Fig.~\ref{fig1}. The intersection between $\T$ and $\S_1$ is the spacelike two-surface $\B_1$, and $\B_2$ is the intersection
between $\T$ and $\S_2$. This is one of the cases that were first
considered by Hayward \cite{Hayward:1993my}.

To avoid confusion we must carefully specify the notation employed on
each hypersurface segment and the joints between them. To set the
stage we consider only $\T$ and the future surface $\S \equiv \S_2$, 
which intersect at $\B \equiv \B_2$; the past surface will be added at
a later stage.  

The spacelike hypersurface $\S$ has intrinsic coordinates $y^a$, a future-directed
unit normal vector $n_\alpha$ with $\epsilon = n^\alpha n_\alpha=-1$, and a
set of tangent vectors $e^\alpha_a = \partial x^\alpha/\partial
y^a$. The induced metric is $h_{ab}$, the extrinsic curvature is
$K_{ab}$, and the boundary quantity introduced in Eq.~(\ref{deltaA})
is again denoted $\dslash A^a$. The inverse metric on $\S$ is
expressed as    
\begin{equation} 
g^{\alpha\beta} = -n^\alpha n^\beta + h^{ab} e^\alpha_a e^\beta_b.   
\end{equation} 
The two-dimensional surface $\B$ can be given an embedding in $\S$. In
this description, it has intrinsic coordinates $\theta^A$, an outward-directed unit
normal vector $r_a$, and a set of tangent vectors 
$e^a_A = \partial y^a/\partial \theta^A$. The induced metric is
$\gamma_{AB}$, and we have the completeness relation  
\begin{equation} 
h^{ab} = r^a r^b + \gamma^{AB} e^a_A e^b_B.  
\end{equation} 
We promote $r^a$ and $e^a_A$ to spacetime vectors according to
$r^\alpha = r^a e^\alpha_a$ and $e^\alpha_A = e^a_A e^\alpha_a$, and
combine the completeness relations to give 
\begin{equation} 
g^{\alpha\beta} = -n^\alpha n^\beta + r^\alpha r^\beta 
+ \gamma^{AB} e^\alpha_A e^\beta_B, 
\end{equation} 
the inverse metric evaluated on $\B$. 

Turning to the timelike hypersurface $\T$, we give it intrinsic
coordinates $z^j$, an outward-directed unit normal vector $s_\alpha$ with 
$\epsilon = s^\alpha s_\alpha=+1$, and a set of tangent vectors 
$e^\alpha_j = \partial x^\alpha/\partial z^j$. The induced metric is
$f_{jk}$, the extrinsic curvature is $L_{jk}$, and the boundary
quantity introduced in Eq.~(\ref{deltaA}) is now denoted 
$\dslash B^j$. The inverse metric on $\T$ is    
\begin{equation} 
g^{\alpha\beta} = s^\alpha s^\beta + f^{jk} e^\alpha_j e^\alpha_k.   
\end{equation} 
The two-dimensional surface $\B$ can also be given an embedding in
$\T$. In this description, it has the same intrinsic coordinates
$\theta^A$, but the outward-directed unit normal vector\footnote{This normal $m^j$ is in the tangent space of $\T$ and ``outward-directed'' from this boundary surface. This also means that it is a future-directed timelike unit vector.} is now $m_j$, and the tangent vectors are $e^j_A = \partial z^j/\partial \theta^A$. The induced metric is $\gamma_{AB}$, and we have the completeness relation 
\begin{equation} 
f^{jk} = -m^j m^k + \gamma^{AB} e^j_A e^k_B.  
\end{equation} 
We promote $m^j$ and $e^j_A$ to spacetime vectors according to
$m^\alpha = m^j e^\alpha_j$ and $e^\alpha_A = e^j_A e^\alpha_j$, and 
combine the completeness relations to give 
\begin{equation} 
g^{\alpha\beta} = -m^\alpha m^\beta + s^\alpha s^\beta 
+ \gamma^{AB} e^\alpha_A e^\beta_B\,, 
\end{equation} 
an alternative expression for the inverse metric evaluated on $\B$. 

Each pair $\{ n^\alpha, r^\alpha \}$ and $\{ m^\alpha, s^\alpha \}$ 
forms a set of (mutually orthogonal) unit normals on $\B$.  Each pair
can be used as a two-dimensional vector basis, and the bases are
related by a spacetime boost. For example, we may write 
\begin{equation} 
n^\alpha = \cosh\eta\, m^\alpha + \sinh\eta\, s^\alpha, \qquad 
r^\alpha = \sinh\eta\, m^\alpha + \cosh\eta\, s^\alpha 
\end{equation} 
for some boost parameter $\eta$. A consequence of these relations is 
\begin{equation} 
m^\alpha = \frac{1}{\cosh\eta} n^\alpha 
- \frac{\sinh\eta}{\cosh\eta} s^\alpha, \qquad 
r^\alpha = \frac{\sinh\eta}{\cosh\eta} n^\alpha  
+ \frac{1}{\cosh\eta} s^\alpha,
\labell{mr-vs-ns} 
\end{equation} 
which expresses $m^\alpha$ (the normal to $\B$ embedded in $\T$) and
$r^\alpha$ (the normal to $\B$ embedded in $\S$) in terms of
$n^\alpha$ (the normal to $\S$) and $s^\alpha$ (the normal to $\T$). 

Now from Eq.~(\ref{deltaS-nonnullfinal}) we have that 
\begin{equation} 
\int_{\S} \dslash v^\mu\, d\Sigma_\mu 
= \delta \biggl( 2 \int_{\S} K\, \sqrt{h} d^3y \biggr) 
- \oint_{\B} r_a \dslash A^a\,\sqrt{\gamma} d^2\theta,
\end{equation} 
where we have inserted the relation $dS_a = r_a 
\sqrt{\gamma} d^2\theta$ for the surface element on $\B$. The same 
equation also produces  
\begin{equation} 
\int_{\T} \dslash v^\mu\, d\Sigma_\mu 
= \delta \biggl( -2 \int_{\T} L\, \sqrt{-f} d^3z \biggr) 
+ \oint_{\B} m_j \dslash B^j\,\sqrt{\gamma} d^2\theta, 
\end{equation} 
where this time we used the relation $dS_j = -m_j  
\sqrt{\gamma} d^2\theta$ for the surface element. Following the
notation introduced above, we are using $L$ to denote the trace of the
extrinsic curvature on $\T$. Combining these two terms gives  
\begin{equation} 
\int_{\S + \T} \dslash v^\mu\, d\Sigma_\mu 
= \delta \biggl( 2 \int_{\S} K\, \sqrt{h} d^3y 
-2 \int_{\T} L\, \sqrt{-f} d^3z \biggr) 
- \oint_{\B} \dslash C\,\sqrt{\gamma} d^2\theta, 
\labell{S+T} 
\end{equation} 
where $\dslash C := r_a \dslash A^a - m_j \dslash B^j$.   

To evaluate the joint term on $\B$, we recall from Eq.~(\ref{deltaA}) that
$\dslash A^a = +e^a_\alpha n_\beta \delta g^{\alpha\beta}$ and
$\dslash B^j = -e^j_\alpha s_\beta \delta g^{\alpha\beta}$. This gives 
\begin{align} 
\dslash C &= \bigl( r_a e^a_\alpha n_\beta 
+ m_j e^j_\alpha s_\beta \bigr) \delta g^{\alpha\beta} 
\nonumber \\ 
&= -\bigl( r^a e^\alpha_a n^\beta 
+ m^j e^\alpha_j s^\beta \bigr) \delta g_{\alpha\beta} 
\nonumber \\ 
&= -\bigl( r^\alpha n^\beta + m^\alpha s^\beta \bigr) 
\delta g_{\alpha\beta}  
\nonumber \\ 
&= +\frac{\sinh\eta}{\cosh\eta} \bigl( -n^\alpha n^\beta 
+ s^\alpha s^\beta \bigr) \delta g_{\alpha\beta}   
- \frac{1}{\cosh\eta} \bigl( n^\alpha s^\beta 
+ s^\alpha n^\beta \bigr) \delta g_{\alpha\beta},
\end{align} 
where Eq.~(\ref{mr-vs-ns}) was used in the last step. 

On the other hand, we can vary the equation 
\begin{equation} 
\sinh\eta = g^{\alpha\beta} n_\alpha s_\beta 
\labell{eta-def} 
\end{equation} 
using Eq.~(\ref{deltan}) for $\delta n_\alpha$ and an analogous
relation for $\delta s_\beta$. Simple algebra then returns 
\begin{equation} 
\delta \eta = -\frac{\sinh\eta}{2\cosh\eta} \bigl( -n_\alpha n_\beta  
+ s_\alpha s_\beta \bigr) \delta g^{\alpha\beta}   
+ \frac{1}{2\cosh\eta} \bigl( n_\alpha s_\beta 
+ s_\alpha n_\beta \bigr) \delta g^{\alpha\beta},
\end{equation} 
and we conclude that $\dslash C = 2\delta \eta$. Incorporating this
in Eq.~(\ref{S+T}), we arrive at 
\begin{equation} 
\int_{\S + \T} \dslash v^\mu\, d\Sigma_\mu 
= \delta \biggl( 2 \int_{\S} K\, \sqrt{h} d^3y 
-2 \int_{\T} L\, \sqrt{-f} d^3z 
- 2 \oint_{\B} \eta\, \sqrt{\gamma} d^2\theta \biggr)\,.
\labell{S+T-final} 
\end{equation} 
Hence the full boundary term for the hypersurface
$\S + \T$ includes the Hayward term \cite{Hayward:1993my}, proportional to the boost parameter $\eta$, at the joint $\B$. 

The joint term in Eq.~(\ref{S+T-final}) involves the parameter
required to boost between the normal $n^\alpha$ to $\T$ and the normal
$s^\alpha$ to $\S$, \ie 
$\eta := \mbox{arcsinh}(n \cdot s)$, with $n \cdot s :=
g_{\alpha\beta} n^\alpha s^\beta$. It is useful to give a simpler 
expression for $\eta$, one which will be adapted to other types of
joints below. For this purpose, we introduce a basis of null vectors
$k^\alpha$ and $\bar{k}^\alpha$; we take $k^\alpha$ and
$\bar{k}^\alpha$ to be incoming and outgoing with respect to
$\T$, respectively\footnote{That is, both null vectors are future-directed, \ie
  $k\cdot n<0$ and $\bar k\cdot n<0$, and then we choose $k\cdot s<0$
  and $\bar k\cdot s>0$.}, and we temporarily  
normalize them so that $k \cdot \bar{k} = -1$. 
 In terms of this basis, we have 
\begin{equation} 
n^\alpha = \frac{1}{2A} k^\alpha + A\, \bar{k}^\alpha, \qquad 
s^\alpha = -\frac{1}{2B} k^\alpha + B\, \bar{k}^\alpha, 
\end{equation} 
where $A := -n\cdot k > 0$ and $B := -s \cdot k > 0$. With these expressions, we 
can easily show that $\sinh\eta = \frac{1}{2} (A/B - B/A)$, so that  
\begin{equation} 
\eta = \ln(-n \cdot k) - \ln(-s \cdot k). 
\labell{eta-ln1} 
\end{equation}   
Noting that $A = 1/(2\bar{A})$ and $B = 1/(2\bar{B})$ with 
$\bar{A} := -n \cdot \bar{k}$ and $\bar{B} := s \cdot \bar{k}$,
$\eta$ can alternatively be expressed as 
\begin{equation} 
\eta = -\ln(-n \cdot \bar{k}) + \ln(s \cdot \bar{k}).  
\labell{eta-ln2} 
\end{equation}   
These expressions reveal that $\eta$ is independent of the
normalization of the null vectors $k^\alpha$ and $\bar{k}^\alpha$, as
it should be. The normalization condition $k \cdot \bar{k} = -1$,
which facilitated the computations producing to Eqs.~(\ref{eta-ln1}) and
(\ref{eta-ln2}), can therefore be relaxed; the expressions are valid
for arbitrarily normalized null vectors. 

At this stage we introduce the past surface $\S_1$ and construct the closed 
hypersurface $\partial \V$. The previous analysis can again be applied to determine the boundary term on $\S_1$ and the joint $\B_1$ where the former intersects with $\T$. However, we must alter some  
signs to account for the fact that while $d\Sigma_\mu$ is
outward-directed on all of the segments comprising $\partial \V$, we take $n^\alpha$ to be future-directed on both $\S_1$ and $\S_2$.  The final result is
\begin{equation} 
\int_{\partial \V} \dslash v^\mu\, d\Sigma_\mu 
= \delta \biggl( 2 \int_{\S_2} K\, \sqrt{h} d^3y 
- 2 \int_{\T} L\, \sqrt{-f} d^3z 
- 2 \int_{\S_1} K\, \sqrt{h} d^3y 
- 2 \oint_{\B_2} \eta\, \sqrt{\gamma} d^2\theta 
+ 2 \oint_{\B_1} \eta\, \sqrt{\gamma} d^2\theta 
\biggr). 
\end{equation} 
Hence Eq.~(\ref{boundaryterm-def}) implies 
\begin{equation} 
S_{\partial \V} = -2 \int_{\S_2} K\, \sqrt{h} d^3y 
+ 2 \int_{\T} L\, \sqrt{-f} d^3z 
+ 2 \int_{\S_1} K\, \sqrt{h} d^3y 
+ 2 \oint_{\B_2} \eta\, \sqrt{\gamma} d^2\theta 
- 2 \oint_{\B_1} \eta\, \sqrt{\gamma} d^2\theta, 
\labell{Sboundary-TS} 
\end{equation} 
and we have reproduced Hayward's expression for the complete boundary
action when $\partial\V$ consists of the union of a timelike surface
$\T$, a past spacelike surface $\S_1$, and a future spacelike surface
$\S_2$. The boost parameter $\eta$ that appears in the integrals over
$\B_1$ and $\B_2$ is defined by Eq.~(\ref{eta-def}), and given more
explicitly by Eqs.~(\ref{eta-ln1}) and (\ref{eta-ln2}).  

\subsection{Rules for timelike and spacelike joints 
\labell{sec:TSjoints} }

\begin{figure}
\includegraphics[width=.9\linewidth]{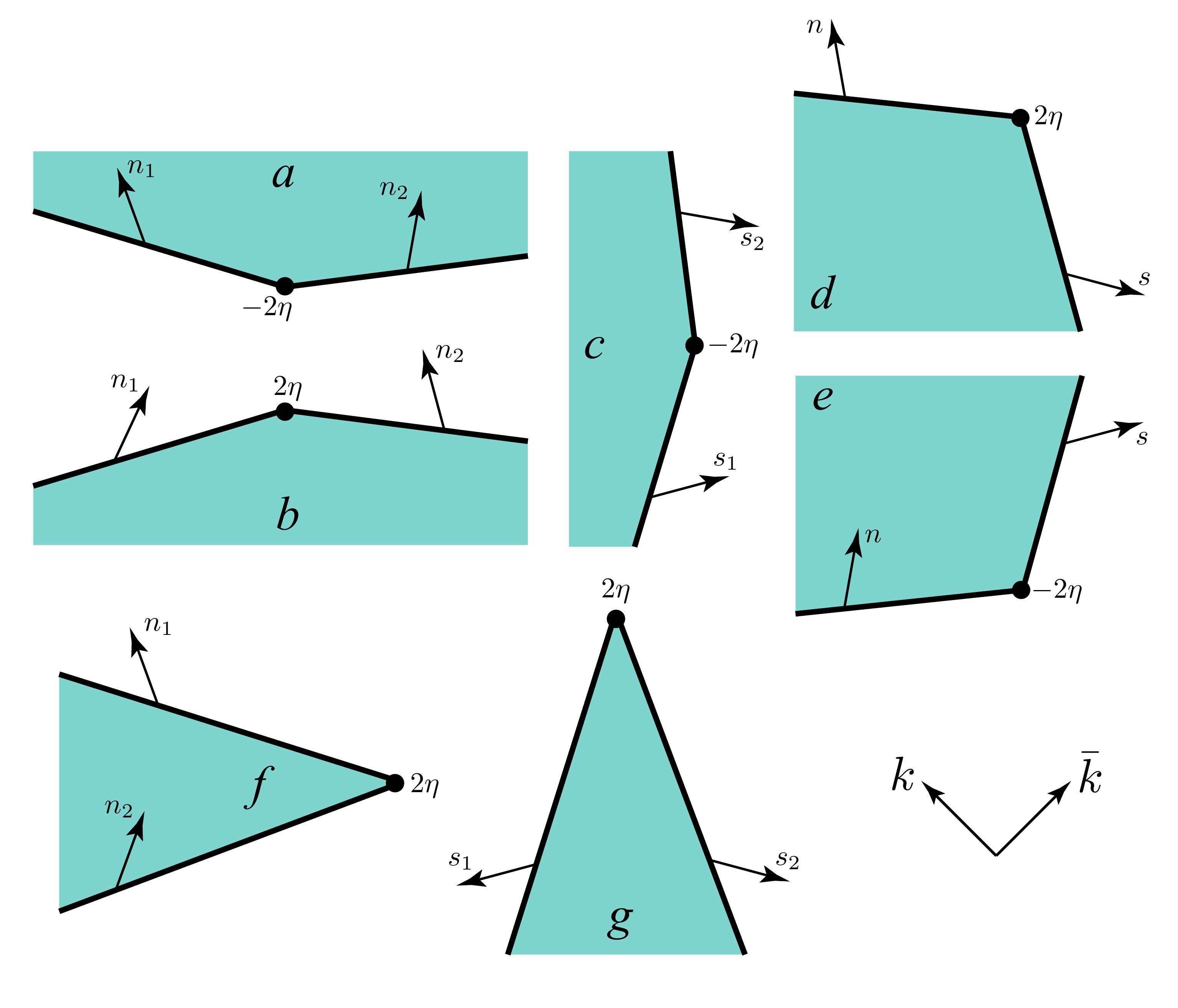}
\caption{Joint terms in the boundary action. In panel $a$, a past
  boundary is broken at $\B$ into two spacelike segments of normal 
  $n_1^\alpha$ and $n_2^\alpha$; the contribution from $\B$ to the
  boundary action is $-2\oint_\B \eta\, dS$. In panel $b$, a future
  boundary is broken into two spacelike segments; the contribution to
  the boundary action is $2\oint_\B \eta\, dS$. In panel $c$, a
  timelike boundary is broken into two timelike segments of normal
  $s_1^\alpha$ and $s_2^\alpha$; the contribution to the boundary
  action is $-2\oint_\B \eta\, dS$. In panel $d$, a timelike boundary
  of normal $s^\alpha$ is joined at $\B$ to a future, spacelike
  boundary of normal $n^\alpha$; the contribution to the boundary
  action is $2\oint_\B \eta\, dS$. In panel $e$, a timelike boundary
  is joined to a past boundary, with contribution 
  $-2\oint_\B \eta\, dS$. In panel $f$, two spacelike boundaries are
  joined, with contribution $2\oint_\B \eta\, dS$. Finally, two
  timelike boundaries are joined in panel $g$, with contribution 
  $2\oint_\B \eta\, dS$. In all panels the shaded region represents
  the interior of $\V$. The figure also shows the null vectors 
  $k^\alpha$ and $\bar{k}^\alpha$, which are introduced in the main
  text. \labell{fig:TSjoints} }
\end{figure} 

The considerations of Sec.~\ref{sec:closed-nonnull} can easily be
adapted to other types of joints between timelike and spacelike
boundary segments. A number of relevant cases are illustrated in
Fig.~\ref{fig:TSjoints}; a more complete set of situations was presented
in Hayward's original work \cite{Hayward:1993my}. For all these cases,
the boost parameter $\eta$ can be expressed in terms of the
projections of the normal vectors in the directions of the null
vectors $k^\alpha$ and $\bar{k}^\alpha$ introduced previously.  

In the situation depicted in panel $a$ of Fig.~\ref{fig:TSjoints}, we
have a past boundary broken at the two-surface $\B$ into two
spacelike segments of normal  $n_1^\alpha$ and $n_2^\alpha$. In this
case, the contribution from $\B$ to the boundary action is 
$-2\oint_\B \eta_a\, dS$, where $dS := \sqrt{\gamma}\, d^2\theta$ is a
surface element on $\B$, and where the boost parameter is given by   
\begin{equation} 
\eta_a = \ln(-n_1 \cdot k) - \ln(-n_2 \cdot k) 
= -\ln(-n_1 \cdot \bar{k}) + \ln(-n_2 \cdot \bar{k})\,. \labell{bend1}
\end{equation}  
In panel $b$, the past boundary is replaced by a future boundary, and 
the contribution to the boundary action is $2\oint_\B \eta_b\, dS$
with $\eta_b = \eta_a$. 

In the situation illustrated in panel $c$ of Fig.~\ref{fig:TSjoints}, we
have two timelike segments of normals $s_1^\alpha$ and $s_2^\alpha$
joined together at $\B$. In this case the contribution to the boundary
action is $-2\oint_\B \eta_c\, dS$, with  
\begin{equation} 
\eta_c = \ln(-s_1 \cdot k) - \ln(-s_2 \cdot k) 
= -\ln(s_1 \cdot \bar{k}) + \ln(s_2 \cdot \bar{k})\,. \labell{bend2}
\end{equation}  

Panel $d$ represents the situation examined in detail in
Sec.~\ref{sec:closed-nonnull}, which features a timelike boundary
of normal $s^\alpha$ joined at $\B$ to a future, spacelike boundary of
normal $n^\alpha$. In this case the contribution to the boundary
action is $2\oint_\B \eta_d\, dS$, with
\begin{equation} 
\eta_d = \ln(-n \cdot k) - \ln(-s \cdot k) 
= -\ln(-n \cdot \bar{k}) + \ln(s \cdot \bar{k})\,.\labell{bend3} 
\end{equation}  
In panel $e$, the future boundary is replaced by a past boundary, and
the contribution to the boundary action becomes $2\oint_\B \eta_e\, 
dS$, with $\eta_e = \eta_d$.    

In the situation depicted in panel $f$, we have two spacelike segments
of normal $n_1^\alpha$ and $n_2^\alpha$ joined together at $\B$. The
joint gives a contribution $2\oint_\B \eta_f\, dS$ to the boundary
integral, with  
\begin{equation} 
\eta_f = \ln(-n_1 \cdot k) - \ln(-n_2 \cdot k) 
= -\ln(-n_1 \cdot \bar{k}) + \ln(-n_2 \cdot \bar{k}). 
\end{equation}  
Finally, in panel $g$, we have two timelike segments of normal
$s_1^\alpha$ and $s_2^\alpha$, and the joint contribution is
$2\oint_\B \eta_g\, dS$, with   
\begin{equation} 
\eta_g = \ln(s_1 \cdot k) - \ln(-s_2 \cdot k) 
= -\ln(-s_1 \cdot \bar{k}) + \ln(s_2 \cdot \bar{k}). 
\end{equation}  

A summary of the general rules for the construction of joint terms for
intersections of spacelike and/or timelike boundary segments, as well
as all of the other boundary terms in the gravitational action, appear
in Appendix \ref{action-manual}. Note that our presentation of these
joint terms differs somewhat from that
originally given in \cite{Hayward:1993my,1994PhRvD..50.4914B}; 
our results, however,  are in precise agreement with those earlier
works. Our construction also provides an explicit prescription for the
sign of these terms, which was left ambiguous there. 

\subsection{Closed hypersurface: Null and spacelike segments 
\labell{sec:closed-null} }

In this section, we form a closed hypersurface $\partial \V$ by
combining null and spacelike hypersurfaces. Cases in which null
segments are joined to timelike hypersurfaces can be treated along the
same lines, but we shall not describe such a construction here. However,
the appropriate joint terms in the gravitational action for these situations will be described in Sec.~\ref{sec:Njoints}.

\subsubsection{Past light cone truncated by spacelike segments} 

\begin{figure} 
\includegraphics[width=.5\linewidth]{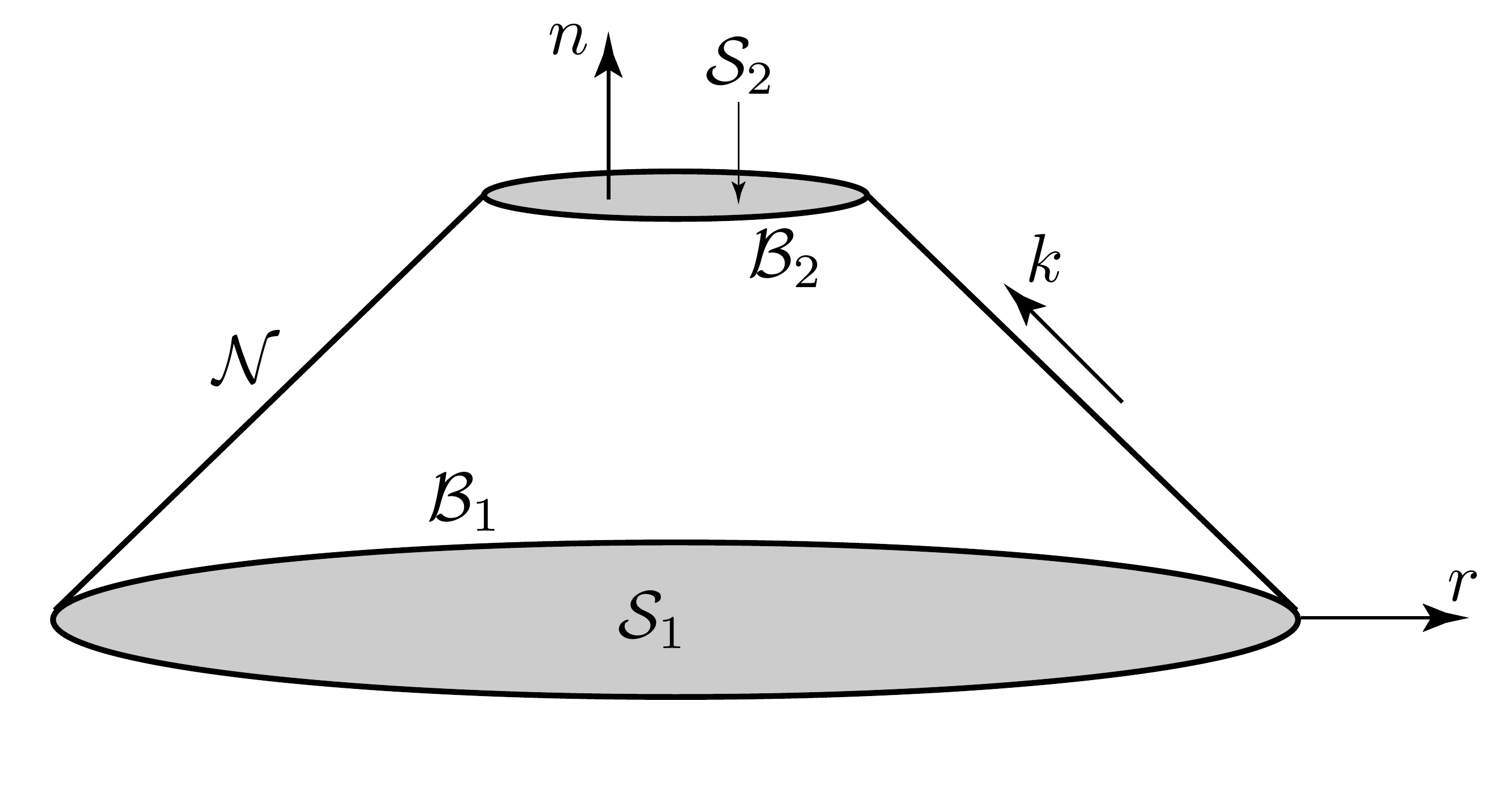}
\caption{A closed hypersurface $\partial \V$ consisting of a past
  spacelike surface $\S_1$, a truncated past null cone $\N$, and a
  future spacelike surface $\S_2$. 
\labell{fig2}} 
\end{figure} 

We begin by joining a truncated past light cone $\N$ to two spacelike
segments $\S_1$ (in the past) and $\S_2$ (in the future), as illustrated in
Fig.~\ref{fig2}. The intersection between $\N$ and $\S_1$ is the
two-surface $\B_1$, and $\B_2$ is the intersection between $\N$ 
and $\S_2$. From Eq.~(\ref{deltaS-nonnullfinal}), we have that 
\begin{equation} 
\int_{\S_2} \dslash v^\mu\, d\Sigma_\mu 
= \delta \biggl( 2 \int_{\S_2} K\, \sqrt{h} d^3y \biggr) 
- \oint_{\B_2} r_a \dslash A^a\,\sqrt{\gamma} d^2\theta,
\labell{S2term} 
\end{equation} 
where we inserted the relation $dS_a = r_a 
\sqrt{\gamma} d^2\theta$ for the surface element on $\B_2$. 
According to the conventions introduced in Sec.~\ref{sec:nonnull},
$d\Sigma_\mu \propto \partial_\mu \Phi$, with the function $\Phi$
increasing toward the future of the hypersurface. Because this 
coincides with the direction out of $\V$, we have that
$d\Sigma_\mu$ is correctly oriented on $\partial\V$. From 
Eq.~(\ref{deltaS-nonnullfinal}) we also get 
\begin{equation} 
\int_{\S_1} \dslash v^\mu\, d\Sigma_\mu 
= \delta \biggl( -2 \int_{\S_1} K\, \sqrt{h} d^3y \biggr) 
+ \oint_{\B_1} r_a \dslash A^a\,\sqrt{\gamma} d^2\theta,
\labell{S1term} 
\end{equation} 
with the change in sign accounting for the fact that the outward
direction now coincides with the past of $\S_1$. On the other hand,
Eq.~(\ref{deltaS-nullfinal}) gives  
\begin{equation} 
\int_{\N} \dslash v^\mu\, d\Sigma_\mu =  
\delta \biggl( 2 \int_{\N} \kappa\, \sqrt{\gamma} 
d^2\theta d\lambda \biggr) 
- \oint_{\B_2} \delta a\, \sqrt{\gamma} d^2\theta
+ \oint_{\B_1} \delta a\, \sqrt{\gamma} d^2\theta, 
\labell{Nterm} 
\end{equation} 
with the same convention that $d\Sigma_\mu \propto \partial_\mu \Phi$,
with $\Phi$ increasing toward the future of $\N$, which coincides with
the exterior of $\V$.   

Combining these expressions produces 
\begin{align} 
\int_{\partial\V} \dslash v^\mu\, d\Sigma_\mu  &= 
\delta \biggl( 2 \int_{\S_2} K\, \sqrt{h} d^3y
+ 2 \int_{\N} \kappa\, \sqrt{\gamma} d^2\theta d\lambda 
-2 \int_{\S_1} K\, \sqrt{h} d^3y \biggr) 
\nonumber \\ & \quad \mbox{} 
- \oint_{\B_2} \bigl( \delta a + r_a \dslash A^a \bigr)\,
\sqrt{\gamma} d^2\theta 
+ \oint_{\B_1} \bigl( \delta a + r_a \dslash A^a \bigr)\,
\sqrt{\gamma} d^2\theta.  
\end{align} 
Below we shall show that $r_a \dslash A^a = \delta a$ when a truncated
past light cone is joined to a segment of spacelike hypersurface. This
remarkable property allows us to write  
\begin{equation} 
\int_{\partial\V} \dslash v^\mu\, d\Sigma_\mu  
= \delta \biggl( 2 \int_{\S_2} K\, \sqrt{h} d^3y
+ 2 \int_{\N} \kappa\, \sqrt{\gamma} d^2\theta d\lambda 
-2 \int_{\S_1} K\, \sqrt{h} d^3y 
- 2\oint_{\B_2} a\, \sqrt{\gamma} d^2\theta 
+ 2\oint_{\B_1} a\, \sqrt{\gamma} d^2\theta \biggr),   
\end{equation} 
and to identify the boundary action  
\begin{equation} 
S_{\partial\V} = -2 \int_{\S_2} K\, \sqrt{h} d^3y
- 2 \int_{\N} \kappa\, \sqrt{\gamma} d^2\theta d\lambda 
+ 2 \int_{\S_1} K\, \sqrt{h} d^3y 
+ 2\oint_{\B_2} a\, \sqrt{\gamma} d^2\theta 
- 2\oint_{\B_1} a\, \sqrt{\gamma} d^2\theta\,.    
\labell{SV-fig2} 
\end{equation} 
We shall also show that in Eq.~(\ref{SV-fig2}), $a$ must be of the
form 
\begin{equation} 
a = \ln(-n \cdot k) + a_0\,, 
\end{equation} 
where $n^\alpha$ is the unit normal to $\S_1$ or $\S_2$, $k^\alpha$ is 
the null normal to $\N$, $n \cdot k$ is their inner product, and $a_0$ is
an arbitrary quantity that satisfies $\delta a_0 = 0$.   

Reiterating the statements made near the end of 
Sec.~\ref{subsec:null-boundary-term}, we observe that in general, the
boundary action is ill-defined because it depends on the choices made
for the parameter $\lambda$ and function $a_0$.   

\subsubsection{Future light cone truncated by spacelike segments} 

\begin{figure} 
\includegraphics[width=.5\linewidth]{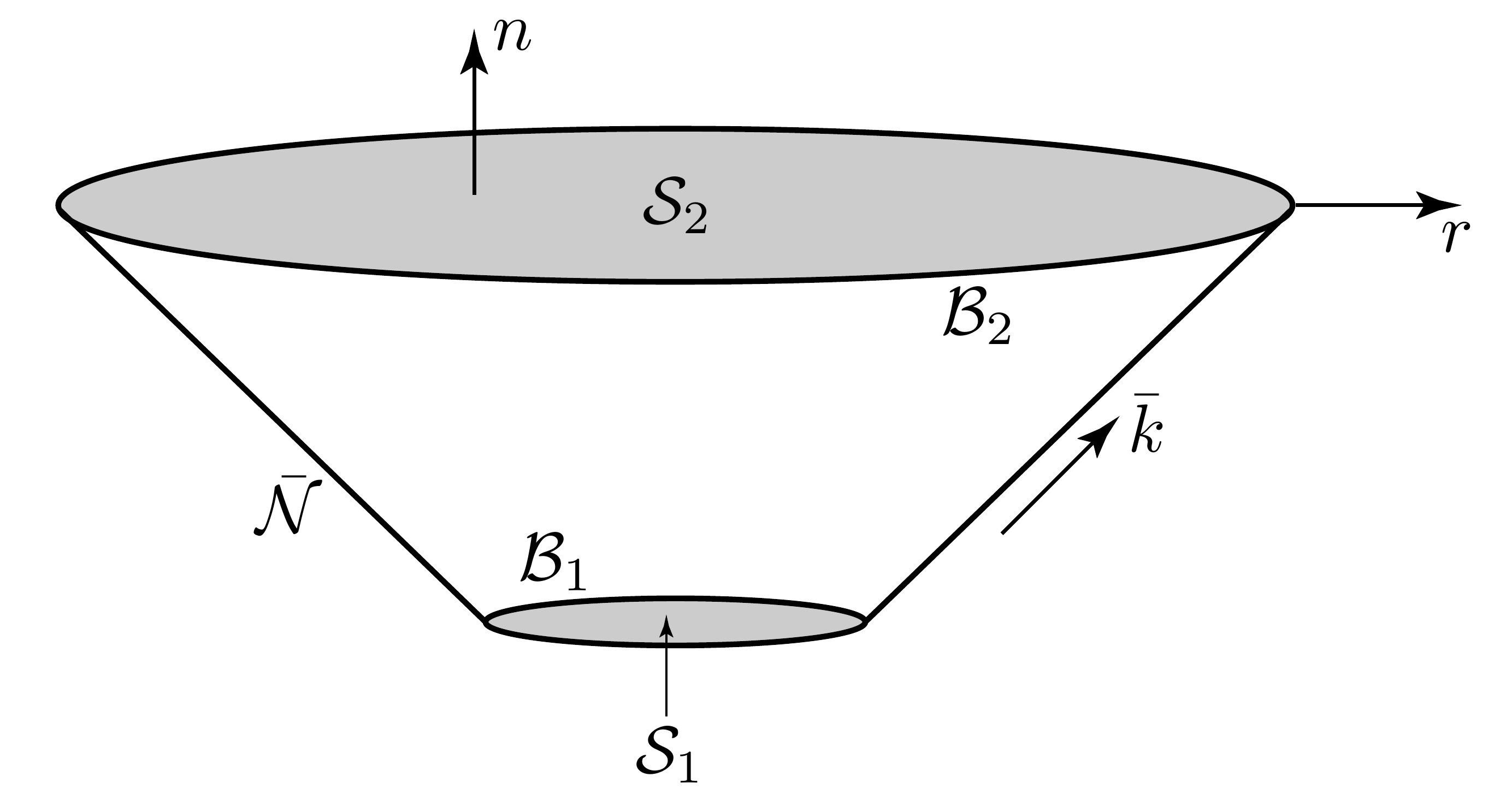}
\caption{A closed hypersurface $\partial \V$ consisting of a past
  spacelike surface $\S_1$, a truncated future null cone $\bar{\N}$,
  and a future spacelike surface $\S_2$. 
\labell{fig3}} 
\end{figure} 

Next we form a closed hypersurface $\partial \V$ by joining a
truncated future light cone $\bar{\N}$ to two spacelike segments
$\S_1$ (in the past) and $\S_2$ (in the future); see
Fig.~\ref{fig3}. The intersection between $\bar{\N}$ and $\S_1$ is the  
two-surface $\B_1$, and $\B_2$ is the intersection between $\bar{\N}$  
and $\S_2$. 

The contributions to the $\partial\V$ integral coming from $\S_2$ and
$\S_1$ are still given by Eqs.~(\ref{S2term}) and (\ref{S1term}),
respectively. The contribution from $\bar{\N}$, however, requires us
to introduce an overall minus sign in Eq.~(\ref{Nterm}), so that 
\begin{equation} 
\int_{\bar{\N}} \dslash v^\mu\, d\Sigma_\mu =  
\delta \biggl( -2 \int_{\bar{\N}} \bar{\kappa}\, \sqrt{\gamma} 
d^2\theta d\bar{\lambda} \biggr) 
+ \oint_{\B_2} \delta \bar{a}\, \sqrt{\gamma} d^2\theta
- \oint_{\B_1} \delta \bar{a}\, \sqrt{\gamma} d^2\theta, 
\labell{Nbarterm} 
\end{equation} 
where quantities with overbars refer to the null generators of the
future light cone. The minus sign accounts for the fact that in the
conventions employed to derive Eq.~(\ref{deltaS-nullfinal}),
$d\Sigma_\mu \propto \partial_\mu \Phi$ with $\Phi$ increasing toward
the future of $\bar{\N}$. This direction corresponds to the interior
of $\V$, and a correct outward orientation for $d\Sigma_\mu$
therefore requires the change of sign.  

Combining these expressions produces 
\begin{align} 
\int_{\partial\V} \dslash v^\mu\, d\Sigma_\mu  &= 
\delta \biggl( 2 \int_{\S_2} K\, \sqrt{h} d^3y
- 2 \int_{\bar{\N}} \bar{\kappa}\, 
  \sqrt{\gamma} d^2\theta d\bar{\lambda}  
-2 \int_{\S_1} K\, \sqrt{h} d^3y \biggr) 
\nonumber \\ & \quad \mbox{} 
+ \oint_{\B_2} \bigl( \delta \bar{a} - r_a \dslash A^a \bigr)\,
\sqrt{\gamma} d^2\theta 
- \oint_{\B_1} \bigl( \delta \bar{a} - r_a \dslash A^a \bigr)\,
\sqrt{\gamma} d^2\theta.  
\end{align} 
Below we shall show that $r_a \dslash A^a = -\delta \bar{a}$ when a
truncated future light cone is joined to a segment of spacelike
hypersurface. This allows us to write 
\begin{equation} 
\int_{\partial\V} \dslash v^\mu\, d\Sigma_\mu  
= \delta \biggl( 2 \int_{\S_2} K\, \sqrt{h} d^3y
- 2 \int_{\bar{\N}} \bar{\kappa}\, 
   \sqrt{\gamma} d^2\theta d\bar{\lambda} 
-2 \int_{\S_1} K\, \sqrt{h} d^3y 
+ 2\oint_{\B_2} \bar{a}\, \sqrt{\gamma} d^2\theta 
- 2\oint_{\B_1} \bar{a}\, \sqrt{\gamma} d^2\theta \biggr),   
\end{equation} 
and to identify the boundary action  
\begin{equation} 
S_{\partial\V} = -2 \int_{\S_2} K\, \sqrt{h} d^3y
+ 2 \int_{\bar{\N}} \bar{\kappa}\, 
   \sqrt{\gamma} d^2\theta d\bar{\lambda}  
+ 2 \int_{\S_1} K\, \sqrt{h} d^3y 
- 2\oint_{\B_2} \bar{a}\, \sqrt{\gamma} d^2\theta 
+ 2\oint_{\B_1} \bar{a}\, \sqrt{\gamma} d^2\theta.    
\labell{SV-fig3} 
\end{equation} 
Further, we shall show that in Eq.~(\ref{SV-fig3}), $\bar{a}$ must be of
the form 
\begin{equation} 
\bar{a} = \ln(-n \cdot \bar{k}) + \bar{a}_0,   
\end{equation} 
where $n^\alpha$ is the unit normal to $\S_1$ or $\S_2$, $\bar k^\alpha$ is
the null normal to $\bar{\N}$, $n \cdot \bar{k}$ is their inner
product, and $\bar{a}_0$ is an arbitrary quantity that satisfies
$\delta \bar{a}_0 = 0$.     

Once more we observe that the boundary action is ill-defined because 
it depends on the choices made for the parameter $\bar{\lambda}$ and
function $\bar{a}_0$.    

\subsubsection{Past and future light cone truncated by spacelike
  segments}  

\begin{figure} 
\includegraphics[width=.5\linewidth]{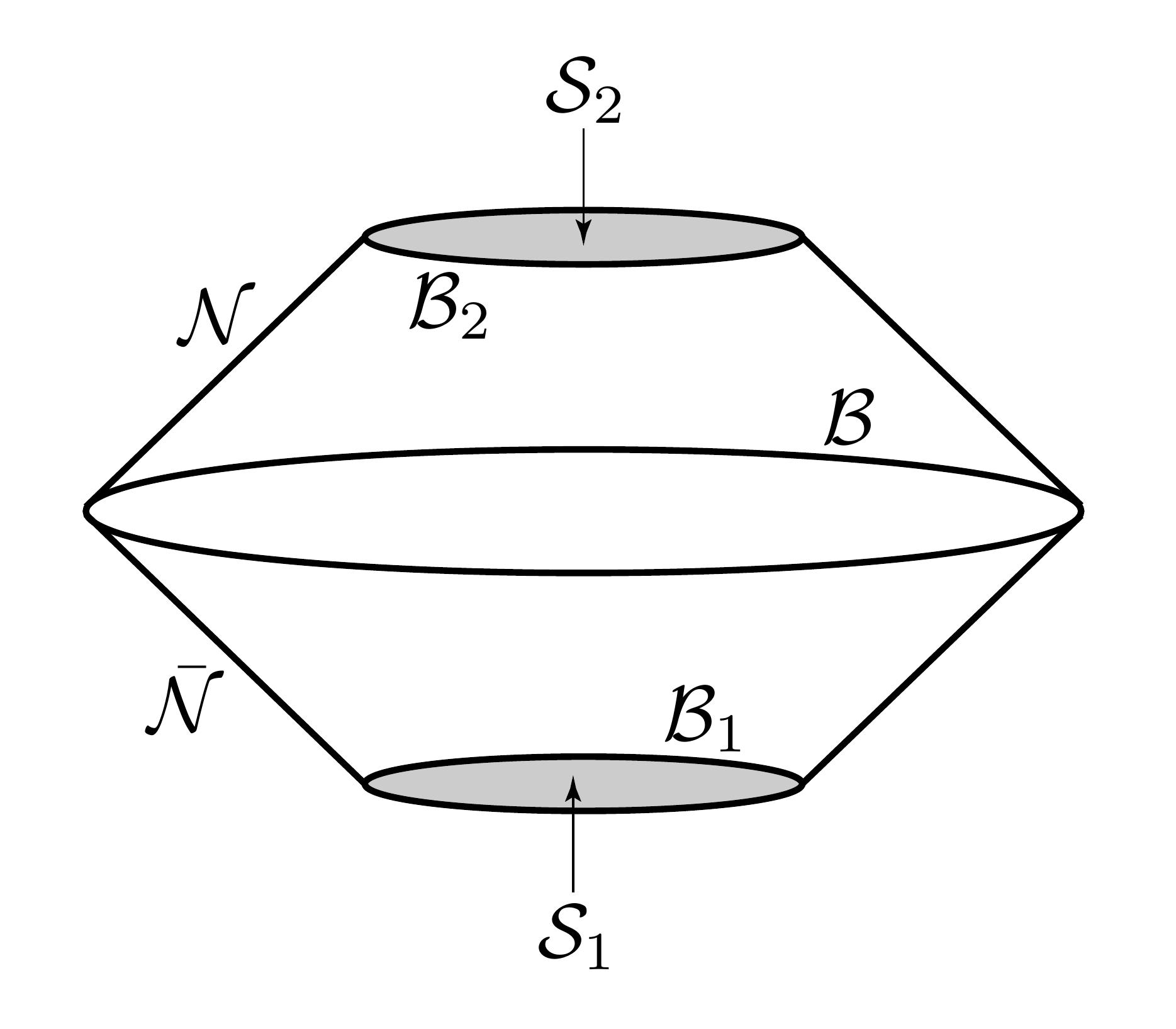}
\caption{A closed hypersurface $\partial \V$ consisting of a past  
  spacelike surface $\S_1$, a truncated past null cone $\N$, a
  truncated future null cone $\bar{\N}$, and a future spacelike
  surface $\S_2$. 
\labell{fig4} }
\end{figure} 

As a final variation on the theme, we form $\partial \V$ by taking the
union of a past spacelike surface $\S_1$, a truncated future null cone
$\bar{\N}$, a truncated past null cone $\N$, and a future spacelike
surface $\S_2$; see Fig.~\ref{fig4}. The intersection between
$\bar{\N}$ and $\S_1$ is the two-surface $\B_1$, the intersection
between $\bar{\N}$ and $\N$ is the two-surface $\B$, and $\B_2$ is the
intersection between $\N$ and $\S_2$.  

With the contributions listed previously we have that 
\begin{align} 
\int_{\partial\V} \dslash v^\mu\, d\Sigma_\mu  &= 
\delta \biggl( 2 \int_{\S_2} K\, \sqrt{h} d^3y
+ 2 \int_{\N} \kappa\, 
  \sqrt{\gamma} d^2\theta d\lambda  
- 2 \int_{\bar{\N}} \bar{\kappa}\, 
  \sqrt{\gamma} d^2\theta d\bar{\lambda}  
-2 \int_{\S_1} K\, \sqrt{h} d^3y \biggr) 
\nonumber \\ & \quad \mbox{} 
- \oint_{\B_2} \bigl( \delta a + r_a \dslash A^a \bigr)\,
\sqrt{\gamma} d^2\theta 
+ \oint_{\B} ( \delta a + \delta \bar{a})\,
\sqrt{\gamma} d^2\theta 
- \oint_{\B_1} \bigl( \delta \bar{a} - r_a \dslash A^a \bigr)\,
\sqrt{\gamma} d^2\theta.  
\end{align} 
We have already stated that $r_a \dslash A^a = \delta a$ on $\B_2$
and $r_a \dslash A^a = -\delta \bar{a}$ on $\B_1$. We may also
show that $\delta \bar{a} = \delta a$ on $\B$, and use this property
to simplify the expression. We note first that on $\B$, the null
vectors $k^\alpha$ and $\bar{k}^\alpha$ satisfy  
\begin{equation} 
k \cdot \bar{k} = -c\,,  
\end{equation} 
where $c$ is a positive scalar field. Next we take the variation of 
$c = -k_\alpha \bar{k}^\alpha$, recalling that 
$\delta \bar{k}^\alpha = 0$ and invoking Eq.~(\ref{deltak}) for
$\delta k_\alpha$; we find that $\delta c = c\, \delta a$. Doing the
same with $c = -\bar{k}_\alpha k^\alpha$, we now find that 
$\delta c = c\, \delta \bar{a}$ and conclude that indeed, 
$\delta \bar{a} = \delta a$. Our manipulations also reveal that
$\delta a = \delta \ln c = \delta \ln(-k \cdot \bar{k})$, so that 
$a = \ln(-k \cdot \bar{k}) + \hat{a}_0$, where $\hat{a}_0$ is an
arbitrary function such that $\delta \hat{a}_0 = 0$.    

With all these results in hand, we find that 
\begin{align} 
\int_{\partial\V} \dslash v^\mu\, d\Sigma_\mu &= 
\delta \biggl( 2 \int_{\S_2} K\, \sqrt{h} d^3y
+ 2 \int_{\N} \kappa\, 
  \sqrt{\gamma} d^2\theta d\lambda  
- 2 \int_{\bar{\N}} \bar{\kappa}\, 
  \sqrt{\gamma} d^2\theta d\bar{\lambda}  
-2 \int_{\S_1} K\, \sqrt{h} d^3y  
\nonumber \\ & \quad \mbox{} 
- 2 \oint_{\B_2} a\, \sqrt{\gamma} d^2\theta 
+ 2 \oint_{\B} a\, \sqrt{\gamma} d^2\theta 
- 2 \oint_{\B_1} \bar{a}\, \sqrt{\gamma} d^2\theta \biggr),   
\end{align} 
and we have identified the boundary action 
\begin{align} 
S_{\partial\V} &= 
-2 \int_{\S_2} K\, \sqrt{h} d^3y
- 2 \int_{\N} \kappa\, 
  \sqrt{\gamma} d^2\theta d\lambda  
+ 2 \int_{\bar{\N}} \bar{\kappa}\, 
  \sqrt{\gamma} d^2\theta d\bar{\lambda}  
+ 2 \int_{\S_1} K\, \sqrt{h} d^3y
\nonumber \\ & \quad \mbox{} 
+ 2 \oint_{\B_2} a\, \sqrt{\gamma} d^2\theta 
- 2 \oint_{\B} a\, \sqrt{\gamma} d^2\theta 
+ 2 \oint_{\B_1} \bar{a}\, \sqrt{\gamma} d^2\theta.    
\labell{SV-fig4} 
\end{align} 
The joint terms all come with specific forms for the integrand: On
$\B_1$ we have that $\bar{a} = \ln(-n_1\cdot \bar{k}) + \bar{a}_0$,
where $n_1^\alpha$ is the unit normal to $\S_1$, on $\B_2$ we have 
that $a = \ln(-n_2\cdot k) + a_0$, where $n_2^\alpha$ is normal to
$\S_2$, and on $\B$ we have instead $a = \ln(-k \cdot \bar{k}) +
\hat{a}_0$.  

Again this boundary action \reef{SV-fig4} is ill-defined because it depends on the choices
made for the parameters $\lambda$ and $\bar{\lambda}$, as well as the
functions $a_0$, $\bar{a}_0$, and $\hat{a}_0$.    

\subsubsection{Proof that $r_a \dslash A^a = \pm \delta a$ 
and $\delta a = \delta \ln(- n \cdot k )$
\labell{subsubsec:proof} }

We now establish that 
\begin{equation} 
r_a \dslash A^a = -\zeta \delta a, \qquad 
\delta a = \delta \ln(-n \cdot k)   
\labell{radA} 
\end{equation} 
on a two-surface $\B$ formed from the intersection of a spacelike 
surface $\S$ and a null surface $\N$. Here, $\zeta = -1$ when $\N$ is
a past light cone, and $\zeta = +1$ when it is a future light
cone. The (future-directed) unit vector $n^\alpha$ is normal to $\S$, $k^\alpha$ is
the (future-directed) normal to $\N$, and $n \cdot k := g_{\alpha\beta}\, n^\alpha k^\beta$ is
their inner product.  

We rely on a system of adapted coordinates $x^\alpha = (\lambda, r,
\theta^A)$ defined in an open domain $\V$ that includes $\N$ and
$\S$. We have that $\lambda$ is a time coordinate, and surfaces of
constant $\lambda$ provide a foliation of $\V$ in spacelike
hypersurfaces; the coordinate is defined such that 
$\lambda = \lambda_0$ on $\S$. We also have that $r$ is constant
on each member of a family of nested hypersurfaces, which can be
either timelike or null; it is such that $r = r_0$ on $\N$. When
intersected with a surface of constant $\lambda$ such  
as $\S$, the hypersurfaces of constant $r$ become nested spheres, and 
$\B$ is also described by $r = r_0$. Finally, the angular coordinates
$\theta^A$ range over the spheres of constant $\lambda$ and $r$. The
coordinates are illustrated in Fig.~\ref{fig5}.   

\begin{figure} 
\includegraphics[width=.5\linewidth]{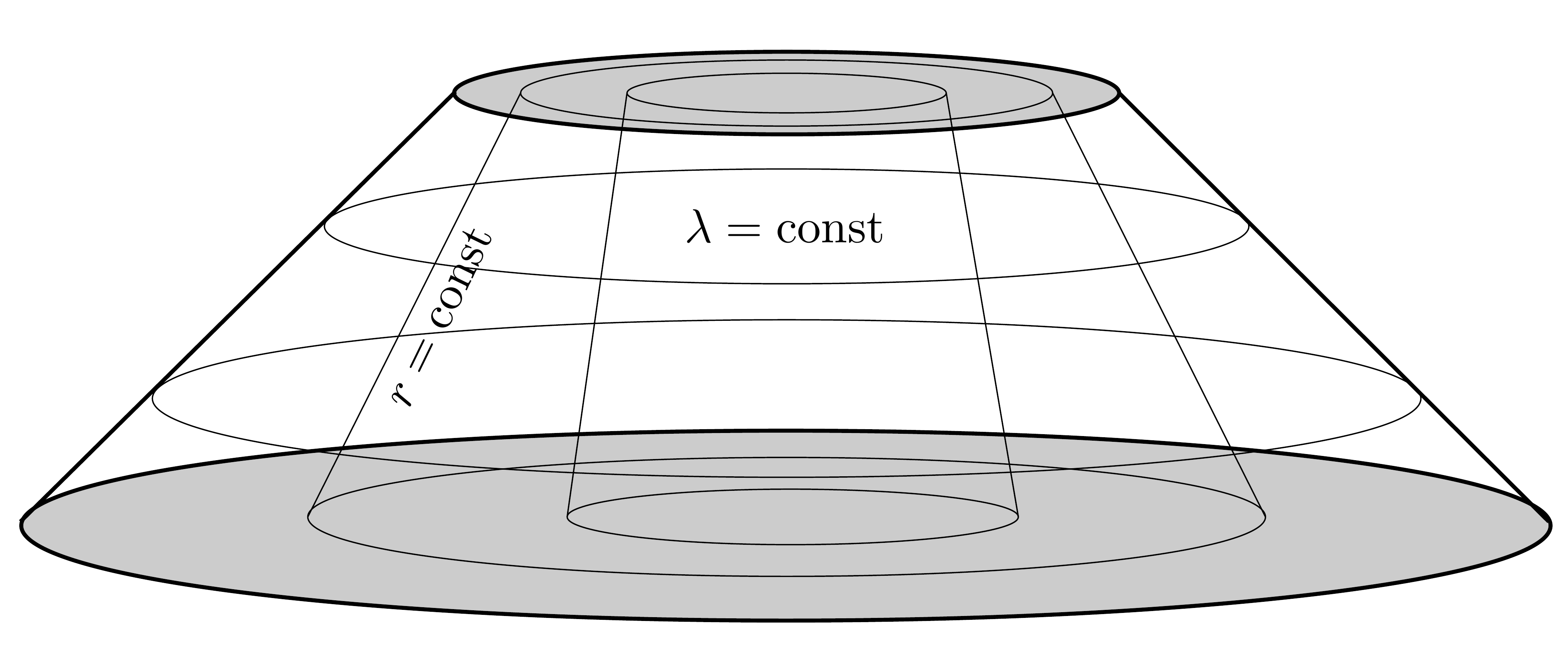}
\caption{Adapted coordinates $(\lambda, r, \theta^A)$ in $\V$.   
\labell{fig5} }
\end{figure} 
 
In these coordinates, $\N$ is the hypersurface $r = r_0$, and
$(\lambda, \theta^A)$ are intrinsic coordinates. When $\N$ is a past
light cone ($\zeta = -1$), $r$ increases toward the future of $\N$,
and when $\N$ is a future light cone ($\zeta = +1$), $r$ increases
toward its past. The null generators are parametrized with $\lambda$,
and the angular coordinates are calibrated to ensure that $\theta^A$
is constant on each generator. We have that $k^\alpha = (1,0,0,0)$,
$k_\alpha = (0,\zeta\alpha,0,0)$ for some scalar $\alpha>0$, and $k_\alpha$
is orthogonal to $e^\alpha_A$. These relations imply that 
$g_{\lambda\lambda} = 0$, $g_{\lambda r} = \zeta\alpha$,
$g_{\lambda A} = 0$, and $g_{AB} = \gamma_{AB}$ on $\N$. 

The hypersurface $\S$ is described by $\lambda = {\rm const}$, and 
$(r,\theta^A)$ are intrinsic coordinates. Its unit normal is 
$n_\alpha = (-1/\beta,0,0,0)$, with $\beta>0$ a scalar field on $\S$, 
and we have that $g^{\lambda\lambda} = -\beta^2$ on $\S$. The
two-surface $\B$ is at $r=r_0$ in $\S$, and $\theta^A$
serve as intrinsic coordinates. Its unit normal is 
$r_a = (1/\gamma,0,0)$, where $\gamma>0$ is a scalar field on 
$\B$, and we have that $h^{rr} = \gamma^2$ on $\B$.  
Because $\B$ is also a surface $\lambda = {\rm const}$ of
$\Sigma$, its induced metric is necessarily $\gamma_{AB}$. 

By virtue of the foregoing results, the spacetime metric evaluated on
$\B$ has the structure 
\begin{equation} 
g_{\alpha\beta} = \left( 
\begin{array}{cccc} 
0 & \zeta\alpha & 0 & 0 \\ 
\zeta\alpha & h_{rr} & h_{r2} & h_{r3} \\ 
0 & h_{r2} & \gamma_{22} & \gamma_{23} \\ 
0 & h_{r3} & \gamma_{23} & \gamma_{33} 
\end{array} 
\right), 
\end{equation} 
and the spatial metric $h_{ab}$ is given by the submatrix that
excludes the first row and column. A key observation is that with
$h_{ab}$ fixed on $\S$ and $\gamma_{AB}$ fixed on $\N$, the only
variable component of the metric is $g_{\lambda r} = \zeta\alpha$. 
Calculation of $g^{\alpha\beta}$ and $h^{ab}$ reveals that 
\begin{equation} 
\beta\gamma = \alpha^{-1}, 
\labell{betagamma} 
\end{equation}
a result that will be required presently. 

We may now proceed with the derivation of Eq.~(\ref{radA}). We first
invoke Eq.~(\ref{deltaA}) and calculate  
\begin{equation} 
r_a \dslash A^a = r_a e^a_\alpha n_\beta \delta g^{\alpha\beta} 
= - r^a e_a^\alpha n^\beta \delta g_{\alpha\beta} 
= -r^r n^\lambda \delta g_{\lambda r}. 
\end{equation} 
Writing $r^r = h^{ra} r_a = h^{rr}/\gamma = \gamma$, 
$n^\lambda = g^{\lambda \alpha} n_\alpha = -g^{\lambda\lambda}/\beta =
\beta$, we may conclude that 
\begin{equation} 
r_a \dslash A^a = -\zeta \beta\gamma\, \delta\alpha 
= -\zeta \frac{\delta\alpha}{\alpha} = -\zeta\, \delta \ln\alpha. 
\end{equation}
The definition of $a := \ln\mu$ is provided by the equation 
$k_\alpha = \zeta \mu\, \partial_\alpha \Phi$, which relates
$k_\alpha$ to the gradient of an arbitrary function $\Phi$ that goes
to zero on $\N$. In our adapted coordinates we can always write 
$\Phi = (r-r_0) \Psi(\lambda, r, \theta^A)$, where $\Psi$ is another
arbitrary function, and conclude that $k_r = \zeta \mu \Psi$. Since
this must be equal to $\zeta \alpha$, we have that 
$\alpha = \mu \Psi$, or $a = \ln\alpha - \ln\Psi$. Because $\Phi$ and
$\Psi$ are fixed during the variation, we have that 
$\delta a = \delta \ln\alpha$, and 
\begin{equation} 
r_a \dslash A^a = -\zeta \delta a. 
\end{equation} 
The first part of Eq.~(\ref{radA}) is thus established. To establish
the second part we observe that 
$n \cdot k = k^\alpha n_\alpha = -1/\beta = -\gamma \alpha$, so that
$\ln(-n \cdot k) = \ln\gamma + \ln\alpha$. But $\gamma^2 = h^{rr}$ is
fixed during the variation, so 
\begin{equation} 
\delta a = \delta\ln\alpha = \delta\ln(-n \cdot k), 
\end{equation}   
as required. Notice that $\delta a$ is now expressed independently of
the adapted coordinates. This relation can be integrated to yield 
\begin{equation} 
a = \ln(-n \cdot k) + a_0, 
\labell{a-vs-a0} 
\end{equation} 
where $a_0$ is an arbitrary scalar field on $\B$ whose variation
$\delta a_0$ is required to vanish. 

The lesson behind the result of Eq.~(\ref{a-vs-a0}) is that while the
piece $\ln(-n \cdot k)$ of $a$ becomes determined when a segment of
null hypersurface is joined to a spacelike segment, the remaining
piece $a_0$ continues to be arbitrary. The first piece 
$\ln(-n \cdot k)$ contains the dependence on the choice of
parametrization, while the second piece $a_0$ contains the dependence
on the choice of $\Phi$.   

\subsection{Rules for null joints
\labell{sec:Njoints} }

\begin{figure}
\includegraphics[width=.9\linewidth]{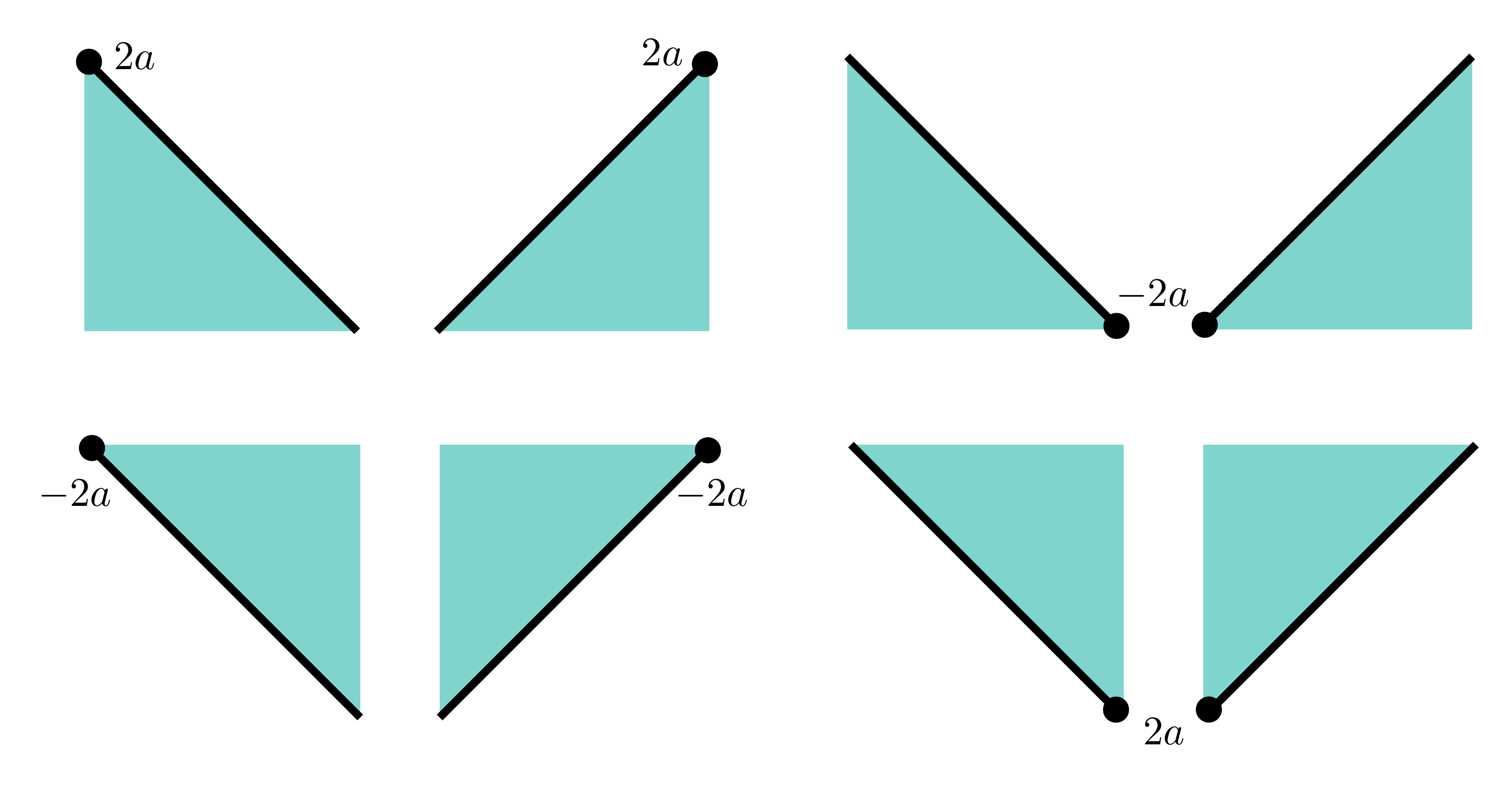}
\caption{Joint terms in the boundary action. The upper-left panel
  displays a null segment joined at $\B$ to another segment (not
  shown) which can be either spacelike, timelike, or null; because 
  $\B$ is a future boundary to the null segment and the outward
  direction is a future direction, the contribution from $\B$ to the
  boundary action is $2 \oint_\B a\, dS$. In the upper-right panel,
  $\B$ is a past boundary, the outward direction continues to be a
  future direction, and the contribution to the boundary action is 
  $-2\oint_\B a\, dS$. In the lower-left panel, $\B$ is
  again a future boundary, but the outward direction is now a past
  direction; for these cases the contribution to the boundary action
  is $-2\oint_\B a\, dS$. In the lower-right panel, $\B$ is a past
  boundary, the outward direction is a past direction, and the
  contribution to the boundary action is $2\oint_\B a\, dS$. In all
  panels the shaded region indicates the interior of $\V$. 
\labell{fig:Njoints} }
\end{figure} 

The considerations of Sec.~\ref{sec:closed-null} can easily be
adapted to other types of joints involving one or two null
hypersurfaces. All possible cases are illustrated in
Fig.~\ref{fig:Njoints}, which displays a null segment and its boundary
$\B$; the (null, timelike, or spacelike) segment to which it is joined
is not shown. When $\B$ is a future boundary and the outward direction
across the null hypersurface coincides with the future direction
(upper-left panel), the contribution to the boundary action is
$2\oint_\B a\, dS$, where $dS := \sqrt{\gamma}\, d^2\theta$ is the
surface element on $\B$. When $\B$ is a past boundary and the outward 
direction still coincides with the future direction (upper-right
panel), the contribution to the boundary action is 
$-2\oint_\B a\, dS$. When $\B$ is a future boundary and the outward
direction coincides with the past direction (lower-left panel), the
contribution to the boundary action is again $-2\oint_\B a\, dS$. And
finally, when $\B$ is a past boundary and the outward direction
coincides with the past direction (lower-right panel), the
contribution to the boundary action is $2\oint_\B a\, dS$.  

When a null segment is joined at $\B$ to a spacelike segment, we have
seen that 
\begin{equation} 
a^{\rm spacelike} = \ln(-n \cdot k) + a^{\rm spacelike}_0, 
\end{equation} 
where $n^\alpha$ is the unit normal to the spacelike segment,
$k^\alpha$ is the null normal, $n \cdot k := g_{\alpha\beta} n^\alpha
k^\beta$ is their inner product, and $a^{\rm spacelike}_0$ is an
arbitrary scalar field on $\B$ required to have a vanishing
variation. When the null segment is joined instead to a timelike
segment, a calculation similar to the one carried out in
Sec.~\ref{subsubsec:proof} would reveal that in this case,  
\begin{equation} 
a^{\rm timelike} = \ln|s \cdot k| + a^{\rm timelike}_0, 
\end{equation} 
where $s^\alpha$ is the unit outward normal to the timelike segment,
and $a^{\rm timelike}_0$ in another arbitrary scalar field with zero
variation. And when the null segment is joined to another null
segment, we have seen that  
\begin{equation} 
a^{\rm null} = \ln(-k \cdot \bar{k}) + a^{\rm null}_0, 
\end{equation} 
where $\bar{k}^\alpha$ is the normal to the second null segment, and
$a^{\rm null}_0$ is yet another arbitrary scalar field with vanishing
variation.  

\subsection{Additivity rules 
\labell{sec:additivity} } 

We will say that an action is additive if the action for the union of
two regions $\V_1$ and $\V_2$ is equal to the sum of the actions for
$\V_1$ and $\V_2$ separately, when the action is evaluated on
field configurations that extend across $\V=\V_1\cup\V_2$. This is not a
property that is typically discussed in the context of classical field
theory, but it was in fact a primary consideration in
\cite{PhysRevD.15.2752}. There, addivity of the gravitational action
was argued to be a requirement for quantum amplitudes (as described by 
path integrals in quantum gravity) to be additive, and this was
presented as a motivation to introduce the Gibbons-Hawking-York
boundary term. However, it was subsequently shown
that when taking into account the contribution of joint terms, the
gravitational action $S = S_{\V} + \S_{\partial \V}$ is not additive
in general \cite{1994PhRvD..50.4914B}. More precisely, obstacles to
additivity arise from timelike joints (at the intersection of
two timelike boundary surfaces). The volume and 
hypersurface terms are all properly additive, and there is no obstacle
to additivity coming from joints (between timelike and/or spacelike
segments) that are entirely spacelike --- the case
considered throughout this paper. 

The consideration of null boundary segments creates
additional obstacles to additivity, due to the arbitrariness
associated with joint 
terms. An example of this situation is provided by Fig.~\ref{fig4},
which can be viewed as the union of Figs.~\ref{fig2} and \ref{fig3},
with the $\S_1$ of Fig.~\ref{fig2} identified with the $\S_2$ of
Fig.~\ref{fig3}. In this case we find that according to
Eq.~(\ref{SV-fig4}), the joint term coming from $\B_1 \equiv \B_2 
\equiv \B$ is given by 
\begin{equation} 
S_\B[\mbox{Fig.~\ref{fig4}}] 
= -2 \oint_{\B} a\,\sqrt{\gamma} d^2\theta, 
\end{equation} 
with $a = \ln(-k \cdot \bar{k}) + \hat{a}_0$. On the other hand,   
Eqs.~(\ref{SV-fig2}) and (\ref{SV-fig3}) imply 
\begin{equation} 
S_\B [\mbox{Fig.~\ref{fig2}} \cup \mbox{Fig.~\ref{fig3}}] 
= -2 \oint_{\B} (a_1 +\bar{a}_2)\, \sqrt{\gamma} d^2\theta,   
\end{equation} 
with $a_1 = \ln(-n_1 \cdot k) + a_{01}$ and 
$\bar{a}_2 = \ln(-n_2 \cdot \bar{k}) + \bar{a}_{02}$; because the
$\S_1$ of Fig.~\ref{fig2} is identified with the $\S_2$ of
Fig.~\ref{fig3}, we have that 
$n^\alpha_1 \equiv n^\alpha_2 \equiv n^\alpha$. To work out the
relation between $a$ and $a_1+\bar{a}_2$, we decompose the null vectors 
$k^\alpha$ and $\bar{k}^\alpha$ in a basis consisting of the mutually
orthogonal unit vectors $n^\alpha$ and $r^\alpha$, the second vector 
pointing out of $\B$. We have $k^\alpha = A( n^\alpha  - r^\alpha)$,   
$\bar{k}^\alpha = \bar{A}( n^\alpha + r^\alpha )$  
for some scalars $A$ and $\bar{A}$, and it follows that 
$n \cdot k = -A$, $n \cdot \bar{k} = -\bar{A}$, and 
$k \cdot \bar{k} = -2 A \bar{A}$. We next find that 
\begin{equation} 
a - (a_1 + \bar{a}_2) = \ln 2 + \hat{a}_0  
- \bigl( a_{01} + \bar{a}_{02} \bigr)\,,  \labell{soso}
\end{equation}    
and observe that the two versions of $S_\B$ disagree unless the
right-hand side happens to vanish. Failure to achieve this would
result in a gravitational action that is not properly additive. 

\begin{figure}
\includegraphics[width=.9\linewidth]{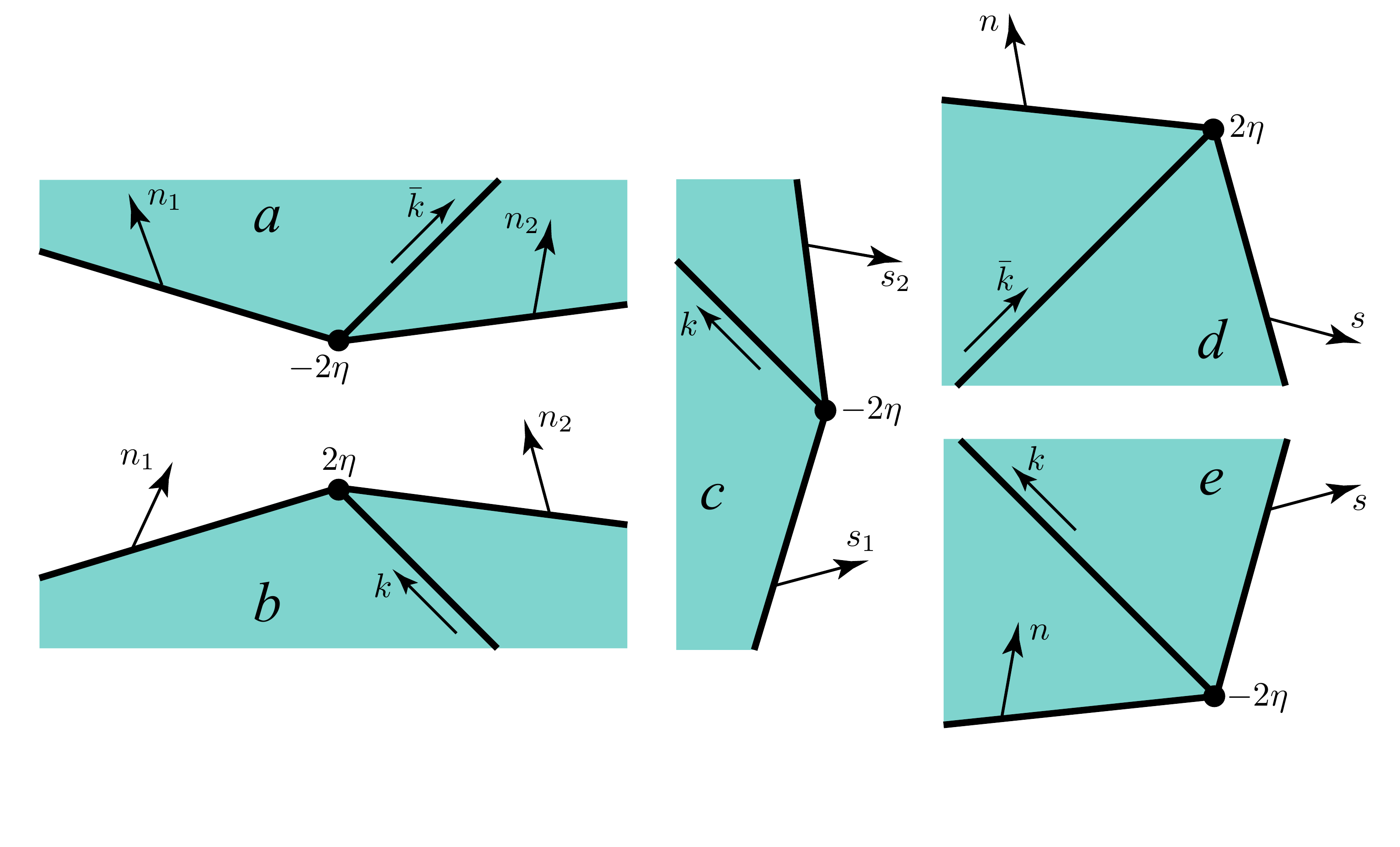}
\caption{Composition of boundary actions. In panels $a$ and $b$, a 
  spacelike/spacelike joint is obtained by taking the union of two
  null/spacelike joints. In panel $c$, a timelike/timelike joint is
  the union of two null/timelike joints. In panels $d$ and $e$, a
  spacelike/timelike joint is the union of a null/spacelike joint and
  a null/timelike joint.}  
\labell{fig:addrule1} 
\end{figure} 

It is possible to exploit the arbitrariness of $\hat{a}_0$,
$a_{01}$, and $\bar{a}_{02}$ to produce a gravitational action which
is additive. That is, we demand additivity for the boundary terms
at spacelike joints arising when null segments intersect
other boundary segments. This requirement, in fact, becomes a
prescription to remove the arbitrariness of these joint terms. For
example, in Eq.~\reef{soso}, the simplest way to achieve addivity is
to set $a_{01} = \bar{a}_{02} = 0$ and $\hat{a}_0 = -\ln2$.  
These choices give us {\it additivity rules for null joints}, which
can be formulated as follows: 
 \begin{itemize} 
\item {\bf spacelike rule:} for a joint between null and spacelike 
  hypersurfaces, assign $a = \ln(-n \cdot k)$, where $k^\alpha$ is the 
  future-directed normal to the null hypersurface (with arbitrary
  normalization), and $n^\alpha$ is the future-directed unit normal to
  the spacelike hypersurface; 
\item {\bf timelike rule:} for a joint between null and timelike  
  hypersurfaces, assign $a = \ln|s \cdot k|$, where $s^\alpha$ is the
  outward-directed unit normal to the timelike hypersurface; 
\item {\bf null rule:} for a joint between two null hypersurfaces, 
  assign $a = \ln(-\frac{1}{2} k \cdot \bar{k})$, where $k^\alpha$ is
  the future-directed normal to the first null hypersurface (with
  arbitrary normalization), and $\bar{k}^\alpha$ is the
  future-directed normal to the second hypersurface (also with
  arbitrary normalization).   
\end{itemize} 
The additivity rules eliminate the arbitrariness of the joint terms,
once a choice of normalization has been made for the null normals.    

We may test the applicability of these rules in a few examples. In
Fig.~\ref{fig:addrule1} we show the first few
examples of intersections between timelike and/or spacelike boundary
segments considered in Fig.~\ref{fig:TSjoints}, but with a null surface
now extending from each joint to subdivide the spacetime region into two
parts. Combining the joint rules in section \ref{sec:TSjoints} with
those above for joints involving null segments, we see in each case
that addivity is indeed satisfied:  

In panel $a$ of Fig.~\ref{fig:addrule1}, the composite figure gives rise
to a joint term $-2\eta$ with $\eta = -\ln(-n_1 \cdot \bar{k}) 
+ \ln(-n_2 \cdot \bar{k})$ from Eq.~\reef{bend1}. On the other hand,
the null joint on the left contributes $2 a_1$, while the one on the
right contributes $-2 a_2$. The spacelike rule makes the assignments 
$a_1 = \ln(-n_1 \cdot \bar{k})$ and $a_2 = \ln(-n_2 \cdot \bar{k})$,
and we recover $-\eta = a_1 - a_2$, as required for the proper
additivity of the gravitational action. In panel $b$, the
spacelike/spacelike joint contributes $2\eta$ with 
$\eta = \ln(-n_1 \cdot k) - \ln(-n_2 \cdot k)$, again from
Eq.~\reef{bend1}. The null joint on the left gives $2 a_1$ with $a_1 =
\ln(-n_1 \cdot k)$, and the one on the 
right gives $-2 a_2$ with $a_2 = \ln(-n_2 \cdot k)$. We have that
$\eta = a_1 - a_2$, and once again the boundary action is additive. 

In panel $c$, we have that the timelike/timelike joint contributes a
boundary term $-2\eta$ with $\eta = \ln(-s_1 \cdot k) 
- \ln(-s_2 \cdot k)$ from Eq.~\reef{bend2}. The null joint on the bottom contributes
$-2a_1$, while the one on the top gives $2 a_2$. The timelike
rule makes the assignments $a_1 = \ln(-s_1 \cdot k)$ and $a_2 =
\ln(-s_2 \cdot k)$, and we find that $-\eta = -a_1 + a_2$, as required
by additivity. In panel $d$, the composite figure comes with a
contribution $2\eta$ from the joint, with 
$\eta = -\ln(-n \cdot \bar{k})  + \ln(s \cdot \bar{k})$ from
Eq.~\reef{bend3}. The null joint on the top contributes $-2a_1$ with
$a_1 = \ln(-n \cdot \bar{k})$, and the one on the bottom contributes $2a_2$ with 
$a_2 = \ln(s \cdot \bar{k})$. We have $\eta = -a_1 + a_2$, and once
more verify that the gravitational action is additive. Finally, in
panel $e$, we have that the contribution from the  
spacelike/timelike joint is $-2\eta$, with  
$\eta =\ln(-n \cdot k) - \ln(-s \cdot k)$, again from
Eq.~\reef{bend3}.  In this case the null joint on the bottom gives
$-2a_1$ with $a_1 = \ln(-n \cdot k)$, while the 
one on the top gives $2a_2$ with $a_2 = \ln(-s \cdot k)$. We have
$-\eta = -a_1 + a_2$, as required by additivity. 

\begin{figure}
\includegraphics[width=.6\linewidth]{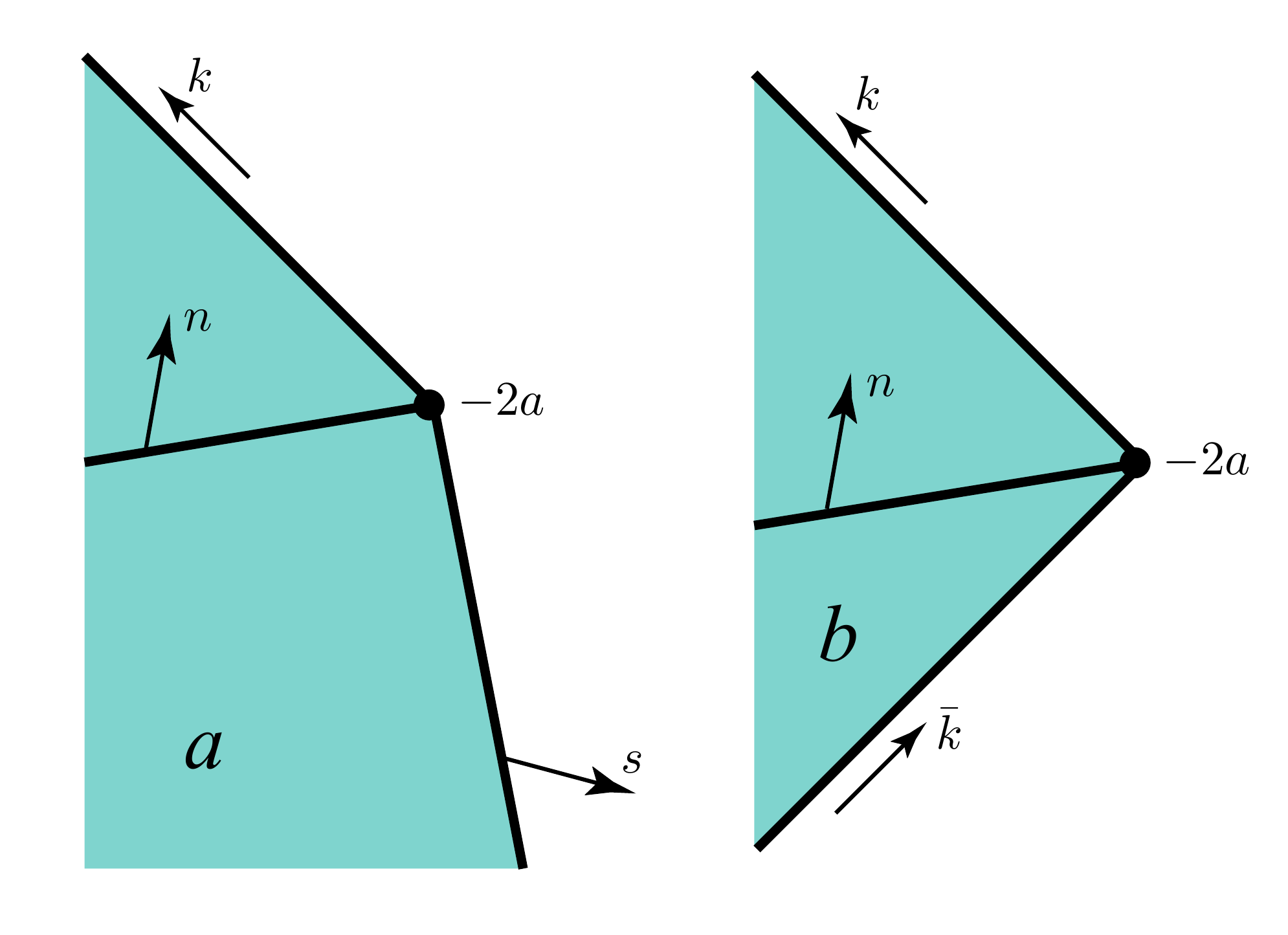}
\caption{Composition of boundary actions. In panel $a$, a
  null/timelike joint is obtained by taking the union of a
  null/spacelike joint with a spacelike/timelike joint. In panel $b$,
  a null/null joint is the union of two null/spacelike joints.}  
\labell{fig:addrule2} 
\end{figure} 

The spacelike and timelike rules can also handle the case depicted  
in panel $a$ of Fig.~\ref{fig:addrule2}. Here the composite figure
describes a null/timelike joint giving rise to a joint term 
$-2a_{\rm comp}$. The null/spacelike joint on the top provides a
contribution $-2 a_{\rm top}$, while the spacelike/timelike joint at
the bottom contributes $2\eta$ with 
$\eta = \ln(-n \cdot k) - \ln(-s \cdot k)$ from Eq.~\reef{bend3}. Here
the timelike rule makes the assignment $a_{\rm comp} = \ln(-s \cdot
k)$, while the spacelike rule gives $a_{\rm top} = \ln(-n \cdot k)$. We have that 
$-a_{\rm comp} = -a_{\rm top} + \eta$, and once more we find that the
rules ensure the proper additivity of the gravitational action. 

The spacelike and timelike rules, however, are not sufficient to handle
the case illustrated in panel $b$ of Fig.~\ref{fig:addrule2}; for this 
and similar cases we require the null rule. The composite figure
represents a null/null joint with joint term $-2 a_{\rm comp}$. The
null/spacelike joint at the top comes with $-2 a_{\rm top}$, while the
joint at the bottom comes with $-2a_{\rm bot}$. The spacelike rule
makes the assignments $a_{\rm top} = \ln(-n \cdot k)$ and 
$a_{\rm bot} = \ln(-n \cdot \bar{k})$. We may decompose $n^\alpha$ in
the null basis provided by $k^\alpha$ and $\bar{k}^\alpha$ and conclude
that $(-n \cdot k) (-n \cdot \bar{k}) = -\frac{1}{2} k \cdot \bar{k}$. 
According to this, we have that $a_{\rm top} + a_{\rm bot} 
= a_{\rm comp}$, as required by additivity. 

Many other examples could be constructed. In all such cases the three 
additivity rules formulated in this section are sufficient to restore
the additivity of the gravitational action when null boundaries are
involved. We not that these additivity rules do nothing, however, to restore
additivity in the problematic case of timelike joints identified by
Brill and Hayward \cite{1994PhRvD..50.4914B}.    
 
\section{Rate of change of the gravitational action for AdS black holes \labell{sec:dSdt} }

In this section we return to the ``complexity equals
action'' conjecture introduced in \cite{Brown:2015bva, Brown:2015lvg}. As described previously, 
this conjecture leads one to consider the action of regions of asymptotically AdS spacetimes with null
boundaries.  Our considerations of boundary terms in Sec.~\ref{contributionstoaction} allow us to 
provide a precise definition for the gravitational action $S = S_{\V} + 
S_{\partial \V}$ when the region $\V$ possesses a boundary with one or
several null segments, which was lacking in \cite{Brown:2015bva,
  Brown:2015lvg}.  Hence we are able to provide a careful examination
of the  results presented there.   

As we have seen, $S$ contains contributions
from $\V$, from the piecewise smooth portions of the boundary
$\partial \V$, and from the joints $\B$ where these portions 
are joined together. However, in general the resulting gravitational action $S$ is
ambiguous for a given spacetime, because the contribution from each
null segment of the boundary depends on an arbitrary choice of
parameterization, and because the contribution from each null joint is
the integral of an arbitrary scalar field $a$. The first source of ambiguity, the one associated
with the choice of parametrization, is naturally tamed by declaring 
that {\it all null segments shall be affinely parametrized}. This choice ensures that $\kappa = 0$ and that the
null segments make no contribution to the gravitational action. 
Then the additivity rules formulated in Sec.~\ref{sec:additivity}
allow us to eliminate (much of) the arbitrariness associated with  the
null joints. We adopt both of these conventions in the following 
calculations, but we must acknowledge that these choices do not
completely eliminate the ambiguities. In particular, 
there remains the freedom to rescale the affine parameter $\lambda$ by
a constant factor on each generator of the null boundaries, which in
turn will rescale the contribution of the corresponding joint
terms. We fix this remaining ambiguity by imposing a fixed
normalization condition of the null normals at the asymptotic AdS
boundary. While this normalization is again an arbitrary
choice, such a condition must be imposed if one is going to compare
the actions of different regions (potentially in different spacetimes)
in a meaningful way.  

We wish to exploit our precise definition of the gravitational action
to calculate how $S$ changes with time when evaluated for a
Wheeler-deWitt patch of a black hole in anti-de Sitter spacetime. This
computation was first presented by Brown et al \cite{Brown:2015lvg}
using an incomplete specification of the action, and the analysis
there might be viewed as questionable. However, we shall show that our
more complete and rigorous methods produce precisely the same answer:
In particular, for a Schwarzschild-anti de Sitter black hole at late times $t$,   
\begin{equation} 
\frac{dS}{dt} = 32\pi G_\mt{N}\,M\,, \labell{house}
\end{equation} 
where $M$ is the total mass-energy assigned to the black hole. We
recall that our convention for the gravitational action $S$ omits the
usual factor of $1/(16\pi G_\mt{N})$, \ie $S = 16\pi G_\mt{N}
I$. Hence, in the more usual convention, this equation 
would read $dI/dt = 2M$, and with Eq.~\reef{house} we have therefore 
reproduced the elegant universal result of \cite{Brown:2015bva,
  Brown:2015lvg}. In fact, our calculations extend the previous
analysis to include black holes with planar and hyperbolic
horizons. We conclude this section by reconsidering the case of
charged black holes in anti-de Sitter  spacetime and again, our
analysis reproduces the results of \cite{Brown:2015bva, Brown:2015lvg}.  

\subsection{Schwarzschild-anti de Sitter spacetime \labell{sec:SadS} }

We express the metric of an $(n+2)$-dimensional SAdS spacetime as follows:
\begin{equation} 
ds^2 = -f(r)\, dt^2 + \frac{dr^2}{f(r)} + r^2\, 
d\Sigma^2_{k,n}\,, \qquad {\rm with}\quad
f(r) = \frac{r^2}{L^2}+k - \frac{\omega^{n-1}}{r^{n-1}} \,.
\labell{metrix}
\end{equation} 
Here, $L$ is the AdS curvature scale and $k= \{+1, 0, -1\}$ denotes
the curvature of the $n$-dimensional line-element $d\Sigma^2_{k,n}$,  
given by
\begin{equation} 
d\Sigma^2_{k,n}=\left\lbrace\begin{matrix}
d\Omega^2_n&=d\theta^2+\sin^2\theta\, d\Omega^2_{n-1}\ \ &\ \  {\rm for\ }k=+1\ \,,\cr
d\ell^2_n&=d\theta^2+\theta^2\, d\Omega^2_{n-1}\qquad&\ \ {\rm for\ }k=0\ \ \ \,,\cr
d\Xi^2_n&=d\theta^2+\sinh^2\theta\, d\Omega^2_{n-1}&\ \  {\rm for\ }k=-1\,.
\end{matrix}\right. \labell{transverse}
\end{equation} 
Here, $d\Omega^2_n$ is the standard metric on a unit $n$-sphere, while
$d\ell^2_n$ is the flat metric on $R^n$ (with dimensionless
coordinates) and  $d\Xi^2_n$ is the metric on an $n$-dimensional
hyperbolic `plane' with unit curvature. In all three cases, the
metric of Eq.~\reef{metrix} is a solution of Einstein's equations with a
negative cosmological constant, \ie  
\be
R_{\mu\nu}-\frac12 R\,g_{\mu\nu} + \Lambda\, g_{\mu\nu}=0 
\qquad {\rm with}\ \ \ \Lambda = -\frac{n(n+1)}{2L^2}\,.
\labell{Eom2}
\ee
Each of these solutions can be represented by the same Penrose
diagram, as shown in Fig.~\ref{fig:WdW}. In particular, the black
holes corresponding to $k=\{+1, 0, -1\}$ have spherical, planar, and
hyperbolic horizons, respectively. Of course, these geometries are
also static with Killing vector $\partial_t$. 

The parameter $\omega$ is related to the position of the
event-horizon $r_{\rm H}$ by\footnote{Let us note that with $k=-1$,
  this mass parameter vanishes when $r_{\rm H}=L$, but a smooth
  horizon remains for smaller values of $r_{\rm H}$ in the range 
  $\frac{n-1}{n+1}\le \frac{r_{\rm H}^2}{L^2}< 1$, in which case the
  mass parameter becomes negative \cite{counter,roberto}. However, in
  this regime, the causal structure of the black hole takes the form shown in
  Fig.~\ref{fig:WdWcharged}, with an outer and an inner horizon. Hence
  the calculation of $dI/dt$ in section \ref{sec:WdW} is restricted to 
  $r_{\rm H}>L$ when $k=-1$. We thank Shira Chapman and Hugo Marrochio
  for this observation.} 
\begin{equation} 
\omega^{n-1} = r_{\rm H}^{n-1}\bigl[ (r_{\rm H}/L)^2 +k\bigr]\,. 
\end{equation} 
The total mass-energy of the spacetime is given by \cite{casimir,counter}
\begin{equation} 
M = \frac{n\,\Omega_{n,k}}{16\pi G_\mt{N}}\, \omega^{n-1}\,, 
\labell{mass-SadS} 
\end{equation} 
where $\Omega_{n,k}$ denotes the (dimensionless) volume of the
corresponding spatial geometry. Hence, for $k=+1$ we have the volume
of a unit $n$-sphere,
$\Omega_{n,+1}=2\pi^{(n+1)/2}/\Gamma\left(\frac{n+1}2\right)$, while
for $k=0$ and $-1$, we implicitly introduce an infrared regulator to
produce a finite volume. 

For our calculations it is useful to introduce the null coordinates $u$ and $v$, defined by 
\begin{equation} 
du := dt + f^{-1}\, dr\,, \qquad 
dv := dt - f^{-1}\, dr\,. \labell{EF-def}
\end{equation} 
Integrating these relations yields the ``infalling" null coordinate $u
= t + r^*(r)$ and the ``outgoing" null coordinate $v = t - r^*(r)$,
where $r^*(r) := \int f^{-1}\, dr$. The metric becomes   
\begin{equation} 
ds^2 = -f\, du^2 + 2\, dudr + r^2\, d\Sigma^2_{k,n}
\end{equation} 
or 
\begin{equation} 
ds^2 = -f\, dv^2 - 2\, dvdr + r^2\, d\Sigma^2_{k,n}
\end{equation} 
when expressed in terms of the null coordinates. For the three choices
$(t,r)$, $(u,r)$, and $(v,r)$ we have that 
\begin{equation} 
\int \sqrt{-g}\, d^{n+2} x = \Omega_{n,k}\, \int r^n dr\, dw, 
\labell{vol-element} 
\end{equation} 
where $w = \{ t, u, v \}$. 

\subsection{Wheeler-deWitt patch  \labell{sec:WdW} }

We consider the Wheeler-deWitt (WdW) patches of a
Schwarzschild-anti-de Sitter spacetime illustrated in
Fig.~\ref{fig:WdW}. As described in Sec.~\ref{sec:intro}, the
corresponding action $S(t_{\rm L},t_{\rm R})$ depends on the choice of
the time slice on the left and right boundaries \cite{Brown:2015bva,
  Brown:2015lvg}. As shown in the figure, the Killing 
vector corresponding to time translations in Eq.~\reef{metrix}
generates an upward (downward) flow in the asymptotic region on the
left (right), and hence the action is invariant upon shifting the time
slices as $S(t_{\rm L}+\delta t,t_{\rm R}-\delta t)
=S(t_{\rm L},t_{\rm R})$. Instead we will
fix the time on the right boundary and only vary the
asymptotic time slice on the left-hand side. In particular, we will
compare the actions for the two WdW patches shown on the left panel of
the figure. For the first, shown in dark color, the time on the left
boundary is $t_0$ and we denote $S(t_0)$ the action evaluated for
this patch. The asymptotic time for the second WdW patch,
shown in light color, is translated slightly with respect to the first
by $\delta t$, and the action evaluated for this patch is denoted 
$S(t_0 + \delta t)$. These two actions contain contributions from the
interior of the corresponding patches, the bounding surfaces, and the
joints between them. Our aim will be to evaluate the difference $\delta S :=
S(t_0+\delta t) - S(t_0)$.  

\begin{figure}
\includegraphics[width=.9\linewidth]{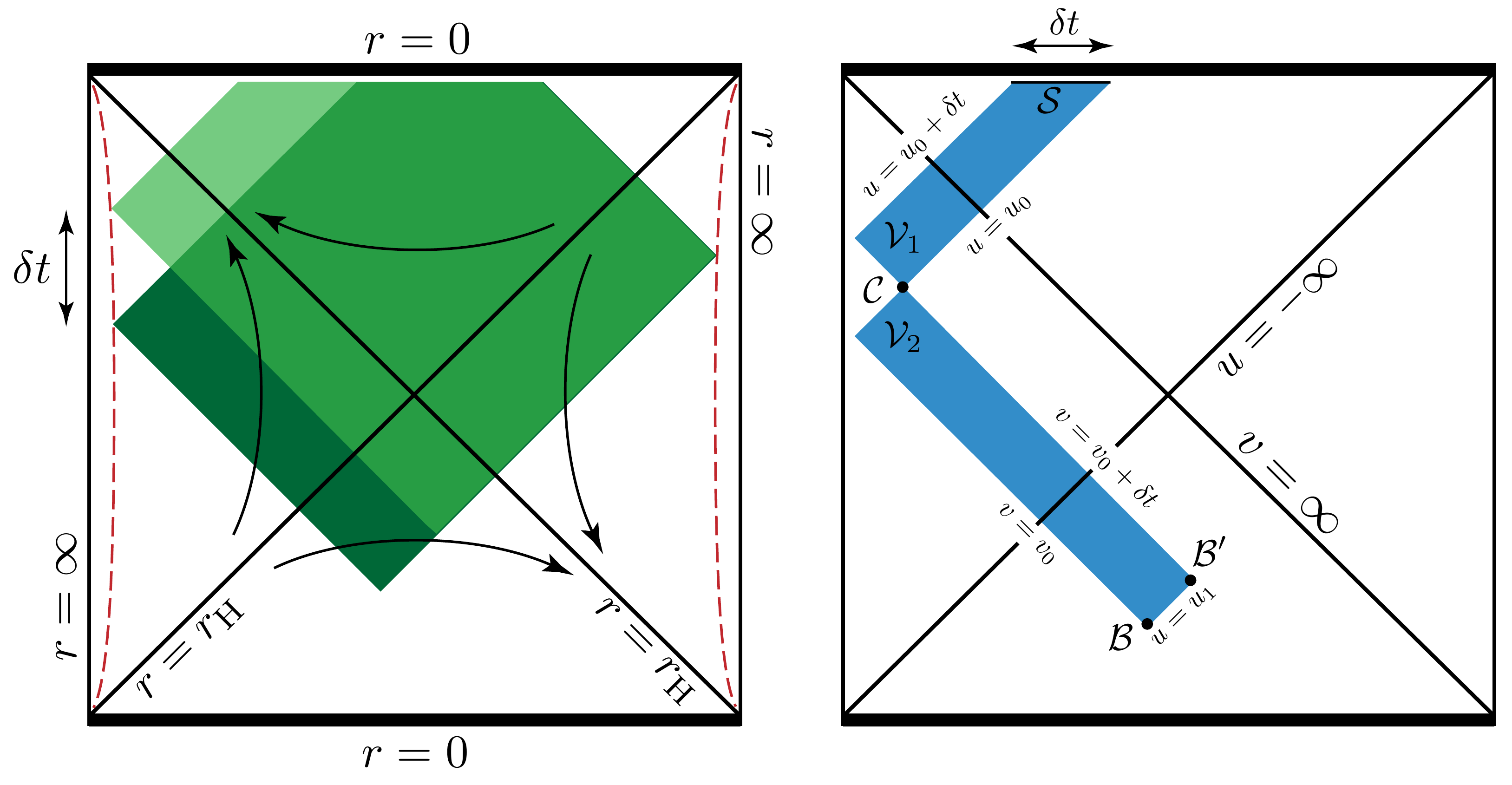}
\caption{Wheeler-deWitt patch of a Schwarzschild-anti de Sitter
  spacetime. On the left panel, the patch at coordinate time $t_0$ is
  shown in dark color, and the patch at time $t_0 + \delta t$ is shown
  in light color. The difference between the two patches is shown on
  the right panel. In the left panel, the curved arrows indicate the
  flow of the Killing vector $\partial_t$ in each of the quadrants of
  the Penrose diagram. Further the red dashed curves indicate the
  cut-off surfaces at  $r=r_\mt{max}$ near the asymptotic AdS
  boundaries. 
\labell{fig:WdW} }
\end{figure} 

We are considering ``late times," and so both patches
reach the spacelike singularity that defines the future boundary of
the Penrose diagram --- but they do not touch 
the past singularity. That is, the patches are bounded by a
spacelike surface near the future singularity at $r=0$, and by four
null segments extending (almost) all the way to the AdS boundaries at 
$r=\infty$; in fact, these must be truncated to regulate the 
gravitational action evaluated for these WdW patches. 
Note that this is a standard issue in holographic calculations
\cite{revue} and the standard procedure is to evaluate  
the quantity of interest with a (timelike) cut-off surface at some
large radius. In particular, one might  
choose the latter to be $r=r_\mt{max}=L^2/\delta$, in which case
$\delta$ plays the role of a short distance  
cut-off in the boundary theory. Typically, holographic calculations are 
employed to evaluate UV-safe quantities, such as correlation
functions, and so one is able to take the limit $\delta \to 0$ at the
end of the calculation \cite{revue}. In particular, in a standard
calculation of the renormalized action, one can introduce a finite set
of boundary counterterms which then yield a finite result in this
limit \cite{counter}. However, such an approach does not yield a
finite result for the action of a WdW patch \cite{prep,prep2}. One
might therefore interpret the divergence in the action here as being related
to the complexity required to establish correlations at arbitrarily
short distance scales in the state of the boundary theory. In this
way, the divergences found here would be similar to those found in
holographic calculations of entanglement entropy \cite{rt1,rt2}. For
simplicity, our approach to regulating the action will be to let the
null sheets defining the boundaries of the WdW patch originate
slightly inside the AdS boundary, \ie, at 
$(t,r)=(t_0,r_\mt{max})$. An alternative approach would be to let the
null sheets originate at $(t,r)=(t_0,\infty)$ but truncate the region
on which we are evaluating the action at the cut-off surface,
$r=r_\mt{max}$. These different choices for the regulator do not
significantly change the result for the gravitational action, and both
approaches yield the same results in the following; we
will return to these issues in \cite{prep,prep2}. 

As noted above, another important ingredient in our computations is 
that the vectors defining the null boundaries of the WdW patch will
be normalized in precisely the same way at the asymptotic AdS
boundary. While in different contexts we may make different choices
for this normalization, here we must fix the normalization
in order to compare the action of different WdW patches in a
meaningful way. For example, in the following we choose 
$k\cdot \hat t_L=-c$, where $k$ is the (future-directed) normal to the
past null boundary of the left-hand side of Fig.~\ref{fig:WdW}, $\hat
t_L=\partial_t$ is the asymptotic Killing vector which is normalized
to describe the time flow in the left boundary theory, and $c$ is an
arbitrary (positive) constant. If instead we allowed the
latter to be a function of time, \ie $k\cdot \hat t_L=c(t)$, we could
produce whatever answer we might desire for the difference $\delta S$
and hence for the time derivative in Eq.~\reef{house}. It is only with a
fixed constant $c$ that a meaningful result is produced; we return
to this issue in section \ref{discuss}. 

Now to compute $\delta S=S(t_0+\delta t)-S(t_0)$, we recall that with
an affine parametrization for each null surface, these make  
no contribution to the action and thus to $\delta S$. Further, we
observe that the left null joint at $r=r_\mt{max}$ for $S(t_0+\delta t)$,
is simply related to the one for $S(t_0)$ by a time translation; 
given our fixed normalization of the corresponding null normals,
the corresponding joint contributions are identical, and they
therefore make no contribution to $\delta S$. A similar conclusion
holds for the joints linking the incoming null segment to the
spacelike surface near the singularity. Consequently the 
computation of $\delta S$ relies only on the pieces illustrated on  
the right panel of Fig.~\ref{fig:WdW}: We have the volume
contributions from the regions $\V_1$ and $\V_2$, the surface
contribution from the spacelike segment $\S$, and the joint
contributions from the $n$-surfaces $\B$ and $\B'$. All told, we have that  
\begin{equation} 
\delta S = S_{\V_1} - S_{\V_2} - 2 \int_{\S} K\, d\Sigma
+ 2\oint_{\B'} a\, dS - 2\oint_{\B} a\, dS, 
\end{equation} 
where $d\Sigma$ is a volume element on $\S$, and $dS$ is a surface
element on $\B$ and $\B'$.  

\subsection{Calculation of $\delta S$ \labell{sec:cal-deltaS} }

We first evaluate the volume contribution 
\begin{equation} 
S_{\V} = \int_{\V} (R - 2\Lambda)\, \sqrt{-g}\, d^{n+2}x 
\end{equation} 
for the regions $\V_1$ and $\V_2$ depicted on the right panel of
Fig.~\ref{fig:WdW}. By virtue of the Einstein field equations, 
$R = 2(n+2) \Lambda/n$, and the integrand is the constant  
\begin{equation} 
R - 2\Lambda = -\frac{2(n+1)}{L^2}. 
\end{equation} 
We begin with the computation for $\V_1$, and next turn to $\V_2$.  

As shown in Fig.~\ref{fig:WdW}, the past and future null boundaries
on the left of the first WdW patch are labelled by $u=u_0$ and
$v=v_0$, respectively. These null boundaries become $u=u_0+\delta t$
and $v=v_0+\delta t$ for the second, shifted WdW patch. Hence the
region $\V_1$ is bounded by the null 
surfaces $u = u_0$, $u = u_0+ \delta t$, $v = v_0 + \delta t$, as
well as the spacelike surface $r = \epsilon \ll r_{\rm H}$. The volume integral is best performed in
the $(u, r)$ coordinate system, which is regular throughout the
region; in this system the surface $v = v_0 + \delta t$ is described
by $r = \rho(u)$, with $\rho(u)$ defined implicitly by $r^*(\rho) = 
\frac{1}{2} (v_0+\delta t-u)$. Making use of
Eq.~(\ref{vol-element}), we have that   
\begin{equation} 
S_{\V_1} = -\frac{2(n+1)}{L^2}\, \Omega_{n,k} \int^{u_0+ \delta t}_{u_0} du
\int_\epsilon^{{\rho}(u)} r^n\, dr 
= -\frac{2\Omega_{n,k}}{L^2} \int^{u_0+ \delta t}_{u_0} du\, 
{\rho}^{n+1}(u)\,,   
\labell{schv1}
\end{equation} 
where we have neglected the $\epsilon^{n+1}$ term that was to be subtracted from 
${\rho}^{n+1}$ in the final integrand.  

As also shown in the figure, the region $\V_2$ is bounded by the null surfaces $u=u_0$, $u=u_1$,
$v=v_0$, and $v=v_0+\delta t$. In this case, the volume integral is most easily
performed in the $(v,r)$ coordinates, in which the surfaces $u=u_{0,1}$ are
described by $r=\rho_{0,1}(v)$, with $r^*(\rho_{0,1}) = \frac{1}{2}(v-u_{0,1})$. Then we have 
\begin{equation} 
S_{\V_2} = -\frac{2(n+1)}{L^2}\, \Omega_{n,k} \int_{v_0}^{v_0+\delta t} dv 
\int_{\rho_1(v)}^{{\rho}_0(v)} r^n\, dr 
= -\frac{2\Omega_{n,k}}{L^2} \int_{v_0}^{v_0+\delta t} dv\, 
\bigl[ {\rho}_0^{n+1}(v) - \rho_1^{n+1}(v) \bigr]\,. 
\end{equation} 

Now we perform the change of variables $u = u_0+v_0 + \delta t - v$ in
the integral for $S_{\V_1}$, combine it with the integral for
$S_{\V_2}$, and notice that the terms involving ${\rho}(u)$ and
${\rho}_0(v)$ cancel out. This cancellation was to be expected,
because the portion of $\V_1$ below the future horizon and the portion
of $\V_2$ above the past horizon have equal volumes, as noted in \cite{Brown:2015lvg}. Taking this
property into account, the radial integral for $S_{\V_1}$ could have
been limited to the interval $\epsilon < r < r_{\rm H}$, and the
integral for $S_{\V_2}$ could have been limited to 
$r_{\B} < r < r_{\rm H}$; the dependence of each term on $r_{\rm H}$
would have similarly cancelled out in the difference $S_{\V_1} - S_{\V_2}$. 
In any event, we are left with 
\begin{equation} 
S_{\V_1} - S_{\V_2} = -\frac{2 \Omega_{n,k}}{L^2} 
\int_{v_0}^{v_0+\delta t} dv\, \rho_1^{n+1}(v), 
\end{equation} 
with the function $\rho_1$ varying from $r_{\B}$ to $r_{\B'}$ as $v$
increases from $v_0 $ to $v_0+ \delta t$. This is a small variation in 
the radius, \ie $r_{\B'} = r_{\B} + O(\delta t)$, and hence the volume
contribution to $\delta S$ is simply  
\begin{equation} 
S_{\V_1} - S_{\V_2} = 
-\frac{2\Omega_{n,k}}{L^2} r_{\B}^{n+1}\, \delta t\,.  
\labell{dS:volume} 
\end{equation} 

We next evaluate the surface contribution to $\delta S$, given by 
$-2 \int_{\S} K\, d\Sigma$, where $\S$ is the boundary segment given by the spacelike
hypersurface $r = \epsilon$. The (future-directed) unit normal to any
surface $r = \mbox{constant}$ inside the future horizon is given by
$n_\alpha = |f|^{-1/2} \partial_\alpha r$. The extrinsic curvature is   then
\begin{equation} 
K = \nabla_\alpha n^\alpha = -\frac{1}{r^n} \frac{d}{dr} 
\Bigl( r^n |f|^{1/2} \Bigr), 
\end{equation} 
and the volume element becomes 
\begin{equation} 
d\Sigma = \Omega_{n,k} \,|f|^{1/2} r^n\, dt 
\end{equation} 
after integrating over the ``angular" variables described by Eq.~\reef{transverse}. Letting $r = \epsilon 
\ll r_{\rm H}$ and then approximating $f \simeq -(\omega/r)^{n-1}$, we find 
that 
\begin{equation} 
-2 \int_{\S} K\, d\Sigma = (n+1) \,\Omega_{n,k}\, \omega^{n-1}\, \delta t\,. 
\labell{dS:surface} 
\end{equation} 
Given the proximity of $\S$ to the spacelike singularity at $r=0$, it
is remarkable that the answer turns out to be finite and independent
of $\epsilon$. This occurs because the divergence in $K$ is precisely 
compensated for by the vanishing of $d\Sigma$. We return to discuss this point in section \ref{discuss}.

We next turn to the joint terms $\pm 2\oint a\, dS$ contributed by
the $n$-surfaces $\B$ and $\B'$. The null rule formulated in
Sec.~\ref{sec:additivity} states that 
\begin{equation} 
a = \ln\bigl( -{\textstyle \frac{1}{2}} k \cdot \bar{k} \bigr), 
\end{equation} 
where $k^\alpha$ is the (future-directed) null normal to the
left-moving null hypersurfaces, \ie on which $v = v_0$ and $v_0+\delta t$, while
$\bar{k}^\alpha$ is the  (future-directed) null normal to the
right-moving surface, on which $u = u_1$.  

Our convention was to choose the vectors $k^\alpha$ and $\bar{k}^\alpha$ to be
affinely parametrized, and suitable expressions are 
\begin{equation} 
k_\alpha = -c\, \partial_\alpha v = -c\, \partial_\alpha(t - r^*)\,, \qquad 
\bar{k}_\alpha = \bar{c}\, \partial_\alpha u 
= \bar{c}\, \partial_\alpha(t+r^*)\,, \labell{roundh}
\end{equation} 
where $c$ and $\bar{c}$ are arbitrary (positive) constants. This
choice implements the asymptotic normalizations $k\cdot \hat t_L=-c$
and $\bar{k}\cdot \hat t_R=-\bar{c}$, where $\hat t_{L,R}$ are the
asymptotic Killing vectors which are normalized to describe the time
flow in the left and right boundary theories, respectively. With these
choices, we have that $k \cdot \bar{k} = 2c\bar{c}/f$, so that  
\begin{equation} 
a = -\ln\biggl( \frac{-f}{c\bar{c}} \biggr)\,. 
\end{equation} 
 With the above expression, we find that 
\begin{equation} 
2 \oint_{\B'} a\, dS - 2 \oint_{\B} a\, dS 
= 2\Omega_{n,k} \bigl[ h(r_{\B'}) - h(r_{\B})\bigr], 
\end{equation} 
where $h(r) := -r^n \ln(-f/c\bar{c})$.  

To express this result in its final form, we perform a Taylor expansion of
$h(r)$ about $r = r_\B$. Because the displacement is in a direction of
increasing $v$, we have that $du = 0$, $dv = \delta t$, and 
$dr = -\frac{1}{2} f\, \delta t$. This gives us 
\begin{equation} 
h(r_{\B'}) - h(r_{\B}) = -\frac{1}{2} f \frac{dh}{dr}\bigg|_{r=r_{\B}} \delta t 
= \frac{1}{2} \biggl[ r^n \frac{df}{dr} + n r^{n-1} f
\ln\biggl(\frac{-f}{c\bar{c}} \biggr) \biggr]\bigg|_{r=r_{\B}} \delta t,
\end{equation} 
and then 
\begin{equation} 
2 \oint_{\B'} a\, dS - 2 \oint_{\B} a\, dS 
= \Omega_{n,k} \biggl[ r^n \frac{df}{dr} + n r^{n-1} f
\ln\biggl(\frac{-f}{c\bar{c}} \biggr) \biggr]\bigg|_{r=r_{\B}} \delta t\,.
\labell{dS:joints} 
\end{equation} 

Combining Eqs.~(\ref{dS:volume}), (\ref{dS:surface}), and
(\ref{dS:joints}), we arrive at 
\begin{equation} 
\delta S = \Omega_{n,k} \biggl[ -\frac{2 r^{n+1}}{L^2} 
+ (n+1) \omega^{n-1} + r^n \frac{df}{dr} 
+ n r^{n-1} f \ln\biggl(\frac{-f}{c\bar{c}} \biggr) 
\biggr]\bigg|_{r= r_\B}\, \delta t 
\end{equation} 
for the change in gravitational action when the left time slice of the WdW 
patch is translated by $\delta t$. Making use of the explicit expression for
$f$, this expression implies that 
\begin{equation} 
\frac{dS}{dt} = 2 n \Omega_{n,k} \omega^{n-1} \biggl[ 
1 + \frac{1}{2} \biggl( \frac{r}{\omega} \biggr)^{n-1} 
f  \ln\biggl(\frac{-f}{c\bar{c}} \biggr) \biggr] \bigg|_{r= r_\B}. 
\labell{dS:final} 
\end{equation} 
When this is evaluated at late times, $r_\B$ approaches $r_{\rm H}$,
$f$ approaches zero, and we see that $dS/dt$ rapidly approaches the asymptotic
constant $2n \Omega_n \omega^{n-1}$. Recalling Eq.~(\ref{mass-SadS})
for the mass-energy of the SAdS spacetime, this is 
\begin{equation} 
\frac{dS}{dt} = 32\pi G_\mt{N}\,M
\end{equation} 
at late times.
In the more usual convention in which the gravitational action is
$I := S/(16\pi G_\mt{N})$, this is $dI/dt = 2M$, precisely the
same result reported in Brown et al
\cite{Brown:2015bva,Brown:2015lvg}.  We might add that the
calculations there focused on the case of spherical black holes, \ie
$k=+1$. Our analysis shows that the same simple result applies also
for planar and hyperbolic horizons, \ie $k=0$ and $-1$.  

\subsection{Comparison with Brown et al \labell{sec:comp}}

It is remarkable that the two very different methods of calculating $dI/dt$
should produce precisely the same outcome, given how the accounting of 
various contributions to the gravitational action differs in each
method. It is interesting to examine in detail how each contribution
to the action appears in the calculation of $dI/dt$ in
\cite{Brown:2015bva,Brown:2015lvg} and compare with our results:  

First, Brown et al implicitly assume that the gravitational action
is additive. They use this property to divide the WdW patches at $t_0$
and $t_0+\delta t$ into various subregions and evaluate $\delta S$ 
in terms of the action evaluated for each of the
subregions. In the end, they essentially focus on two regions,
${\cal V}_1$ and ${\cal V}_2$ on the right panel of
Fig.~\ref{fig:WdW}, but each of these is further divided into the
portion outside of the horizon and that behind the horizon. We note
that only spacelike joints arise in subdividing the WdW patches there;
as we discussed, with appropriate choices for the boundary
terms, the gravitational action will indeed be additive. Further, we
observe that (segments of) the future and past horizons now play the
role of boundary surfaces for these various subregions. These (null)
surfaces did not appear in our calculations because we 
did not subdivide the WdW patches. One may worry that the final
results will depend on choices made, \eg in defining the
parametrization of these null surfaces. To answer this we make two
observations: First, in general, any internal boundary surface will
be common to two neighbouring subregions, so as long as the common
boundary is described consistently in evaluating the action of these
two subregions (\eg they are assigned the same null normal $k^\alpha$),
the corresponding boundary contributions will cancel when the actions
are added to evaluate the full action of the complete WdW patch. Thus,
the choices made in describing such internal boundary surfaces will
never affect the final result. Second, for the particular case
considered here, the internal boundary surfaces are segments
of a Killing horizon, \ie they are stationary null boundary
surfaces. As discussed below Eq.~\reef{barS} and in Appendix
\ref{sec:limit}, such stationary null boundaries are distinguished
because the corresponding contributions to the gravitational action
are not ambiguous. This point will play an important role in the
following.   

In the calculation presented in \cite{Brown:2015bva,Brown:2015lvg},
the authors argue that the time translation symmetry of the geometry
ensures that the contributions to $S(t_0+\delta t)$ and $S(t_0)$ from
the portions of the corresponding WdW patches outside of the horizon
will cancel in the difference $\delta S$. While we agree with this
conclusion, assuming the reasonable choices described above for the
boundary terms, we would like to point out a subtlety having to do
with the boundary contributions coming from the horizon. In
particular, it is {\it not} true that the boundary contribution coming
from the future horizon (or from the past horizon) is identical for
these two exterior regions. Instead, the two actions cancel because the
contribution from the segment between $u_0$ and $u_0+\delta t$ on the
future horizon cancels that from the segment between $v_0$ and
$v_0+\delta t$ on the past horizon. The simplest way to see that these
two contributions match is to note that since the geometry is static,
it is invariant under an inversion of the time coordinate. Hence,
inverting $t$ about the time slice $t_0+\frac12\delta t$ maps these
boundary segments on the two horizons into one another. Further, in
comparing these two null segments, it is important that there is no
ambiguity in their contributions to the gravitation action, since both
are part of a stationary horizon, as discussed above. 

Next, Brown et al consider the portion of ${\cal V}_2$ which lies
behind the past horizon (see Fig.~\ref{fig:WdW}). At late times, the
radial coordinate is essentially constant throughout this region and
so the geometry reduces to the direct product of a constant transverse
space, \ie the $n$-dimensional geometry described by
Eq.~\reef{transverse},  and the exponentially small two-dimensional
geometry extending in the $r$ and $t$ (or $u$ and $v$)
directions. Since the transverse geometry is constant, the authors
argue that by applying the two-dimensional Gauss-Bonnet theorem
\cite{rocky}, this region does not contribute to the time dependence
of the WdW patch, although they acknowledge that there may be subtleties in
this argument related to regulating the gravitational action. Our
construction seems to eliminate this issue or at least, 
relates any question about the UV divergences in the complexity to the
behaviour of the geometry and the WdW patch near the asymptotic
boundary \cite{prep2}. Let us add that the analysis of
\cite{rocky} implicitly introduces a new imaginary contribution to the
null joint terms, which violates the additivity of the gravitational
action. However, these imaginary terms do not affect the result for
$dI/dt$ --- see section \ref{discuss} for further discussion. From our
perspective, it is the proximity of the $u=u_1$ surface to the past
horizon that ensures the cancellation of the corresponding
contributions. For example, if the generators of the past horizon
were affinely parametrized, the only contributions to the
gravitational action of this region\footnote{The integral of the
  Einstein-Hilbert term is negligible because the proper volume of
  this region is exponentially small.} would come from from the joints
on the boundary, \ie  ${\cal B}$, ${\cal B}'$ and the intersections of
$v=v_0$ and $v=v_0+\delta t$ with the past horizon. Then because of
the proximity of the $u=u_1$ boundary and the past horizon, evaluating
the joint terms at ${\cal B}$ and ${\cal B}'$ yields essentially the
same result as those on the horizon, up to an overall sign. We return
to this point below. 

Lastly, Brown et al consider the portion of ${\cal V}_1$ which lies
behind the future horizon (see Fig.~\ref{fig:WdW}) and whose action
then gives the entire result for $\delta S$. There are three
contributions: {\it i}) the volume integral of the Einstein-Hilbert
action; {\it ii}) the boundary integral of the Gibbons-Hawking-York
(GHY) term on a spacelike surface near the singularity at $r=0$; and
{\it iii}) the GHY term evaluated on a spacelike surface just inside the
horizon which is then taken to approach $r=r_{\rm H}$. The first two
contributions also appear in our calculations\footnote{Given our
  presentation of the calculation of the Einstein-Hilbert term above,
  this statement may not be immediately clear. However,  we observe
  that at the end of the calculation we set $r_{\cal B} \simeq
  r_{\rm H}$, and so there is essentially no contribution to the volume
  integral coming from behind the past horizon in ${\cal V}_2$ --- as
  noted in the previous footnote.}, but the third term may seem
suspect in view of our discussion of ambiguities in taking the null limit
of spacelike or timelike surfaces. However, here the limit is taken to
a stationary null surface, and there is no such ambiguity. A
caveat is that there is no ambiguity for the {\it sum} of the
boundary and joint terms evaluated on a stationary null surface in
this way. Brown et al assume that the joint terms cancel between the
two ends of the null segment, which is not a priori clear from our
perspective. On the other hand, a careful analysis along the lines of
those given for the first example in Appendix \ref{sec:limit} shows
that this cancellation indeed occurs.  

Hence at a pragmatic level, the key difference between the two
calculations is as follows: In the Brown et al computation, an
essential contribution to $\delta S$ originates from the segment of
the future horizon (between $u=u_0$ and $u_0+\delta t$), which plays
no role in our calculation as it appears on a surface that is internal
to the WdW patch. On the other hand, our computation features
contributions from the joints, $\B$ and $\B'$, at the bottom of the
WdW patches, which play an inconsequential role 
in \cite{Brown:2015bva,Brown:2015lvg}. However, when all contributions
are combined together, the final result is the same $dI/dt = 2M$ for late times in
both cases. It thus follows that these two distinct contributions
appearing in the different calculations must in fact be the
same. Tracing through the above discussion, we can see the mechanism
for this equality. First, because the future horizon is an internal
boundary, the boundary contribution of the null segment on the future
horizon is the same for the portions of ${\cal V}_1$ inside and
outside of the horizon, up to an overall sign. Then because of the
time inversion symmetry (as well as the time translation symmetry) of
the geometry, the contribution on the future horizon for the exterior
part of ${\cal V}_1$ can be related to that on the past horizon for
the exterior of ${\cal V}_2$. Again the past horizon is an internal
boundary and so the contribution is the same for the corresponding
null segment of the portion of ${\cal V}_2$ inside the
horizon. Lastly, we found that the joint contributions from 
${\cal B}$ and ${\cal B}'$ match the contribution from the horizon
for this portion of ${\cal V}_2$ inside the horizon. Hence through a
series of equalities, we see that  
the boundary contribution from the segment of the future
horizon in the Brown et al computation must be equal to the joint
contributions from the bottom of the WdW patch in our
computation. Therefore the key difference between the two approaches
is largely a matter of accounting, \ie while our computation directly
compared the full WdW patches at $t=t_0$ and $t_0+\delta t$, Brown et
al begin by subdividing the WdW patches and evaluate the action on a
series of subregions. However, as explained above, Brown et al also
make a number of assumptions, beginning with additivity of the
gravitational action, which we have verified with the detailed
considerations in our paper.

\subsection{Extension to charged black holes \labell{sec:charged}}

We next turn our attention to the case of a charged AdS black holes  
in $n+2$ dimensions.\footnote{The results presented in this section
  were independently confirmed by Shira Chapman and Hugo Marrochio
  (private communication).} The line element takes the same form as in
Eq.~\reef{metrix}, but with (see, for example, \cite{Chamblin:1999tk}),   
\begin{equation} 
 f(r) = \frac{r^2}{L^2}+k - \frac{\omega^{n-1}}{r^{n-1}}+\frac{q^2}{r^{2 (n-1)}} \,.
\end{equation} 
The full solution also includes the Maxwell vector potential, which may be written as
\begin{equation}
A_\alpha\, dx^\alpha 
= \sqrt{\frac{n}{2(n-1)}}\left(\frac{q}{r_{\rm H}^{n-1}}- \frac{q}{r^{n-1}}\right)\, dt \, , 
\end{equation}
where the constant term is chosen to ensure that $A_t$ vanishes at the horizon.\footnote{The latter is required for $A=A_\alpha\, dx^\alpha$ to be a well-defined one-form at the bifurcation surface(s).} The combined metric and vector potential then provide a solution for the Einstein-Maxwell equations resulting from the (bulk) action,
\begin{equation} 
S_\V = \int_{\V} \bigl( R - 2\Lambda 
- F_{\alpha\beta} F^{\alpha\beta} \bigr)\, \sqrt{-g}\, d^{n+2} x\,, 
\labell{act2}
\end{equation} 
where $F_{\alpha\beta} = \partial_\alpha A_\beta 
- \partial_\beta A_\alpha$ is the electromagnetic field strength. Recall that $\Lambda=-n(n+1)/(2L^2)$. The ADM mass of this charged solution is given by the same expression as in Eq.~\reef{mass-SadS}, while the  charge of the gauge potential is related to the parameter $q$ by 
\begin{equation}
Q = q\, \sqrt{2 n (n-1)}\, \frac{\Omega_{n,k}}{8 \pi G} \, . 
\end{equation}

As before, we wish to calculate the change $\delta S = S(t_0+\delta t)-S(t_0)$ in the total (gravitational plus electromagnetic) action between the two WdW patches 
displayed in Fig.~\ref{fig:WdWcharged}, where the time slice on the left boundary is shifted slightly by $\delta t$ (and we are considering late times $t_0$). The details of the calculation are virtually identical to those presented in the preceding
subsection, and we can rely on a few key observations to simplify our 
task. First, the asymptotic joints near the left AdS boundary
are related by a time translation, and their contributions cancel out
in $\delta S$. Second, by virtue of the additivity of the action we
can conclude that the boundaries internal to the regions of interest, \eg the event horizon, do not contribute to $\delta S$; only the external boundaries are
relevant. Third, the external boundaries are all segments of null hypersurfaces, which give no contribution to $\delta S$ (or the individual actions) by virtue of our assumption that the generators are affinely parametrized. As a result, inspection of the right panel of Fig.~\ref{fig:WdWcharged} indicates that in addition to
the volume contributions, only the joints $\B$, $\B'$, $\C$, and $\C'$ 
contribute to $\delta S$. 

\begin{figure}
\includegraphics[width=.9\linewidth]{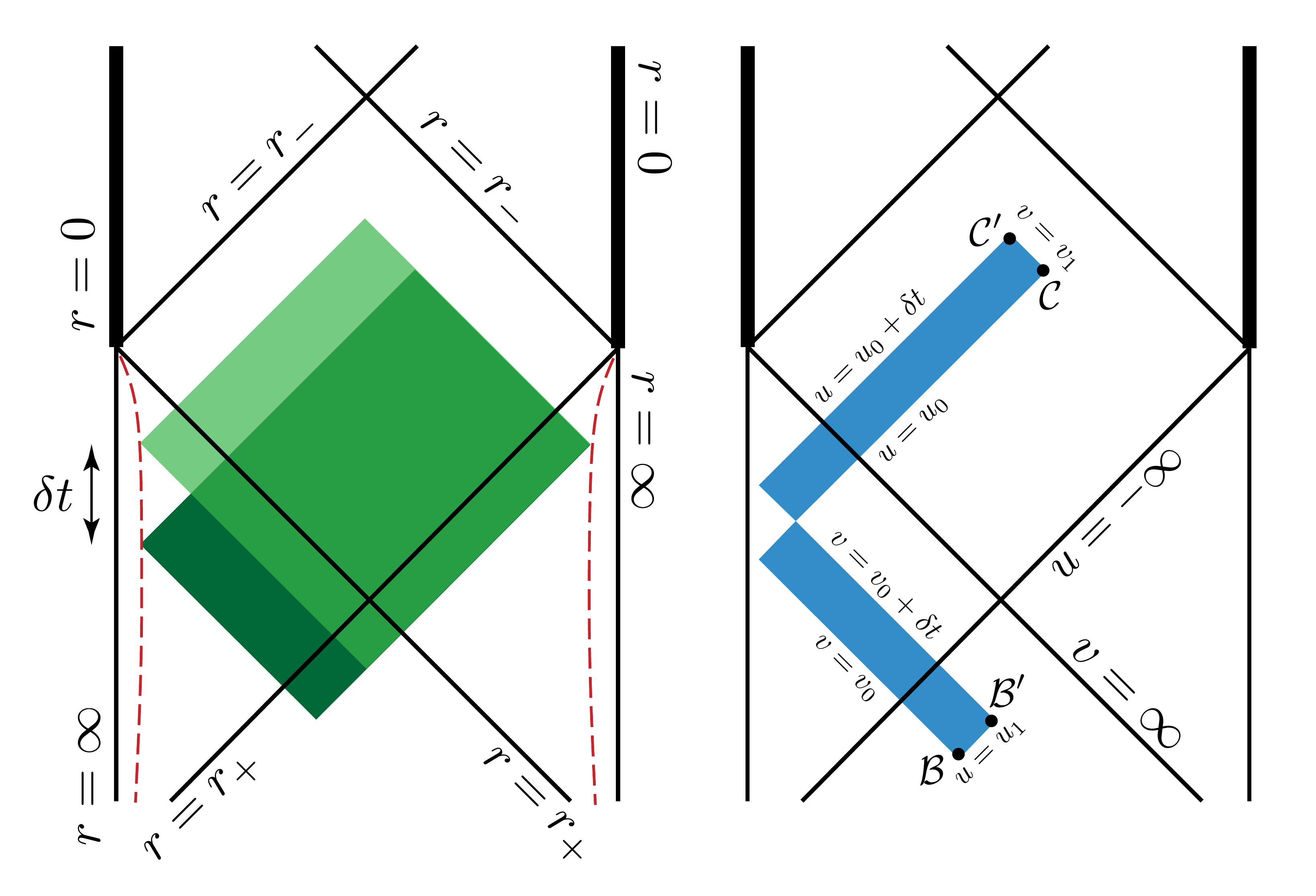}
\caption{Wheeler-deWitt patch of a Reissner-Nordstr\"om-anti de Sitter 
  spacetime. On the left panel, the patch at coordinate time $t_L=t_0$ is
  shown in dark color, and the patch at time $t_L=t_0 + \delta t$ is shown
  in light color. The difference between the two patches is shown on
  the right panel. The red dashed curves on the left panel indicate the cut-off surfaces at  $r=r_\mt{max}$ near the asymptotic AdS boundaries.
\labell{fig:WdWcharged} }
\end{figure} 

We begin with the evaluation of the volume contribution to the action from Eq.~\reef{act2} and use  
\begin{equation} 
R = \frac{2(n+2)}{n} \Lambda 
+ \frac{n-2}{n} F_{\alpha\beta} F^{\alpha\beta} \,,
\end{equation} 
which is a consequence of the Einstein equations. As in the uncharged case,
only the region inside the future horizon contributes to $\delta S$,
because the regions outside the horizon produce canceling
contributions, and the region inside the past horizon contributes a
negligible term at late times. The remaining contribution is then
given by an equation similar to Eq.~\ref{schv1}, with integration
limits given by $r_+$ and $r_-$ at late times. We arrive at  
\begin{equation}
\delta S_{\V} = -2 \Omega_{n,k} \left[\frac{r_+^{n+1} - r_-^{n+1}}{L^2}  
-   \frac{q^2}{r_-^{n-1}} + \frac{q^2}{r_+^{n-1}} 
\right ] \, \delta t \, . 
\labell{chargedvolume1} 
\end{equation}

We also rely on the results presented in the preceding section to
evaluate the joint contributions to $\delta S$. In this case we find
that   
\begin{equation}
\delta S_{{\cal B,B'}} =  \left.\Omega_{n,k}\, r^{n}
  \frac{df}{dr}\right|_{r_+}\delta t\, ,  \qquad 
\delta S_{{\cal C,C'}} = -\left.\Omega_{n,k}\, r^{n}
  \frac{df}{dr}\right|_{r_-} \delta t\, ;  
\end{equation}
the relative sign between the two expressions is a consequence of the  
different signs involved in null-joint terms in the action (as
summarized in Fig.~\ref{fig:Njoints}). The combined joint
contributions therefore  give
\begin{align}
\delta S_{{\cal B,B'}} + \delta S_{{\cal C,C'}} &=
 2\, \Omega_{n,k} \left. \left( \frac{\omega^{n-1}}{2} 
- \frac{q^2 (n-1) }{r^{n-1}} 
+ \frac{r^{n+1}}{L^2} \right)\right|_{r_-}^{r_+} . \nonumber \\
&= 2 \, \Omega_{n,k} \, \left[ \frac{q^2 (n-1)}{r_-^{n-1}} - \frac{q^2 (n-1)}{r_+^{n-1}} 
+ \frac{r_+^{n+1} - r_-^{n+1}}{L^2}\right]  \delta t\, . 
\labell{chargedcorners}
\end{align}

Combining Eqs.~(\ref{chargedvolume1}) and (\ref{chargedcorners}), we
arrive at 
\begin{equation}
\delta S = 2 \, n \, \Omega_{n,k} \left( \frac{q^2}{r_-^{n-1}} 
- \frac{q^2}{r_+^{n-1}}\right) \, \delta t, 
\end{equation}
or 
\begin{equation}
\frac{dI}{dt} = n\,\frac{\Omega_{n,k}}{8 \pi G_\mt{N}} \left( \frac{q^2}{r_-^{n-1}} 
- \frac{q^2}{r_+^{n-1}}\right).  
\end{equation}
This expression agrees with the one reported by Brown et al
\cite{Brown:2015lvg} when $n = 2$ and $k=+1$, that is, for a four-dimensional
spherical charged AdS black hole. Further, let us observe that when $q \to 0$,  $r_-^{n-1} \to q^2/\omega^{n-1}$ 
and $r_+ \to r_{\rm H}$, and hence we recover the expected $dI/dt \to 2M$
for any number of spacetime dimensions. 

\section{Discussion \labell{discuss} }

We have presented a complete analysis of the boundary terms required
in the action functional of general relativity, paying careful
attention to the case of null boundary segments.  As we have seen,
this case requires the introduction of two new classes of boundary
terms, the first on the null segments themselves, and the second on
the associated joint terms where the null boundaries intersect other
segments. For a typical null segment and its associated joint, we
have\footnote{Here we do not concern ourselves with the precise sign
  associated with these contributions to the action --- see section
  \ref{contributionstoaction} or appendix \ref{action-manual}.} 
\beq
-2\,  \int \kappa \, dS d\lambda\  +\  2 \, \oint  a \,dS\,.
\labell{nullb}
\eeq
In the first term, $\kappa$ is defined by Eq.~\reef{kappa-def1},
$k^\beta \nabla_\beta k^\alpha = \kappa\, k^\alpha$, 
where $k^\alpha$ is the (future-directed) null tangent vector along
the boundary segment. With $k^\alpha = {\partial x^\alpha}/{\partial
  \lambda}$ as in Eq.~\reef{ke-def}, $\kappa$ measures the failure of
$\lambda$ to be an affine parameter along the null generators of the
boundary segment. In section \ref{subsubsec:proof}, we showed that the
function $a$ in the joint terms takes the form 
\begin{equation} 
a = \ln|n \cdot k| + a_0, 
\labell{a-vs-a0X} 
\end{equation} 
where $n^\alpha$ is the unit normal to the other boundary segment
forming the joint, and $a_0$ is an arbitrary scalar whose variation
$\delta a_0$ is required to vanish.   

As discussed in the main text, the null boundary terms \reef{nullb}
are ambiguous. In particular, $\kappa$ depends on an arbitrary choice
of parametrization for the null generators, \ie the choice of
$\lambda$.  Further, there are two distinct ambiguities in the
expression \reef{a-vs-a0X} for $a$. First, the piece $\ln|n \cdot k|$
depends on the arbitrary normalization of the null tangent $k^\alpha$,
and this ambiguity is again related to the choice of
$\lambda$.\footnote{Note that this ambiguity remains even if a
  specific prescription is chosen for $\kappa$. That is, if we fix
  $\kappa$ in Eq.~\reef{kappa-def1}, we must still specify the initial
  value for $k^\alpha$ in order to solve this differential equation.}
The second term in Eq.~\reef{a-vs-a0X}, $a_0$, reveals a separate
dependence on the choice of the function $\Phi$ defining the boundary
surface --- see the discussion above Eq.~\reef{k-def}.  Despite these
ambiguities, the variation of the boundary terms on the null segments and null joints is
well-defined and by construction, it cancels the corresponding total
derivative terms coming from the variation of the bulk
action. However, evaluating the gravitational action for a particular
spacetime geometry will yield different numerical values depending on
the different choices in the construction of the boundary terms
\reef{nullb}.  

We might add that there are further ambiguities that are inherent to
any variational problem. For example, one can always add a total
derivative term to the (bulk) action without affecting the equations
of motion, and with a judicious choice, without affecting the vanishing
of the boundary variations. However, in general such a term would
modify the value of the action when it is evaluated on a particular
field configuration. Similarly, one could add boundary terms whose
variation vanishes, \eg because of the boundary conditions imposed on
the fields. The scalar $a_0$ appearing in the null joint terms
\reef{a-vs-a0X} would be an example of such a boundary term. While we
chose this scalar to be a simple constant, the variational principle would remain
intact with more complicated choices, \eg $a_0\propto{\cal R}$ where
${\cal R}$ is the Ricci scalar associated with the induced geometry on
the joint. A further example of the latter class would be the term $\int_\Sigma
\Theta\, \sqrt{\gamma} d^2\theta d\lambda$ introduced on null boundary
segments in \cite{Parattu:2015gga} --- see the discussion below
Eq.~\reef{bond44}. In this paper we have proposed what
we see as the minimal set of boundary terms for the gravitational
action, and we have not considered specious contributions of the above
form.\footnote{Appendix \ref{nocando} introduces an interesting
  ``specious'' boundary term which removes the reparametrization
  ambiguities on the null segments. We hope to discuss this boundary
  ``counterterm'' in greater detail elsewhere \cite{prep}.} 

In the context of the ``complexity equals action" conjecture
\cite{Brown:2015bva, Brown:2015lvg}, the ambiguities described above
may seem to be problematic. However, as 
emphasized in \cite{Brown:2015bva, Brown:2015lvg}, the circuit
complexity of a quantum state is also ambiguous. In particular, it
depends on the choice of initial reference state and specific set of
quantum gates, with which one acts to construct the 
desired state. Further, the precise value of the complexity will
depend on the tolerance that one introduces to describe the accuracy
with which the desired state must be constructed. It would be
interesting to draw a more precise connection between the
ambiguities described here for the circuit complexity and those
described above for the gravitational action.  

Now as we discussed, the ambiguities in the gravitational action can
be tamed with natural prescriptions.  In particular, the
reparametrization ambiguity on the null segments can be mitigated by 
choosing the null generators to be affinely parametrized, and then the
corresponding boundary term simply vanishes. Further, in section
\ref{sec:additivity} the undetermined functions $a_0$ at the null
joints were fixed by demanding additivity for the gravitational
action.\footnote{Implicitly, to produce an additive gravitational
  action, we are assuming that the same null tangents are used on any
  common null segments for neighbouring spacetime regions.} These
choices still leave the freedom to rescale the affine parameter
along any of the null segments by a constant factor. However, in the
context of the AdS/CFT correspondence and evaluating the action on WdW
patches, we can remove this final ambiguity by imposing a normalization
condition on the null normals near the asymptotic AdS boundary; 
see the discussion towards the end of section~\ref{sec:WdW}. This choice
(along with the previous two) allows us to make a meaningful
comparison of the action for different WdW patches. In particular,
following \cite{Brown:2015bva, Brown:2015lvg} we evaluated the
rate of change of the action for the WdW patch of asymptotically
AdS black holes in Sec.~\ref{sec:dSdt} and found the same 
result reported there, \ie $dI/dt = 2M$ (for late times and uncharged
black holes). Our analysis reveals that this simple result applies to 
noncompact horizons (\ie planar and hyperbolic horizons) as well as
spherical horizons.  

We wish to emphasize that the result $dI/dt = 2M$ for
SAdS black holes at late times is very robust, \ie it does not depend
on the specific choices made above to eliminate the ambiguities
associated with null boundary 
terms. Examining each of our choices in reverse order, we see that in 
Eq.~(\ref{roundh}), the asymptotic normalization of the null tangents
was in fact arbitrary, \ie $k\cdot \hat t_L=-c$ and
$\bar{k}\cdot \hat t_R=-\bar{c}$, where $c$ and $\bar{c}$ are
arbitrary constants. Our final result for the rate of
change of the action was independent of these constants
--- see Eq.~\reef{dS:final}. We also see that in the joint contributions
\reef{a-vs-a0X},  $a_0$ could be chosen to be any constant (or in
fact, some function of the intrinsic geometry of the null joint) and
the precise choice would not modify our result for $dI/dt$. The
independence of $a_0$ is a consequence of the Killing symmetry
$\partial_t$ of the black hole spacetime and of the very small
difference in the geometries of the joints $\cal B$ and $\cal B'$ at
late times --- see Fig.~\ref{fig:WdW}. Finally, one can imagine
replacing our choice $\kappa=0$ with $\kappa=\kappa_0$, where
$\kappa_0$ is some arbitrary nonvanishing constant on the null
boundary segments --- in fact, different values of $\kappa_0$ might be
chosen on the different null boundaries. Our result for
$dI/dt$ would again remain unchanged, first because most of the null
boundary contributions would simply cancel as a result of the Killing
symmetry $\partial_t$, and second because at late times, the segment at
$u=u_1$ is almost a stationary surface, \ie this null boundary is very
close to the past horizon at $u=-\infty$ --- again, see Fig.~\ref{fig:WdW}.
We must note that this robustness relies on fixing the
ambiguities with ``reasonable'' choices. For example, if one were to
choose $c=c(t)$ or $\kappa_0=\kappa_0(t)$ above, then one would find
$dI/dt\ne 2M$ and the result would depend on the details of the
selected functions. However, our perspective is that with such arbitrary
(time-dependent) choices, one simply cannot expect to meaningfully
compare the action of WdW patches at different times. The 
significant role of the time-translation Killing symmetry highlighted
above suggests the importance of making detailed studies of
time-dependent scenarios in the future. 

One slightly unsettling feature of the calculation of $dI/dt$ is that
an essential contribution comes from the boundary segment near the
spacelike singularity at $r=0$, \ie the boundary segment $\S$ in
Fig.~\ref{fig:WdW}. This contribution is determined by evaluating the
$K$ term on a regulator surface at $r=\epsilon\ (\ll r_{\rm H})$
and then taking the limit $\epsilon\to0$. It is remarkable that the
result \reef{dS:surface}  turns out to be finite and independent of
$\epsilon$. At a pragmatic level, this occurs because the divergence
in $K$ is precisely compensated for by the vanishing of the volume
element $d\Sigma$. This precise balance relies on the specific
behavior of the metric function $f(r)$ near the singularity, and 
therefore on the assumed validity of the Einstein equations in 
this region of spacetime. Of course, this outcome might be regarded
with suspicion  since UV effects, \eg higher curvature terms arising
as stringy and quantum corrections, are expected to modify the field
equations and spacetime geometry in the vicinity of the
singularity. However, one might argue that these deviations should be
small so long as the regulator scale $\epsilon$ is chosen to be well
above the quantum gravity scale, and hence the evaluation of the action
should be robust.  
As a simple test of this reasoning, one might examine how the rate of
change \reef{dAdt} of the WdW patch action is modified if the
regulator is taken to be small but finite. Here one finds 
\beq
\delta\left(\frac{dI}{dt}\right) = -k\,\frac{n \Omega_{n,k}}{8\pi G_\mt{N}} \epsilon^{n-1}\,.
\labell{eeek}
\eeq
Hence for a spherical horizon (\ie $k=+1$), the regulator corrections
reduce the rate, which seems to align with the conjecture of
\cite{Brown:2015bva, Brown:2015lvg} that there should be a bound
$dI/dt\le 2M$. However, the rate increases for a hyperbolic horizon
(\ie $k=-1$) and the result may seem to contradict those
expectations.  We should add that preliminary investigations
\cite{cai77,dan99} into extending these calculations to (classical)
higher curvature theories have cast doubt on the simple argument given
above. It is certainly a question which deserves further study.  

Future studies of the ``complexity equals action'' conjecture should
 examine less symmetric situations, as well.  With less symmetry, caustics and
 crossings will generically appear on the boundary of the WdW patch, \ie
 there will be points where the null generators of the boundary of the WdW patch cross each other and the boundary fails to be smooth.\footnote{To clarify our nomenclature, we use ``caustic'' to refer to the
 situation where the crossing null rays were only infinitesimally
 separated in the transverse directions on the boundary, and ``crossing''
 for the case where the null rays were initially widely separated.}
 Indeed, this situation will arises generically whenever one seeks the action of a
 spacetime region defined in terms of future- or past-sets.
 
 Our intuition is that locally such crossings will typically have the
 geometry of a spacelike joint formed by the intersection of two null
 surfaces, \eg see~\cite{PhysRevD.52.6982}.  We have not proven that
 this is the only generic case, but it is at least clear that such
 null-null joints form a broad class of stable crossings.  In this
 case, our results would allow the evaluation of the corresponding
 gravitational action, in that one would simply include an additional
 joint contribution with $a = \ln(-\frac{1}{2} k \cdot \bar{k})$,
 where $k^\alpha$ and $\bar{k}^\alpha$ are the null normals on either
 side of the joint.
 
 A new feature would be that these crossings may terminate at a caustic
 or at another crossing of lower dimension, for example at a ``corner" or
 joint of codimension-three, \ie the simultaneous intersection of three
 segments of the boundary surface.  In such cases, one might
 need to include an additional contribution to the action from the corner
 or caustic where the null-null joint terminates -- see further discussion 
of such corner contributions below.  
To evaluate the boundary contribution for a caustic, one approach would be to ``regulate'' the geometry as follows:
 First, introduce an additional timelike boundary surface which cuts the
 endpoints of the crossing out of the boundary and then remove this
 regulator surface so that the endpoints reappear in a limit.  It seems
 that the regulated geometry in such an approach would also typically
 involve a codimension-three joint.  This provides some
 motivation to study the boundary terms (if any are needed) for such
 higher codimension corners, as discussed in the next paragraph.

Our discussion in this paper has focused on the possibility of
spacelike joints, or intersections between pairs of boundary
surfaces. The case of timelike joints where two timelike boundary
surfaces meet was also considered 
in \cite{Hayward:1993my,1994PhRvD..50.4914B}. 
Of course, our analysis could be further generalized to consider more
complicated intersections involving more than two boundary
surfaces. For example, in $d$ spacetime dimensions, a volume with
planar boundary surfaces, \ie a $d$-dimensional polyhedron, would have
joints where pairs of boundaries intersect (as considered here)
but also ``corners'' where three, four and up to $d$ boundary surfaces
intersect simultaneously. In principle, a more complete analysis would
include the possibility of additional boundary terms for each of these
different types of intersections. Of course, this generalization would
need to be carried out for corners involving only timelike and
spacelike surfaces first, before proceeding to cases involving  null
boundary surfaces as well.  

When the spacetime signature is Euclidean, there is only one kind of
boundary segment and hence only one kind of joint to be considered.
As originally described in \cite{JR} (see also \cite{Hayward:1993my}), 
the joint terms in the gravitational action take the familiar form
$2\oint \eta\, dS$ in this context, where $\eta=\pi-\theta$ with
$\theta$ denoting the dihedral angle of the joint. When the spacetime
signature is continued to Lorentzian, it is  not immediately obvious
how the dihedral angles should be defined, but a prescription for
doing so was given in \cite{Sorkin:1974pya,Sorkin:1975ah} and
applied there to the definition of the Regge action. This
continuation can be used to recover the Hayward terms
\cite{Hayward:1993my,1994PhRvD..50.4914B} for spacelike joints that we
have examined in this paper.  However, there is one interesting small
difference. The integrand $\eta$ of the spacelike joint terms which
one obtains from this continuation differs  from that given earlier in
section~\ref{sec:TSjoints} (see also Appendix \ref{action-manual}) by
an imaginary constant. It appears likely that a similar procedure
could also be used to understand the new joint contributions found in
this paper for intersections involving null boundary
segments. However, we expect that the integrand $a$ in these
joint terms would also acquire an imaginary piece. This would
correspond to making an alternative choice of the constants
$\hat{a}_0$, $a_{01}$ and $\bar{a}_{02}$ in section
\ref{sec:additivity}. With these new choices, the gravitational action
(specifically its imaginary part) would lack the additivity that
motivated our choices above, but on the other hand the two cases of
timelike and spacelike joints would now resemble each other more 
closely.\footnote{Of course, as noted previously, the standard
  prescription for timelike joints
  \cite{Hayward:1993my,1994PhRvD..50.4914B} is incompatible with
  additivity of the gravitational action, and the same issue extends
  to the Euclidean setting. Hence it is not surprising that the
  continuation considered here leads to a result that is incompatible
  with additivity. However, let us add
  the following observation: One might consider examining the boundary
  terms for timelike joints exclusively from the perspective of
  requiring a good variational principle, as in section
  \ref{sec:closed-nonnull}. In this framework, one would find that
  there is the freedom to add an arbitrary scalar $\eta_0$ to the
  joint term, as long as $\delta\eta_0=0$. Additivity of
  the gravitational action would be restored if one were to choose
  $\eta_0=-\pi$. The deficiency of this prescription is that the
  timelike joint cannot be modeled as a limiting sequence of smooth
  timelike surfaces, with the joint emerging at the end of the
  limit. In this approach, the joint term emerges as a 
  delta-function contribution in the extrinsic curvature of the usual
  Gibbons-Hawking-York boundary term and yields the standard
  prescription for $\eta$ in the boundary term on a timelike
  joint. This straightforward geometric construction was the original
  approach adopted in \cite{Hayward:1993my}. However, we expect that
  an appropriate analytic continuation of the proposed ``additive"
  prescription would also yield a vanishing imaginary contribution for
  the boundary terms on spacelike joints.} These imaginary
contributions to the action can be ignored if one is interested only in infinitesimal variations 
of the action.\footnote{Further, these contributions would not modify
  our results of $dI/dt$.} However, they become relevant in
considering topology change in quantum gravity \cite{1997CQGra..14..179L}. One also arrives
at new insights into the Bekenstein-Hawking entropy by retaining this
imaginary contribution to the action
\cite{Neiman:2012fx,Neiman:2013ap,Neiman:2013lxa,Neiman:2013taa}. These
imaginary contributions also play a role in a new derivation of
holographic entanglement entropy \cite{Dong:2016hjy}.  
 

\acknowledgments 
We would like to thank Dorit Aharonov, Ivan Booth, Adam Brown, Shira
Chapman, Bartek Czech, Stephen Green, Hugo Marrochio, Henry Maxfield,
Don Marolf, Yasha Neiman, Krishnamohan Parattu, Dan Roberts, Joseph
Samuel, Sumati Surya, Lenny Susskind, Brian Swingle and Ying Zhao for
useful discussions. Research at Perimeter Institute is supported by the  
Government of Canada through the Department of Innovation, Science and
Economic Development and by the Province of Ontario through the
Ministry of Research \& Innovation. All authors are supported by NSERC
Discovery grants. In addition, LL and RCM are supported by 
research funding from the Canadian Institute for Advanced
Research, and RCM acknowledges support from the Simons Foundation 
through the ``It from Qubit'' collaboration.  

\appendix
\section{Ambiguities in the null limit from  timelike surfaces
\labell{sec:limit} }

In Sec.~\ref{contributionstoaction}, we described the contributions
to the gravitational action arising from boundary segments and joints
between them. In particular, we showed that the contributions from
null segments and null joints are in general ambiguous. The question
arises as to whether these ambiguities can be resolved by interpreting
a null hypersurface as the limit of a sequence of timelike or
spacelike surfaces. In this section, we show that such a limiting
procedure is also generically ambiguous. An exception to this general rule
arises when the null limit is a stationary surface; in this case a
unique limit exists.\footnote{Of course, this result is related to the
  discussion at the end of Sec.~\ref{subsub:reparam}, where we found
  that the action was invariant under reparametrizations for a null
  boundary segment which is stationary.} Further, we demonstrate
that the Hayward joint terms that appear in such a limiting procedure 
yield a divergent result. For concreteness we shall consider the
specific case of a sequence of timelike hypersurfaces that is made to
approach a null limit. Also, for the sake of simplicity, we restrict
our attention to a few simple examples involving (i) a static and
spherically-symmetric spacetime, (ii) the Kerr spacetime, and (iii)
the radiative Vaidya spacetime.  

In all cases we evaluate the boundary action $S_{\partial \V}$ on a
segment of $\partial \V$ that consists of a timelike hypersurface $\T$
truncated by spacelike hypersurfaces $\S_1$ and $\S_2$ to the past
and future, respectively. We take $\T$ to be an inner boundary to $\V$, and
we ignore the contribution to $S_{\partial \V}$ that comes from the
outer boundary. In fact, we shall also ignore the contributions from
$\S_2$ and $\S_1$, but retain the joint term at $\B_2$, the
two-surface of intersection between $\T$ and $\S_2$, as well as the
joint term at $\B_1$, the two-surface of intersection between $\T$ and
$\S_1$. Selecting the corresponding terms from Eq.~(\ref{Sboundary-TS}) gives  
\begin{equation} 
S = 2 \int_{\T} L\, \sqrt{-f} d^3z 
+2\oint_{\B_2} \eta\, \sqrt{\gamma} d^2\theta 
- 2\oint_{\B_1} \eta\, \sqrt{\gamma} d^2\theta\,.  
\labell{S-examples} 
\end{equation} 
We recall the notation employed in Sec.~\ref{sec:closed-nonnull}:
Coordinates $z^j$ are placed on the timelike hypersurface $\T$, which
possesses an intrinsic metric $f_{jk}$ and an extrinsic curvature
$L_{jk}$, while coordinates $\theta^A$ are placed on $\B_1$ and
$\B_2$, which possess an intrinsic metric $\gamma_{AB}$; the vector
$s^\alpha$ is normal to $\T$ and points toward smaller values of $r$
on the inner surface,\footnote{The convention established in
  Sec.~\ref{contributionstoaction} was that the normal $s^\alpha$ to a
  timelike boundary should point out of the volume of interest.}
$n^\alpha$ is normal to $\S_1$ and $\S_2$ (pointing to the future),
and the boost parameter $\eta$ is defined by $\sinh\eta := n_\alpha s^\alpha$.   

\subsection{Static, spherically-symmetric spacetime
\labell{subsec:static} }

For the first set of examples we consider a spacetime with metric  
\begin{equation} 
ds^2 = -g\, dv^2 + 2dvdr + r^2 d\Omega^2, 
\labell{sssmetric} 
\end{equation} 
in which $v$ is an advanced-time coordinate, and $d\Omega^2 := d\theta^2 
+ \sin^2\theta\, d\phi^2$. We take $g=g(r)$ to be an arbitrary
function of $r$. The surface $\T$ is described by $r=R(v)$,
in which $R(v)$ is an arbitrary function of $v$. Its normal is  
\begin{equation} 
s_\alpha = -(G - 2\dot{R})^{-1/2} [-\dot{R},1,0,0], 
\labell{sssnormal} 
\end{equation} 
in which an overdot indicates differentiation with respect to $v$, and
$G := g(r=R)$. The induced metric is 
\begin{equation} 
f_{jk} dz^j dz^k = -(G-2\dot{R})\, dv^2 + R^2\, d\Omega^2, 
\end{equation} 
so that the corresponding volume element is $\sqrt{-f} d^3z = (G-2\dot{R})^{1/2}
R^2\, dv\, d\Omega$, with $d\Omega := \sin\theta\, d\theta d\phi$. The
trace of the extrinsic curvature is  
\begin{equation} 
L = -\frac{ \frac{1}{2} (G-3\dot{R}) G' + \ddot{R}}{(G-2\dot{R})^{3/2}}   
- \frac{2 (G-\dot{R})}{R (G-2\dot{R})^{1/2} },
\end{equation} 
where $G' := dG/dR$. 

The surfaces $\S_1$ and $\S_2$ are both described by an equation of
the form $v = r + \mbox{const}$, and their normal vector is   
\begin{equation} 
n_\alpha = (2-g)^{-1/2} [-1,1,0,0]. 
\end{equation} 
The inner product of $n_\alpha$ and $s_\alpha$ evaluated at
$\S_1$ or $\S_2$ is 
\begin{equation} 
\sinh\eta = -\frac{G-\dot{R}-1}{\bigl[(2-G)(G-2\dot{R})\bigr]^{1/2}}, 
\end{equation} 
so that 
\begin{equation} 
\eta = \frac{1}{2}\ln \frac{2-G}{G-2\dot{R}}\,.\labell{juice} 
\end{equation} 
In these equations, $G$ and $\dot{R}$ are evaluated at either $v=v_1$
or $v=v_2$, the values of $v$ at $\S_1$ and $\S_2$, respectively. 

These results imply that 
\begin{equation} 
S = -4\pi \int_{v_1}^{v_2} 
\biggl[ \frac{ (G-3\dot{R}) G' + 2\ddot{R}}{G-2\dot{R}}   
+ \frac{4 (G-\dot{R})}{R} \biggr] R^2\, dv 
- 4\pi R^2 \ln \frac{G-2\dot{R}}{2-G}\biggr|^{v_2}_{v_1}. 
\end{equation} 
Now the null limit is achieved by letting $\dot{R} \to \frac{1}{2} G$, and
the expression reveals that the limit diverges in general. 

As a specific example, we may consider the sequence of timelike 
hypersurfaces defined by $\dot{R} = \frac{1}{2}(1-\epsilon) G$ with
$\epsilon \to 0$. In this case, $S$ becomes 
\begin{equation} 
S_1 = -4\pi \int_{v_1}^{v_2} \biggl( \frac{1}{2} G' + \frac{2G}{R}
\biggr) R^2\, dv 
- 4\pi R^2 \ln \frac{\epsilon G}{2-G} \biggr|^{v_2}_{v_1}
+ O(\epsilon), 
\end{equation} 
which diverges logarithmically as $\epsilon \to 0$. However, we note
that in the special case that $R(v_2) = R(v_1)$, the individual
divergences of the joint terms at $v_1$ and $v_2$ will cancel to leave
a finite action. As another example, we take the sequence  
$\dot{R} = \frac{1}{2} G - \epsilon$ with $\epsilon \to 0$. In this
case  
\begin{equation} 
S_2 = -4\pi \int_{v_1}^{v_2} \biggl( G' + \frac{2G}{R}
\biggr) R^2\, dv 
- 4\pi R^2 \ln \frac{2\epsilon}{2-G} \biggr|^{v_2}_{v_1}
+ O(\epsilon), 
\end{equation} 
which also diverges logarithmically unless $R(v_2) = R(v_1)$. Even
when this condition is imposed to eliminate the logarithmic
divergence, the finite terms in $S_1$ and $S_2$ do not agree with each
other. We must conclude that the null limit does not exist, and so
this limiting procedure cannot provide a unique prescription for the
surface action of a null boundary segment.   

An exception to this conclusion arises when the limiting null surface
is stationary, as in the case of a Killing horizon. To recognize this exception,
we consider the sequence of timelike hypersurfaces described by 
\begin{equation} 
R(v) = r_0 \bigl[ 1 + \epsilon b(v) \bigr], \qquad \epsilon \to 0, 
\end{equation} 
where $r_0$ denotes the radial position of a Killing horizon, \ie $g(r=r_0)=0$ in Eq.~\reef{sssmetric}, and
$b(v)$ is an arbitrary function of $v$. Because $R(v)$ is close to
$r_0$ we may simplify our computations by Taylor-expanding 
$G := g(r=R)$ about its zero value at $r_0$; this gives 
$G = 2\epsilon \kappa r_0 b(v)$, in which $\kappa := \frac{1}{2}
dg/dr|_{r=r_0}$. Making these substitutions reveals that 
\begin{equation} 
S = -4\pi r_0^2 \int_{v_1}^{v_2} \frac{dB}{dv}\, dv 
- 4\pi r_0^2 \ln \bigl[ \epsilon r_0 (\kappa_0 b 
- \dot{b}) \bigr] \biggr|^{v_2}_{v_1} + O(\epsilon), 
\end{equation} 
where $B := 2\kappa v - \ln(\kappa b - \dot{b})$. Again the individual joint terms diverge but these logarithmic divergences cancel when combined in the action, and the above expression simplifies
\begin{equation} 
S = -8\pi r_0^2 \kappa (v_2 - v_1) + O(\epsilon)\,.\label{hoje} 
\end{equation} 
In this case, we observe that the null limit is actually finite and
independent of the arbitrary function $b(v)$. The limit is therefore
well-defined, and in fact, $\lim_{\epsilon\to 0} S$ agrees with the
expression of Eq.~(\ref{S-stationary}), which applies to any
stationary null hypersurface and is invariant under
reparametrizations.  

We observe that the limit \reef{hoje} does not differ by some
residual $O(1)$ constant from the result \reef{S-stationary}
calculated with our prescription, but rather they agree precisely. One
may see this precise agreement as further motivation for our choice of
setting $a^{\rm spacelike}_0=0$ in section \ref{sec:additivity}, where
we fixed our final prescription for the null joint contributions. 

\subsection{Kerr spacetime
\labell{subsec:kerr} }

The latter conclusion in not an artifact of our restriction to
spherically-symmetric spacetimes. A similar calculation carried out
for the specific case of a Kerr spacetime reveals that when a sequence   
of timelike surfaces is made to approach the event horizon of a Kerr
black hole,  
\begin{equation} 
S = -4\pi \frac{r_+^2-a^2}{r_+} (v_2-v_1) + O(\epsilon), 
\labell{S-Kerr} 
\end{equation} 
where $r_+$ denotes the radius of the event horizon (in
Boyer-Lindquist coordinates) and $a$ is the black hole's angular
momentum per unit mass. With 
\begin{equation} 
\kappa = \frac{r_+-M}{r_+^2+a^2} = 
\frac{r_+^2-a^2}{2r_+(r_+^2+a^2)}, \qquad 
{\cal A} = 4\pi (r_+^2 + a^2) 
\end{equation} 
standing for the surface gravity and event-horizon area of a Kerr
black hole, respectively, we once more recover
Eq.~(\ref{S-stationary}) in the limit $\epsilon \to 0$.  

We can outline the calculations producing Eq.~(\ref{S-Kerr}) as
follows: We begin with the Kerr metric as in 
Eq.~(5.55) of the {\it Toolkit} \cite{2004rtmb.book.....P}, written in terms
of coordinates $v$ and $\psi$ that are regular at the event
horizon. For $\T$, we adopt the sequence of timelike hypersurfaces
described by $r = R(v) = r_+[1+\epsilon b(v)]$ with $\epsilon \to 0$
and $b(v)$ arbitrary, and we take $\S_1$ and $\S_2$ to be described by
$v = r + \mbox{constant}$. We find that $2L\sqrt{-f}$ can be expressed
as $\partial B/\partial v$ for some function $B(v,\theta)$ that satisfies
$B + 2\eta \sqrt{\gamma} = (r_+^2-a^2) v\sin\theta/r_+$ at $v=v_1$
(on $\B_1$) and $v=v_2$ (on $\B_2$). These results guarantee that the
integral over $\T$ is equal to new boundary terms at $\B_1$ and $\B_2$
that mostly cancel out the original terms coming from $\eta$; what
remains gives rise to Eq.~(\ref{S-Kerr}).  

\subsection{Vaidya spacetime
\labell{subsec:vaidya} }

In this section we consider the Vaidya spacetime, which describes a 
black hole formed by the accretion of null dust. The metric is again
given by Eq.~(\ref{sssmetric}), with the specific choice 
$g(v,r) = 1 - 2m(v)/r$, where $m(v)$ a time-dependent mass 
function. For the sake of simplicity we adopt the specific model
described in Problem 5.2.7 of the {\it Toolkit}
\cite{2004rtmb.book.....P}, for which the mass function is given by  
\begin{equation} 
m(v) = \left\{ 
\begin{array}{ll} 
0 & \quad v < 0 \\ 
v/16 & \quad 0 < v < v_0 \\ 
v_0/16 & \quad v > v_0 
\end{array} 
\right. , 
\end{equation} 
where $v_0$ is a constant. The spacetime is flat when $v<0$, accretion
begins at $v=0$ and causes the mass to increase linearly, and
accretion ends at $v=v_0$, when the black hole has acquired a mass
$v_0/16$.  

Restricting our attention to the interval $0 < v < v_0$, we find that
the radial null geodesics in this spacetime satisfy the differential
equations   
\begin{equation} 
\frac{dv}{dr} = 0 \quad \mbox{(incoming light rays)}, \qquad 
\frac{dv}{dr} = \frac{2}{g} = \frac{16 r}{8r-v} \quad  
\mbox{(outgoing light rays)}. 
\end{equation} 
The generic solution to the outgoing-ray equation can be expressed in 
the parametric form 
\begin{equation} 
v(\lambda) = 4c(2-\lambda) e^\lambda, \qquad 
r(\lambda) = c(1-\lambda) e^\lambda, 
\end{equation} 
where $c$ is a constant, and the parameter $\lambda$ ranges over a
subset of the interval $-\infty < \lambda < 1$. An exceptional
solution to the equation is  
\begin{equation} 
r = v/4; 
\labell{outgoing-ray-special} 
\end{equation}   
these light rays originate from the singularity at $v=0$, $r=0$, which
is therefore momentarily naked. The generators of the event horizon
are identified with the outgoing light rays that become stationary at
$v=v_0$, and join smoothly with the surface $r = 2m_0 = v_0/8$ beyond
$v=v_0$; these light rays have $c = v_0/8$. 

We examine a sequence of timelike surfaces $\T$ that approaches the 
null hypersurface described by Eq.~(\ref{outgoing-ray-special}). We choose the
sequence to be described by $r = R(v)$ with  
\begin{equation} 
R(v) = v/4 + \epsilon b(v), \qquad \epsilon \to 0. 
\end{equation} 
We let $b(v)$ be an arbitrary function of the advanced-time $v$, but we
assume that $b > 0$. We also take $\dot{b} := db/dv < 0$, to ensure
that the hypersurfaces are timelike when $\epsilon > 0$. The unit
normal $s_\alpha$ and the intrinsic metric $f_{jk}$ take the same
expressions as in Sec.~\ref{subsec:static}, but with $G$ now standing
for $1 - v/(8R)$. The computation of $L_{jk}$ requires a few changes
to account for the $v$-dependence of the mass function. We again take
the spacelike hypersurfaces $\S_1$ and $\S_2$ to be described by
equations of the form $v = r + \mbox{const}$, and the unit normal
$n_\alpha$ can be imported without change from
Sec.~\ref{subsec:static}. 

We compute $L$ and $\eta$, expand in powers of $\epsilon$, and 
insert within Eq.~(\ref{S-examples}). After simplification we find
that the boundary action becomes 
\begin{equation} 
S = -\frac{\pi}{4} \Biggl\{ 
\int_{v_1}^{v_2} \Bigl[ 7v - v^2 \frac{d}{dv} 
\ln\bigl(b-v \dot{b} \bigr) \Bigr]\, dv 
+ v^2 \ln \frac{4\epsilon \bigl( b-v\dot{b} \bigr)}{3v} 
\biggr|^{v_2}_{v_1} \Biggr\} + O(\epsilon), 
\end{equation} 
with the integral representing the contribution from $\T$, while the
boundary terms represent the joint contributions from the
intersections with $\S_1$ and $\S_2$. Again we see that the individual
joint terms yield a divergent result in the null limit. Integration by
parts allows us to rewrite the above expression as  
\begin{equation} 
S = -\frac{\pi}{8} \biggl\{ 4 \int_{v_1}^{v_2} v 
\ln \bigl( b-v\dot{b} \bigr)\, dv 
+ v^2 \biggl( 7 + 2 \ln \frac{4\epsilon}{3v} \biggr)
\biggr|^{v_2}_{v_1} \Biggr\} + O(\epsilon). 
\end{equation} 
Hence the divergent joint terms still yield a logarithmic divergence
in the action when $\epsilon \to 0$, and the limit is not
defined. Even if we set this divergence aside and examine the finite
terms, we see that the integral depends on the detailed behavior of
the function $b(v)$, but that the joint terms are independent of
$b(v)$. This shows that the limit would be ill-defined even if the
logarithmic divergence could be regularized: the limit to $r = v/4$
depends on how the null hypersurface is approached.  

In Sec.~\ref{subsec:static}, we also saw the logarithmic divergence
survive in $\lim_{\epsilon \to 0} S$ when the limit was to a
{\it nonstationary} null hypersurface. We note that the nonstationary
nature of the hypersurface appears to be a key aspect for the
appearance of this divergence in the action; whether or not the
spacetime itself is stationary appears to be unimportant.

\section{Counterterm for the null boundary action
\labell{nocando}}

One might ask whether the dependence of the gravitational action on
the parametrization of the null generators can be eliminated by adding
an additional ``counterterm'' to the boundary action
$S_\Sigma(\mbox{joined})$, as given by Eq.~(\ref{S-joined})? We recall
that the change to $S_\Sigma(\mbox{joined})$ under a reparametrization 
is given by Eq.~(\ref{barS}), 
\begin{equation} 
\bar{S}_\Sigma(\mbox{joined}) = S_\Sigma(\mbox{joined}) 
+ 2\int_\Sigma \Theta\, \beta\, \sqrt{\gamma} d^2\theta d\lambda, 
\end{equation} 
where $e^{-\beta} := \partial{\bar{\lambda}}/\partial{\lambda}$. Remarkably, the 
answer to this question is in the affirmative. 

Any functional of the hypersurface's intrinsic geometry can be added
to $S_\Sigma(\mbox{joined})$ without affecting the variational
principle, and we seek a counterterm of the suitable form  
\begin{equation}   
\Delta S_\Sigma = \int_\Sigma {\cal L} \sqrt{\gamma}\, d^2\theta\,
d\lambda\,, 
\end{equation} 
where ${\cal L}$ is a function constructed from scalars that
characterize the intrinsic geometry of the hypersurface. We may
consider a number of such scalars, for example, $\Theta$, $R$, and 
$B_{AB} R^{AB}$, where $R_{AB}$ is the Ricci tensor constructed from
$\gamma_{AB}$, and $R$ the corresponding Ricci scalar. For simplicity, let us assume that
${\cal L}$ is a function of only $\Theta$, and ignore more exotic
possibilities. This changes according to $\bar{\Theta} = e^\beta
\Theta$ under a reparametrization, and so the proposed counterterm becomes 
\begin{equation} 
\Delta \bar{S}_\Sigma = \int_\Sigma e^{-\beta} 
{\cal L}\bigl( e^\beta\Theta \bigr)\, 
\sqrt{\gamma}\, d^2\theta\, d\lambda\,. 
\end{equation} 
Can a judicious choice for ${\cal L}(\Theta)$ ensure the invariance of
$S_\Sigma(\mbox{joined}) + \Delta S_\Sigma$?  

It is easy to check that with  
\begin{equation} 
{\cal L} = -2 \Theta\, ( \ln|\Theta| + c )\,, 
\labell{Lsol} 
\end{equation} 
where $c$ is an arbitrary constant, the change in the counterterm is given by  
\begin{equation} 
\Delta \bar{S}_\Sigma = \Delta S_\Sigma 
- 2 \int_\Sigma \Theta \beta\, \sqrt{\gamma}\, d^2\theta\, d\lambda\,, 
\end{equation} 
so that 
\begin{equation} 
\bar{S}_\Sigma(\mbox{joined}) + \Delta \bar{S}_\Sigma  
= S_\Sigma(\mbox{joined}) + \Delta S_\Sigma.  
\end{equation} 
With this counterterm, therefore, the boundary action becomes  
invariant under a reparametrization of the null generators. 

To see how Eq.~(\ref{Lsol}) was obtained, take $\beta$ to be 
infinitesimal, perform a Taylor expansion of the transformed boundary
action, and obtain 
\begin{equation} 
\bar{S}_\Sigma(\mbox{joined}) + \Delta \bar{S}_\Sigma 
= S_\Sigma(\mbox{joined}) + \Delta S_\Sigma 
+ \int_\Sigma \beta \biggl( 2\Theta 
+ \Theta \frac{d{\cal L}}{d\Theta} - {\cal L} \biggr) 
\sqrt{\gamma} d^2\theta d\lambda\,. 
\end{equation} 
Then to have invariance of the action for an arbitrary $\beta$, we require 
\begin{equation} 
\Theta \frac{d{\cal L}}{d\Theta} - {\cal L} 
+ 2\Theta = 0,  
\end{equation} 
and the solution to this differential equation is Eq.~(\ref{Lsol}).

\section{Action User's Manual 
\labell{action-manual}}

We include a summary of how to evaluate the gavitational action with
all its relevant contributions. We write the gravitational action as 
\begin{eqnarray} 
S_{\V} &:=&    \int_{\V} (R -2 \Lambda) \, \sqrt{-g} \,dV \nonumber \\
	 && +  2\, \Sigma_{T_i}  \int_{\partial \V_{T_i}} K \, d\Sigma
	 +2\, \Sigma_{S_i} {\rm sign}({S_i}) \int_{\partial \V_{S_i}} K \, d\Sigma    
          - 2\, \Sigma_{N_i} {\rm sign}({N_i})  \int_{\partial \V_{N_i}} \kappa \, dS d\lambda  \nonumber \\
	 & & +  2 \,\Sigma_{j_i} {\rm sign}({j_i})  \oint   \eta_{j_i} \, dS   
             +  2 \,\Sigma_{m_i}  {\rm sign}({m_i}) \oint  a_{m_i} \,dS
\end{eqnarray}
where we have arranged contributions from the bulk, surfaces, and
joints in the first, second and third lines
respectively. For bookkeeping, spacelike, timelike and null boundary
surfaces are labeled by ${S_i}$, ${T_i}$ and $N_i$,
respectively. Joints formed by an intersection
involving no null segments are denoted by $j_i$, while those with at
least one null segment are denoted by $m_i$. 

The expressions for surface and joint contributions are sensitive to
the conventions adopted. We have chosen conventions whereby the
timelike vectors normal to spacelike boundary segements are always
directed towards the future, the null vectors tangent to null boundary
segments are always directed towards the future, and spacelike vectors
normal to timelike boundary segments always point out away from the
volume of interest.  Consequently, for {\bf surface contributions}, 
the following signs must be accounted for:
\begin{itemize}
\item For spacelike boundaries, 
 ${\rm sign}({S_i}) = 1 (-1)$ if the spacetime volume for which we are
 evaluating the action lies to the future (past) of the boundary
 segment, \ie the  normal vector points into (out of) the region of
 interest. 
\item For null boundaries,  ${\rm sign}(N_i) = 1 (-1)$ if the volume
  of interest lies to the future (past) of the null segment.  
\end{itemize}

\vspace{0.5cm}

The {\bf joint contributions}, discussed in
sections~\ref{sec:TSjoints}, \ref{sec:Njoints}, and
\ref{sec:additivity} are summarized in a rather straightforward way
below. While our description of the contributions coming from joints
between spacelike and/or timelike surfaces might appear to differ from
that given in \cite{Hayward:1993my,1994PhRvD..50.4914B}, our results
are in fact in precise agreement with those earlier works, and our
summary provides an explicit prescription for the sign of these terms,
which was previously left ambiguous.\\ 

\noindent $\Diamond$ {\em Joints formed by the intersection of spacelike surfaces:}\\
As in the main text, we denote the (future-directed) timelike unit
normal to each hypersurface as $n^\alpha_i$ with $i=1,2$. For each
boundary segment we  introduce a spacelike unit vector $p^\alpha_i$ which is
in the tangent space of the corresponding segment, orthogonal to the
joint, and points outward from the segment.  
Then the contribution from the corresponding joint can be written as
\begin{equation}
 \eta_{j_i} = \ln|(n_1 + p_1) \cdot n_2|\,. 
\end{equation}
Further, ${\rm sign}({j_i})= +1$ if $n^\alpha_1$ is directed out of
the volume of interest, and ${\rm sign}({j_i})= -1$ otherwise. (We
note that these rules are sensitive to which boundary segment is
labeled $S_1$ and which $S_2$ --- \eg consider interchanging the
labels in Fig~\ref{fig:TSjoints}f. \\ 

\noindent $\Diamond$ {\em Joints formed by the intersection of timelike surfaces:}\\
Let the spacelike unit normal to each hypersurface be given by $s^\alpha_i$
(with $i=1,2$ and $s^\alpha_i$ is chosen to point out of the volume of
interest), and at the joint, introduce two timelike unit vectors $p^\alpha_i$
which are tangent to the corresponding segment, orthogonal to the
joint, and point outward from their segment. The joint contribution
can  be written as 
\begin{equation}
 \eta_{j_i} = \ln|(s_1 + p_1) \cdot s_2|\,. 
\end{equation}
Further, ${\rm sign}({j_i})= -1$ in all cases.\\

\noindent $\Diamond$ {\em Joints formed by the intersection of a spacelike and a timelike surface:}\\
Assuming the (outward-directed spacelike) unit normal to the timelike
surface is given by $s^\alpha$, the (future-directed) timelike unit normal  
to the spacelike hypersurface is given by $n^\alpha$ and the spatial unit
vector orthogonal to the joint in the latter boundary segment is given
by $p^\alpha$, the contribution from the corresponding joint is
\begin{equation}
 \eta_{j_i} = \ln|(n + p) \cdot s|\,. 
\end{equation}
Further, ${\rm sign}({j_i})= +1$ if $n^\alpha$ is directed out of the
volume of interest, and ${\rm sign}({j_i})= -1$ otherwise.\\ 

\noindent $\Diamond$ {\em Joints formed by the intersection of at least one null surface:}\\
Assuming the null vector $k^\alpha$ is future directed and tangent to the null surface, and
the intersecting surface has normal vector $n^\alpha$ if spacelike,
$s^\alpha$ if timelike, or (future directed) null tangent vector $\bar k^\alpha$, we have
\begin{equation}
 a =
\left\{
	\begin{array}{ll}
		 \ln|k\cdot n| & \mbox{for a spacelike intersecting surface}\,, \\
		  \ln|k\cdot s|  & \mbox{for a timelike intersecting surface}\,, \\
		 \ln|k\cdot \bar k/2| & \mbox{for a null intersecting surface}\,.
	\end{array}
\right.
\end{equation}
Further, ${\rm sign}({m_i})= +1$ if the spacetime volume of interest
lies to the future (past) of the null segment and the joint lies at
the past (future) end of the segment; and ${\rm sign}({m_i})= -1$
otherwise -- see Fig.~\ref{fig:Njoints}. 



%

\end{document}